\documentclass[]{aa}

\usepackage{graphicx}
\usepackage{natbib}
\bibpunct{(}{)}{;}{a}{}{,} 

\usepackage{amsmath,amssymb}
\usepackage{url}
\usepackage{bm}
\usepackage[normalem]{ulem}
\usepackage{hyperref}
\usepackage{color,soul}

\usepackage{txfonts}
\usepackage{stfloats}
\newcommand{\chandra}{{\it Chandra}}
\usepackage[dvipsnames]{xcolor}

\newcommand{\bq}{\begin{equation}}
\newcommand{\eq}{\end{equation}}

\DeclareMathAlphabet{\mathitbf}{OML}{cmm}{b}{it}

\newcommand{\kms}{\ensuremath{\mathrm{km}\,\mathrm{s}^{-1}}}

\makeatletter

\newcommand{\Rmnum}[1]{\expandafter\@slowromancap\romannumeral #1@}
\makeatother

\newcommand{\msun}{\ensuremath{\mathrm{M}_{\odot}}}

\newcommand{\msunperyr}{\ensuremath{\msun\,\mathrm{yr}^{-1}}}

\providecommand{\arcsec}{\ensuremath{^{\prime\prime}}}

\newcommand{\pion}{\textsc{pion}}

\usepackage{newtxtext,newtxmath}

\begin{document}

\title{Thermal emission from bow shocks II: }
\subtitle{3D magnetohydrodynamic models of $\zeta$~Ophiuchi}

\author{Samuel Green\inst{1,2} \and Jonathan Mackey\inst{1,2} \and Patrick Kavanagh\inst{1} \and Thomas J.~Haworth\inst{4} \and Maria Moutzouri\inst{1,2,3} \and Vasilii V.~Gvaramadze\inst{5,6}
        }
\offprints{green@cp.dias.ie}
\institute{
  Dublin Institute for Advanced Studies, Astronomy \& Astrophysics Section, 31 Fitzwilliam Place, Dublin 2, Ireland
  \and
  Centre for AstroParticle Physics and Astrophysics, DIAS Dunsink Observatory, Dunsink Lane, Dublin 15, Ireland
  \and
  School of Physics, University College Dublin, Belfield, Dublin 4, Ireland
  \and
  Astronomy Unit, School of Physics and Astronomy, Queen Mary University of London, London E1 4NS, UK
  \and
  Sternberg Astronomical Institute, Lomonosov Moscow State University, Universitetskij Pr.~13, Moscow 119992, Russia
  \and
  Space Research Institute, Russian Academy of Sciences, Profsoyuznaya 84/32, 117997 Moscow, Russia
}

\date{Draft 26.05.2022 / Received DD Month 2022 / Accepted 26 May 2022}

\abstract{The nearby, massive, runaway star $\zeta$ Ophiuchi has a large bow shock detected in optical and infrared light and, uniquely among runaway O stars, diffuse X-ray emission is detected from the shocked stellar wind. 
Here we make the first detailed computational investigation of the bow shock of $\zeta$ Ophiuchi, to test whether a simple model of the bow shock can explain the observed nebula, and to compare the detected X-ray emission with simulated emission maps.
We re-analysed archival {\it Chandra} observations of the thermal diffuse X-ray emission from the shocked wind region of the bow shock, finding total unabsorbed X-ray flux in the 0.3-2 keV band corresponding to a diffuse X-ray luminosity of $L_\mathrm{X}=2.33^{+1.12}_{-1.54}\times10^{29}$~erg~s$^{-1}$, consistent with previous work. The diffuse X-ray emission arises from the region between the star and the bow shock. 3D magnetohydrodyanmic simulations were used to model the interaction of the star's wind with a uniform interstellar medium (ISM) using a range of stellar and ISM parameters motivated by observational constraints. Synthetic infrared, H$\alpha$, soft X-ray, emission measure, and radio 6\,GHz emission maps were generated from three simulations, for comparison with the relevant observations.
Simulations where the space velocity of $\zeta$ Ophiuchi has a significant radial velocity produce infrared emission maps with opening angle of the bow shock in better agreement with observations than for the case where motion is fully in the plane of the sky.
All three simulations presented here have X-ray emission fainter than observed, in contrast to results for NGC\,7635. The simulation with the highest pressure has the closest match to X-ray observations, with flux level within a factor of 2 of the observational lower limit, and emission weighted temperature of $\log_{10} (T_\mathrm{A}/\mathrm{K})=6.4$, although the morphology of the diffuse emission appears somewhat different.
Observed X-ray emission is of a filled bubble brightest near the star whereas simulations predict brightening towards the contact discontinuity as density increases. This first numerical study of the bow shock and wind bubble around $\zeta$~Ophiuchi uses a relatively simple model of a uniform ISM and a ideal-magnetohydrodynamics, and can be used as a basis for comparing results from models incorporating more physical processes, or higher resolution simulations that may show more turbulent mixing.}

%


\maketitle




\section{Introduction}
\label{sec:intro}



Stellar-wind bubbles form around massive stars because the ram pressure of the radially expanding wind pushes the interstellar medium (ISM) outwards \citep{weaver}.
Studying these bubbles provides a constraint on the mass-loss rate, $\dot{M}$, of the driving star(s) by the simple argument that stronger winds drive larger stellar-wind bubbles \citep{gvaramadze2012zeta}.
For stars moving with respect to the surrounding ISM the constraint becomes stronger because the bubble relaxes quickly to a stationary state \citep[e.g.][]{MacVanWoo91, 2006ApJS..165..283A, 1998A&A...338..273C}, even for subsonic motion \citep{mackey2016detecting}.

Runaway massive stars make up $\sim$25$\%$ of all OB stars \citep{Gie87, 1993ASPC...35..207B}.
When these runaway massive stars move supersonically through the ISM they develop an extensive bow shock, produced by the compression of gas and dust from the parent star and the ISM \citep{GulSof79, VanMcC88}.
Intense radiation from the star in question heats and ionizes this gas and dust. This makes these large-scale structures easily observable in optical nebular lines (e.g., H$\alpha$) and mid-infrared (mid-IR) emission \citep{VanMcC88, 2014meyer, 2016MNRAS.456..136A}.
Well-known examples are are $\zeta$~Ophiuchi \citep{GulSof79, gvaramadze2012zeta}, Vela X-1 \citep{KapVanAug97, 2018MNRAS.474.4421G}, and BD+43$^\circ$\,3654 \citep{VanNorDga95, 2007A&A...467L..23C}.

$\zeta$~Ophiuchi is one of the closest O-type stars to Earth at a distance of 135$\pm$3 pc \citep{Gaia2021}.
Its bow shock was discovered in optical narrow-band images \citep{GulSof79} and infrared \citep{VanMcC88} observations.
It is one of the best studied cases of a bow shock around a runaway massive star and became one of the most iconic images from the \emph{Spitzer Space Telescope} \citep{Werner2004}.
Even with its proximity to Earth and its brightness, so far there have been no detailed 2D or 3D simulations of this wind bubble and bow shock.

\begin{table}
	\caption{Summary of the parameters of $\zeta$~Ophiuchi. References: (1) \citet{1989ApJS...69..527H}; (2) \citet{gvaramadze2012zeta}; (3) \citet{Gaia2021}; (4) \citet{2001MNRAS.327..353H}; (5) \citet{2018AN....339...46Z}.} 
	\label{tab:zeta}
	\centering
	\begin{tabular}{l c c} 
		\hline
		Parameter & Value & Refs. \\ [0.5ex] 
		\hline
		Effective  &  & \\
		Temperature ($T_{\ast}$) & 30\,500 K & (1) \\
		Wind velocity ($v_\infty$) & 1500 km\,s$^{-1} $ & (2) \\
		Stellar surface rotation &  & \\
		@ equator, ($v_\mathrm{rot}$) & 400 km\,s$^{-1}$  & (4) \\
		Mass-loss rate ($\dot{M}$) & $2.2^{+4.4}_{-1.1} \times 10^{-8}$ M$_\odot$ yr$^{-1}$ & (2) \\
		Distance ($d$) & $135\pm3$ pc & (3) \\
		Peculiar transverse & & \\
		velocity ($v_{\rm tr}$) & $29\pm2 \, \kms$ & (3) \\ 
		Peculiar radial  &  &  \\ 
		velocity ($v_{\rm r}$) & $-$2.5 or +24.7\,km\,s$^{-1}$ & (3,5) \\
		\hline
	\end{tabular}
\end{table}

$\zeta$~Ophiuchi is thought to have originated in a binary system and have been ejected from this system when its binary companion exploded as a supernova \citep{1993ASPC...35..207B, 2010MNRAS.402.2369T, 2020MNRAS.498..899N}.
During the binary interaction before the supernova event, mass transfer occurs from the primary to the secondary star, leading to the observed rapid rotation and enrichment of He and N.
The significant space velocity of $\zeta$~Ophiuchi \citep{2020MNRAS.498..899N} was imparted when the binary system became unbound after the primary exploded. 

\citet{Toala2016} detected diffuse X-ray emission in the vicinity of $\zeta$~Ophiuchi using \textit{Chandra} observations \citep{Weiss1996}.
They concluded that their detection was thermal X-ray emission and its morphology is consistent with expectations from radiation-hydrodynamic models of wind bubbles produced by moving stars \citep{2015A&A...573A..10M}.
This is so far the only detection of thermal X-ray emission from a stellar wind bubble around an isolated runaway O star.

\citet{Green} investigated the Bubble Nebula, NGC7635, with two-dimensional (2D) hydrodynamic simulations, arguing that the nebula is a bow shock produced by wind-ISM interaction driven by the runaway star BD+60$^\circ$2522.
They made predictions of diffuse thermal X-ray emission, finding that this could be detectable with current instruments.
\citet{2020MNRAS.495.3041T} have since shown that the predicted flux is a factor of $\gtrsim10$ too high, via a non-detection in observations with \textit{XMM-Newton}.
To investigate this discrepency further we here apply our numerical model to the bow shock of $\zeta$~Ophiuchi, to see if our method also overpredicts the thermal X-rays for a bubble with detected emission.

Up until now most simulations of bow shocks \citep{MacVanWoo91, 1998A&A...338..273C, 2014meyer, 2015A&A...573A..10M, Green} and wind bubbles \citep{1996A&A...305..229G, FreHenYor03, 2011ApJ...737..100T} have been 2D because of computational requirements.
Only in the last decade have 3D hydrodynamical (HD) and magnetohydrodynamical (MHD) simulations become feasible.
These were first used to model bow shocks around cool stars with slow winds, where computational requirements are less severe \citep{2007ApJ...660L.129W, 2007MNRAS.382.1233W, 2012A&A...541A...1M, 2021MNRAS.506.5170M}.
Very recently this has also been applied to hot stars with fast winds \citep{2020MNRAS.493.4172S, 2021PION, 2021arXiv210403748B}, also for wind bubbles expanding around young stars in molecular clouds \citep{2021MNRAS.501.1352G, 2015MNRAS.454..238W}.
In \citet{2021PION} we developed and tested a method for 3D, nested-grid, MHD simulations of bow shocks around runaway stars that are rotating and magnetised, based on established methods from Heliospheric modelling \citep{PogZanOgi04}.

Building on our results from 2D hydrodynamical simulations of NGC 7635, here we investigate the bow shock of $\zeta$~Ophiuchi with 3D MHD simulations.
Our aim is to determine whether models can produce X-ray emission comparable with the detected diffuse emission around $\zeta$~Ophiuchi, while simultaneously matching the infrared emission.
In Sect. 2, we review infrared observational data on the bow shock formed by $\zeta$~Ophiuchi and re-analyse \textit{Chandra} archival observations to investigate the thermal diffuse X-ray emission from the nebula formed by $\zeta$~Ophiuchi.
In Sect.~3 - 3.1, we present the numerical methods and simulation setup.
We describe our three simulations in Sect. 3.1.1 - 3.1.3 and describe the features of each simulation.
In 3.2 the global properties of the simulation X-ray emission are described.
In Sect.~ 4 we use the three simulations to produce synthetic infrared, H$\alpha$, soft X-ray, emission measure, and radio 6GHz emission maps and compare them with the relevant observations.
We discuss our results in Sect.\,\ref{sec:discusion} and conclude in Sect.\,\ref{sec:conclusions}.



\section{Observational data}
\label{sec:zeta_obs}

In Table~\ref{tab:zeta} we show a summary of some relevant physical parameters of $\zeta$~Ophiuchi. 
The velocity with which $\zeta$~Ophiuchi is moving through the ISM is somewhat uncertain, and new measurements lead to some revision \citep{2020MNRAS.498..899N} compared with the peculiar velocity estimate, $v_\star = 26.5$\,km\,s$^{-1}$, calculated by \citet{gvaramadze2012zeta}.
To calculate the peculiar transverse velocity ($v_\mathrm{tr}$) in the plane of the sky, and the peculiar radial velocity ($v_\mathrm{r}$) we follow the method described in \citet{Green} but using the \emph{Gaia} EDR3 data.
We use an updated distance to the Galactic Centre of $8.15\pm0.11$\,kpc \citep{2021ARep...65..498B} and the circular Galactic rotation velocity of 240 km\,s$^{-1}$ \citep{2009ApJ...705.1548R}.
The Solar peculiar velocity \citep{2010MNRAS.403.1829S} is taken to be $(U_\odot, V_\odot, W_\odot) = (11.1, 12.2, 7.3)$ km\,s$^{-1}$.
Using this we obtain the values of $v_\mathrm{tr}$ and $v_\mathrm{r}$ in Table~\ref{tab:zeta}.
For $v_\mathrm{r}$ we use the Heliocentric radial velocity quoted in \citet{gvaramadze2012zeta} and obtain $v_\mathrm{r}=-2.5$\,km\,s$^{-1}$.
Recently, however, \citet{2018AN....339...46Z} reviewed previous radial velocity measurements and provide an updated value corresponding to  $v_\mathrm{r} = 24.7$ km\,s$^{-1}$, i.e., the vector of the peculiar velocity is inclined by an angle of about 40 degrees to the plane of sky.
If correct, this radial velocity would imply a factor of 1.3 higher peculiar velocity, $v_\star = \sqrt{v_\mathrm{r}^2+v_\mathrm{tr}^2} = 38$ km\,s$^{-1}$.

All of the parameters shown in Table~\ref{tab:zeta} are relevant because they are used as initial conditions in our simulations, but there is disagreement in the literature on the values of $\dot{M}$ and $v_{\rm r}$. 
Table 2 from \citet{gvaramadze2012zeta} gives estimates for the $\dot{M}$ for $\zeta$~Ophiuchi using a variety of established methods in the literature and the values range from $10^{-7} - 10^{-9}$ M$_\odot$ yr$^{-1}$.
Table 1 from \citet{2018AN....339...46Z} gives estimates for the $v_{\rm r}$ for $\zeta$~Ophiuchi using a variety of established methods in the literature and the values range from -35.0 - +30.0 \kms.
$\dot{M}$ and $v_{\rm r}$ are important parameters that affect our modelling of the $\zeta$~Ophiuchi bow shock with both being able to affect the size and brightness.
Until these values are constrained better, this makes it difficult to accurately model a bow shock.

\subsection{Infrared observations}
\label{sec:zeta_oir_obs}

\begin{figure}
	\centering
	\includegraphics[height=.37\textwidth]{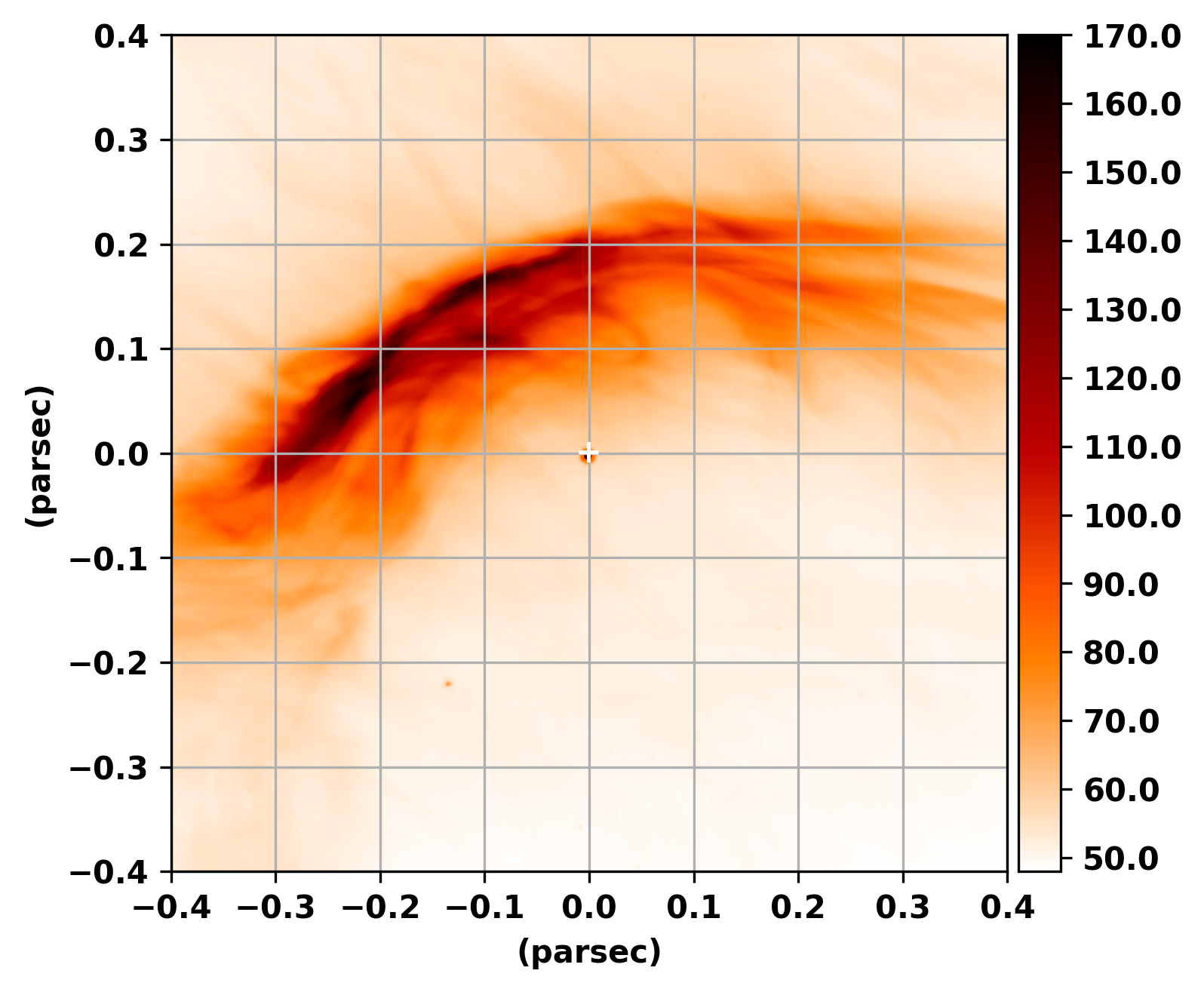}
	\caption{\emph{Spitzer Space Telescope} 24\,$\mu$m observational data of $\zeta$~Ophiuchi in units of MJy ster$^{-1}$. The star is at the origin with coordinates in parsecs relative to the position of the star, for a distance of 135\,pc.}
	\label{fig:obs_dust}    
\end{figure}


Fig.~\ref{fig:obs_dust}  shows 24$\mu$m observational Infrared data from the \emph{Spitzer Space Telescope}.
This 24$\mu$m image was obtained from the NASA/IPAC infrared science archive\footnote{http://irsa.ipac.caltech.edu/}.
The bow shock was observed on 2007-04-11 as part of Program Id.~30088 (PI:
A.~Noriega-Crespo), using the Multiband Imaging Photometer for \textit{Spitzer}
(MIPS; \citealt{2004ApJS..154...25R}). 
At 24$\mu$m the angular resolution is 6 arcsec.
The maximum brightness of the 24$\mu$m data is 170 MJy ster$^{-1}$. 
We estimated the background emission observed in the 24$\mu$m \textit{Spitzer} data of 47 MJy ster$^{-1}$.


The \emph{standoff distance}, $R_0$, is given by \citep{1970DoSSR.194...41B}: 
\begin{equation}
	R_0= \sqrt{\frac{\dot{M}v_\infty}{4\pi\rho_\text{ISM}(v^2_{\ast}+c_\text{s}^2)}} \, ,
	\label{standoff}
\end{equation}
where $\dot{M}$ is the mass-loss rate of the stellar wind, $v_\infty$ is the wind terminal velocity, $\rho_\text{ISM}$ is the density of the ISM, $v_\star$ is the star's space velocity, and c$_s$ is the sound speed in the ISM. By using the visible bow shock size in observational data it is possible to measure a value for $R_0$ and in turn estimate the value of other variables in equation 5.1.
In \citet{gvaramadze2012zeta} this $R_0$ is measured from IR observations to be $R_0=0.16$\,pc for d=112pc.
From this \citet{gvaramadze2012zeta} derived $\dot{M}=2.2\times10^{-8}\,\msunperyr$, by using extra constraints on the local gas density from the size of the H~\textsc{ii} region around $\zeta$ Oph.
\citet{GulSof79} derived a very similar value $\dot{M}=2.3\times10^{-8}\,\msunperyr$ by using radio and H$\alpha$ estimates of the ISM gas density. 

Using the observational data presented in Fig.~\ref{fig:obs_dust} we obtain $R_\mathrm{CD}=202$\,arcsec, and $R_\mathrm{FS}=405$\,arcsec. 
Using the distance of 112\,pc as in \citet{gvaramadze2012zeta} this gives $R_\mathrm{CD}=0.11$\,pc and $R_\mathrm{FS}=0.22$\,pc.
Witht the updated \emph{Gaia} EDR3 distance of 135\,pc, we obtain $R_\mathrm{CD}=0.13$\,pc and $R_\mathrm{FS}=0.26$\,pc. The relative thickness of the bow shock, $\frac{\Delta{R}}{R} = \frac{R_{FS}-R_{CD}}{R_{FS}}$, is 0.5.
For all figures, including Fig.~\ref{fig:obs_dust}, we use $d=135$\,pc to convert angular to linear scales.

\begin{figure}
	\centering
	\includegraphics[height=.32\textwidth]{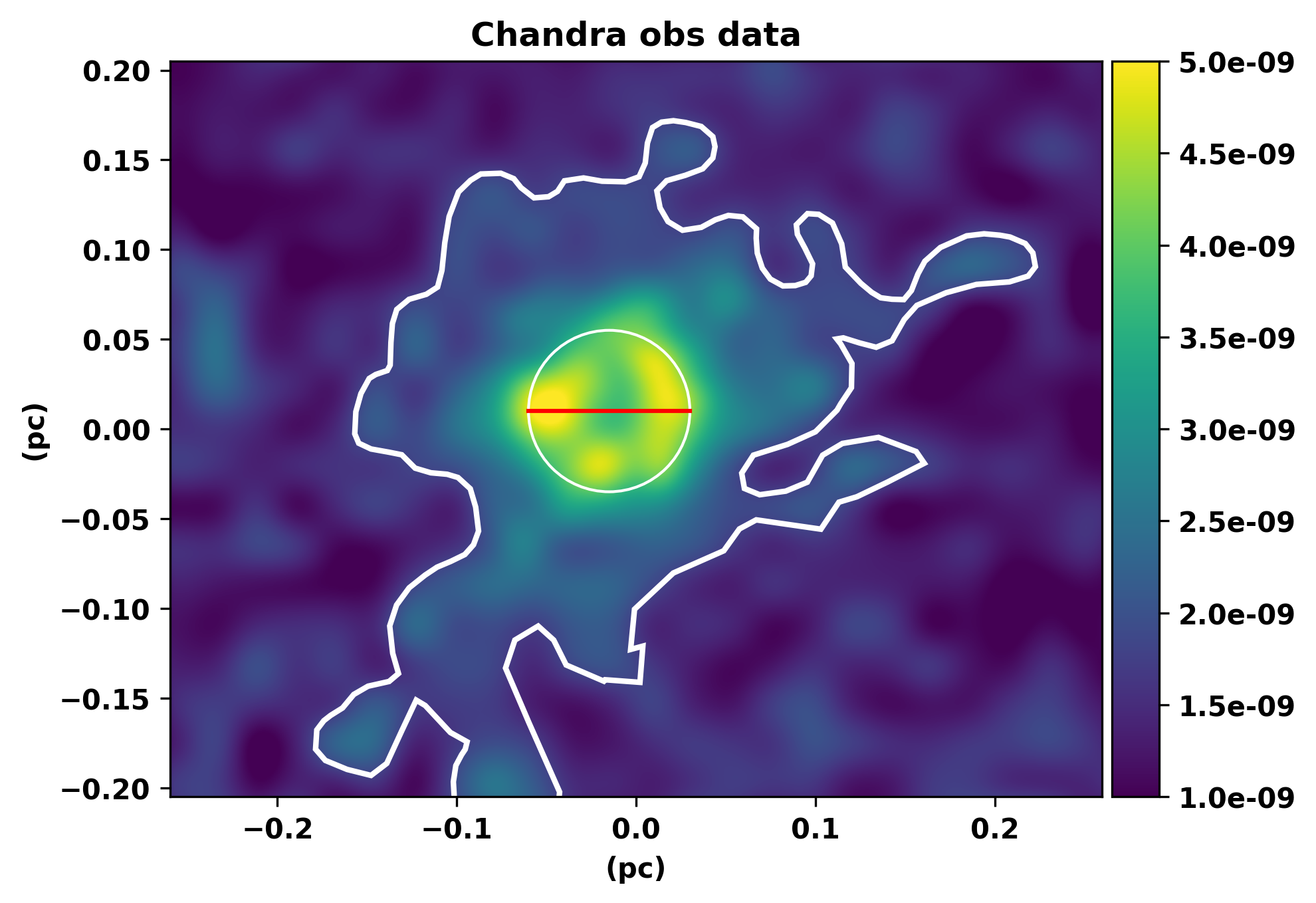}
	\caption{{\it Chandra} diffuse X-ray emission around $\zeta$~Ophiuchi in units of counts s$^{-1}$ pixel$^{-1}$ with a 3$\sigma$ contour line (0.3 - 2 KeV) overlayed. Coordinates are in parsecs relative to the position of the star assuming the distance to the star is 135\,pc.  The star is at the centre of the white circle and its emission has been subtracted from the image (see Sect. 2.2 for details).}
	\label{xray_obs}
\end{figure}

\subsection{X-ray observations} \label{sec:xray-obs}

To compare the observed diffuse X-rays around $\zeta$~Ophiuchi to our simulations, 
we used the \chandra\ observations presented in \citet{Toala2016}. This \chandra\ observation was performed on 2013 July~3 (Obs. ID 14540, PI L.~M. Oskinova) with the Advanced CCD Imaging Spectrometer I-array \cite[ACIS-I,][]{Garmire2003} 
as the primary instrument. We reduced and analysed the \chandra\ observations using the
CIAO~v4.12.1\footnote{See \url{http://cxc.harvard.edu/ciao/}} \citep{Fru2006}
software package with CALDB~v4.9.3\footnote{See
	\url{http://cxc.harvard.edu/caldb/}}. The dataset was reduced using the
contributed script \texttt{chandra\_repro}, resulting in a flare-filtered and dead time-corrected exposure time of 72.1~ks. 

As noted by \citet{Toala2016}, $\zeta$~Ophiuchi is such a bright X-ray emitter that it resulted in photon pile-up at the source location and also a readout streak along the ACIS-I detector column (their Fig.~2, left). A readout streak can have adverse effects on both source detection and diffuse emission analysis so we removed the readout streak before proceeding with our analysis. We did this by identifying the streak region using the \texttt{acis\_streak\_map} CIAO task, extracting a streak background spectrum from a streak-parallel background region free of point sources using \texttt{dmextract}, and correcting the streak using \texttt{acisreadcorr} with the streak region, background spectrum and the sky X and Y position of $\zeta$~Ophiuchi. 

With the readout streak removed, we produced an image of the diffuse X-ray emission by broadly following the CIAO analysis thread `An Image of Diffuse Emission'\footnote{\url{https://cxc.cfa.harvard.edu/ciao/threads/diffuse_emission/}}, with some minor differences. We identified point sources in the field using the CIAO task \texttt{wavdetect} with a point spread function (PSF) map generated using \texttt{mkpsfmap} with an encircled counts fraction (ECF) of 0.95 in ACIS energy range of 0.3--8~keV. The resulting source regions can be used to excise the point sources from the image and synthetically fill the source regions with realistic background using the \texttt{dmfilth} task. However, given the brightness of $\zeta$~Ophiuchi, we treated this source differently as counts in the wings of its PSF could easily contaminate any faint diffuse emission in its surroundings. We replaced the $\zeta$~Ophiuchi source region in the \texttt{wavdetect} source list with a conservative region determined for an ECF of 0.99 and multiplied by a factor of 2. Unfortunately, the large `hole' left when excising $\zeta$~Ophiuchi and the counts still present in its extended PSF wings caused problems for \texttt{dmfilth}, i.e. the background used to fill the hole comprised mostly source counts in its extended PSF wings. Therefore, any apparent structure at or inside the $\zeta$~Ophiuchi excision region should be interpreted as nothing more than an artefact of the \texttt{dmfilth} task. 

We then generated a counts image of the diffuse emission with 2$\arcsec$ bins in the 0.3--2~keV range, before creating the final exposure corrected and adaptively smoothed image using the \texttt{dmimgadapt} task using a minimum smoothing scale of $10\arcsec$. The image region around $\zeta$~Ophiuchi is shown in Fig.~\ref{xray_obs}. It is clear from the image that the counts in the wings of the $\zeta$~Ophiuchi PSF are dominating the apparent diffuse emission close to the star. However, this appears to be immersed in a larger scale, faint emission extending in a roughly southeast-northwest direction. We interpret this as diffuse emission associated with the mixing region as detected by \citet{Toala2016}. We defined its extent as regions where the 0.3--2~keV count rate per pixel rises to 3$\sigma$ above the average background, determined from nearby source free regions. This is shown by the contour in Fig.~\ref{xray_obs}. 

To correctly derive a clean, diffuse surface brightness image to compare to our synthetic X-ray emission maps would require the separation of photons from the $\zeta$~Ophiuchi PSF and diffuse emission components, which is not possible using an imaging analysis. Therefore, we performed a spectral analysis, where the $\zeta$~Ophiuchi and diffuse emission could be fitted as separate components. This allowed us to quantify the diffuse flux contribution, and hence the average surface brightness within the extended emission contour. 

The extended and background spectra, and weighted response files were created using the CIAO task \texttt{specextract}. These were fitted using XSPEC \citep{Arnaud1996} version 12.11.1 with abundance tables set to those of \citet{Wilms2000}, photoelectric absorption cross-sections set to those of \citet{Bal1992}, and atomic data from ATOMDB~3.0.9\footnote{\url{http://www.atomdb.org/index.php}}. Detected point sources were masked. Additional masking was applied to the region around $\zeta$~Ophiuchi to reduce the contribution of the photons in the PSF wings to the spectra, shown by the white exclusion region in Fig.~\ref{xray_obs}. 

When fitting X-ray spectra of faint extended X-ray emission, careful consideration and treatment of the background is necessary. The background components present in all \chandra\ spectra can be split into two categories: the astrophysical X-ray background (AXB) from photons passing through the telescope optical system, and the detector, non-X-ray background (NXB). To constrain the NXB, comprising high-energy particles, fluorescent emission, etc., we used the ``stowed" background data and analytical model for the ACIS-I CCDs \citep[see][]{Bartalucci2014}. We reprocessed and reprojected the stowed event file to match the $\zeta$~Ophiuchi observation using the CIAO tasks \texttt{acis\_process\_events} and \texttt{reproject\_events}, respectively. Stowed spectra were extracted from the same extended emission and background regions on the detector which were used to constrain the NXB contributions in each. 

The AXB typically comprises four or fewer components \citep{Snowden2008,Kuntz2010}, namely the unabsorbed thermal emission from the Local Hot Bubble (LHB), absorbed cool and hot thermal emission from the Galactic halo (CGH, HGH, respectively), and an absorbed power law representing unresolved background active galactic nuclei (AGN). Detailed descriptions of the components and their model parameters are given in, e.g., \citet{Kavanagh2020}. The halo and AGN components are also absorbed by material in the Galaxy. The foreground Galactic absorption component was fixed at a column density of $N_{\rm{H}}=6.0\times10^{21}$~cm$^{-2}$ based on the \citet{Dickey1990} HI maps, determined using the HEASARC $N_{\rm{H}}$ Tool\footnote{\url{http://heasarc.gsfc.nasa.gov/cgi-bin/Tools/w3nh/w3nh.pl}}. To constrain the AXB in the region of $\zeta$~Ophiuchi, we fitted our source and background spectra simultaneously in XSPEC, linking and scaling the AXB model parameters as necessary, and using the stowed spectra fits to constrain the NXB in each. We found that the CGH component was not statistically required so we omitted this component from the final model fits. This is unsurprising as the high foreground absorption through the Galaxy effectively absorbs the cool halo emission. 

For the extended emission region, we assumed that the spectrum comprises real diffuse emission as well as a contribution from photons in the $\zeta$~Ophiuchi PSF wings. The X-ray spectral parameters were already well constrained in the analysis of \citet{Toala2016} using both \chandra\ and {\it Suzaku} data so we adopt the same model, i.e. a thermal plasma plus power law (\texttt{apec+powerlaw} in XSPEC) for $\zeta$~Ophiuchi with the ratio of the normalisations set to 0.93  and a thermal plasma (\texttt{apec}) for the diffuse emission, with temperature/photon index fixed. 

\begin{figure*}
	\centering
	\includegraphics[height=.33\textwidth]{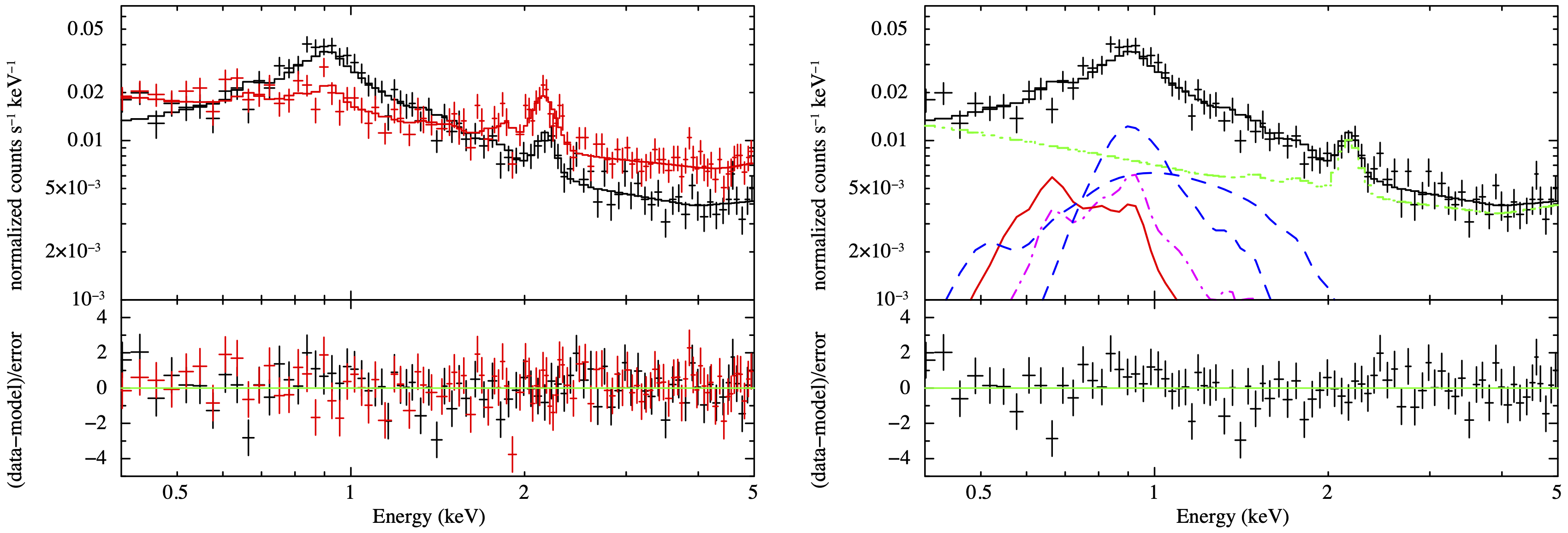} \\
	\caption{{\it Left}: \chandra\ X-ray spectra of the extended emission region (black) and the background (red). The best fit models are shown by the solid lines and the fit residuals are shown in the lower panel. {\it Right}: The extended emission spectrum only with the additive model components shown. The green dash-dot-dot line represent the combined NXB components constrained using the stowed observations, the dashed blue lines shows the $\zeta$~Ophiuchi emission components, the magenta dash-dot lines mark the AXB components, and the red solid line shows the diffuse emission component.}
	\label{xray_spec}
	
\end{figure*}

The best-fit results are given in Table~\ref{chandra_spec_results} and and the spectra are shown in Fig.~\ref{xray_spec}. It is clear from our results and Fig.~\ref{xray_spec} that in spite of our efforts to reduce the contamination by the $\zeta$~Ophiuchi PSF photons, they still contribute significantly to the spectrum (blue dashed lines in Fig.~\ref{xray_spec}). Nevertheless, the diffuse emission component is observed (red line in Fig.~\ref{xray_spec}). We determined its de-absorbed flux in the 0.3--2~keV range to be $1.07^{+0.51}_{-0.71}\times10^{-13}$~erg~cm$^{-2}$~s$^{-1}$ which is equivalent to a luminosity of $2.33^{+1.12}_{-1.54}\times10^{29}$~erg~s$^{-1}$ for $d = 135pc$. Taking into account the sky area of the spectral extraction region, this corresponds to an average surface brightness inside the contour in Fig.~\ref{xray_obs} of $2.04^{+0.82}_{-1.35}\times10^{-18}$~erg~cm$^{-2}$~s$^{-1}$~arcsec$^{-2}$.   

\begin{table*}
	\caption{X-ray spectral fit results. See Sect. 2.2 for details. The quoted AXB normalisations are for the extended emission spectrum. The numbers in parentheses are the 90\% confidence intervals.}
	\begin{center}
		\label{chandra_spec_results}
		\begin{tabular}{llr}
			\hline
			Component & Parameter & Value\\
			\hline
			\hline
			\multicolumn{3}{c}{AXB}\\
			\hline
			\multicolumn{3}{c}{ }\\
			
			\texttt{apec} (LHB) & k$T$ & 0.1 (fixed) \\
			& norm ($10^{-5}$~cm$^{-5}$) & 1.45~(<9.23)  \\
			
			\multicolumn{3}{c}{ }\\
			
			\texttt{tbabs} & $N_{\rm{H,Gal}}$ ($10^{22}$ cm$^{-2}$) &   0.53 (fixed)  \\
			
			\multicolumn{3}{c}{ }\\
			
			\texttt{apec} (HGH) & k$T$ & 0.19 (0.12--0.24) \\
			& norm ($10^{-4}$~cm$^{-4}$) & 4.66 (1.90--82.29) \\
			
			\multicolumn{3}{c}{ }\\
			
			\texttt{powerlaw} (AGN) & $\Gamma$ & 1.47 (fixed) \\
			& norm ($10^{-6}$~ph~keV$^1$~cm$^2$~s$^1$) & 5.74 (fixed) \\
			
			\hline
			\multicolumn{3}{c}{Extended emission}\\
			\hline
			\multicolumn{3}{c}{ }\\
			
			\texttt{tbabs} & $N_{\rm{H,Gal}}$ ($10^{22}$ cm$^{-2}$) &  0.06 (fixed)  \\
			
			\multicolumn{3}{c}{ }\\
			
			\texttt{apec} ($\zeta$~Ophiuchi) & k$T$ & 0.80 (fixed) \\
			& norm$_{\rm{\texttt{apec}}}$ ($10^{-5}$~cm$^{-5}$) & 2.86 (2.52--3.22)  \\
			
			\multicolumn{3}{c}{ }\\
			
			\texttt{powerlaw} ($\zeta$~Ophiuchi) & $\Gamma$ & 3.05 (frozen) \\
			& norm$_{\rm{\texttt{pl}}}$ ($10^{-5}$~ph~keV$^1$~cm$^2$~s$^1$) & ($1.075\times$norm$_{\rm{\texttt{apec}}}$) \\
			
			\multicolumn{3}{c}{ }\\
			
			\texttt{apec} (diffuse) & k$T$ & 0.20 (fixed) \\
			& norm ($10^{-5}$~cm$^{-5}$) & 8.70 (2.94--12.82) \\
			
			\multicolumn{3}{c}{ }\\
			
			Fit statistic & \texttt{$\chi^{2}$/dof}  & 266.38/268 \\
			
			\multicolumn{3}{c}{ }\\
			\hline
		\end{tabular}
	\end{center}
\end{table*}%

\begin{figure}
	\centering
	\includegraphics[height=.45\textwidth]{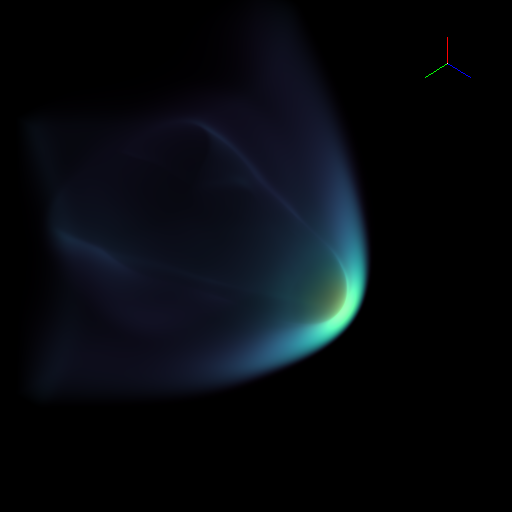} \\
	\includegraphics[height=.25\textwidth]{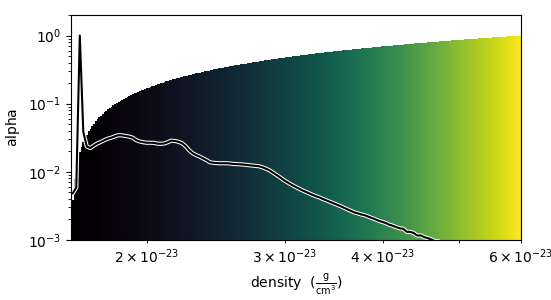}
	\caption{Top: 3D volumetric density image from the Z01 $\zeta$~Ophiuchi simulation in Sect. 3.1.1, with the colour scale showing gas density. Bottom: Transfer function over-plotted on the colour scale showing the alpha value assigned to each density value for the top image. The alpha parameter describes the opacity of the gas as a function of density.}
	\label{fig:3D}
\end{figure}

\begin{table*}
	\centering
	\caption{Stellar and interstellar medium parameters used for simulations Z01, Z02, and Z03.}References: (1) \citet{2001MNRAS.327..353H}; (2) \citet{2001MNRAS.327..353H}; (3) \citet{BagFosLan15}
	\label{tab:sims_zeta}
	\begin{tabular}{l r r r r r}
		\hline
		Parameter & Z01 values & Z02 values & Z03 values & Units & Refs. \\ [0.5ex]
		\hline
		Mass-loss rate ($\dot{M}$) & 1.46 $\times 10^{-8}$ & $1.5 \times 10^{-8}$ & $3\times 10^{-8}$ & M$_\odot$ yr$^{-1}$ & (1) \\
		Wind velocity ($v_\infty$) & 1500 & 1500 & 1500 & km\,s$^{-1}$ & (1) \\
		Stellar surface rotation & & & & & \\
		@ equator, ($v_{rot}$) & 400 & 400 & 400 & km s$^{-1}$ & (2) \\
		Velocity of star, $v_\star$ & 26.5 & 40 & 40 & $\kms$ & \\
		Surface split-monopole & & & & & \\
		field strength, (|$\bm{B}$|) & 1 & 1 & 1 & G & (3) \\
		ISM B-field, ($\bm{B}_0$) & [5, 8.66, 1] & [5, 8.66, 1] & [5, 8.66, 1] & $\mu$G & \\
		ISM density, ($n_\text{i}$) & 8 & 4 & 8 & cm$^{-3}$ & \\
		\hline
	\end{tabular}
	
\end{table*}

\begin{figure*}
	\centering
	\includegraphics[height=.5\textwidth]{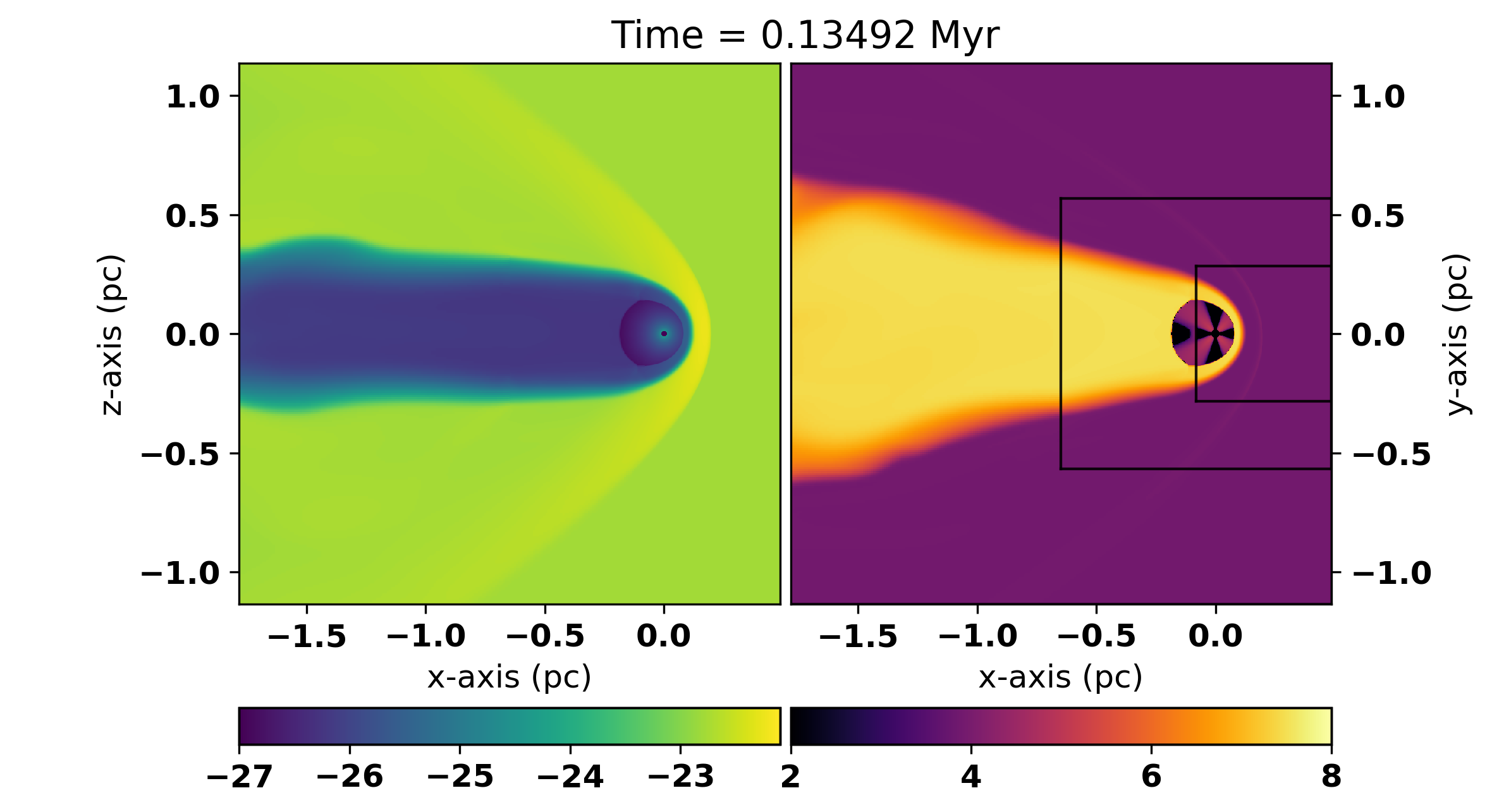}
	\caption{Left: Plots of the Log$_{10}$ gas density (g cm$^{-3}$) of simulation Z01, slice through the ($x$-$z$) plane at y = 0. Right: Plots of the Log$_{10}$ gas temperature (K), slice through the ($x$-$y$) plane at z = 0. The borders of the nested-grid levels are shown in black. Star is at the origin.}
	\label{fig:den}    
\end{figure*}

\section{Computational methods and simulation properties}
\label{sec:meth}

We use the \pion{} radiation-MHD code \citep{2021PION} to model $\zeta$~Ophiuchi as an O star moving through the interstellar medium and emitting a stellar wind. \pion{} solves the MHD equations on a computational grid in the $(x,y,z)$ 3D space. The mass, energy, momentum densities, and magnetic vector $\bm{B}$ are defined at the centre of each computational cell, and evolved with time according to the ideal MHD equations. The $\nabla \cdot \bm{B} = 0$ constraint is maintained using a modified version of the \citet{2002JCoPh.175..645D} divergence-cleaning method with some improvements from \citet{DerWinGas18}, plus the \citet{PowRoeLin99} source terms. Further details of the MHD implementation can be found in \citet{2021PION}.

For numerical convenience, a reference frame in which the star is stationary and located at the origin $(x, y, z)=(0,0,0)$ of a rectangular cuboid is chosen. The ISM flows past the star in the negative $x$-direction, interacting with the stellar wind as it does so. A passive scalar variable is used to distinguish between the ISM and wind gas. For the sake of simplicity, the ISM is assumed to be homogeneous. 
A split-monopole magnetic field is also set up at the star where its rotation winds up the magnetic field into a Parker spiral aligned to the z-axis.

Static mesh-refinement, in the form of a multiple nested grid, has been used in our simulations. This allows us to model the apex of the bow shock at high resolution, and regions further from the star at lower-resolution (e.g. the tail of the bow-shock and wake).

We used the Monte Carlo radiation transport and hydrodynamics code TORUS \citep{2019A&C....27...63H} to calculate a synthetic infrared emission map (at wavelength 24$\mu$m, in MJy ster$^{-1}$) from our simulations. The procedure used is described in \citep{mackey2016detecting, Green} but modified to work for 3D nested-grid simulations as described in \citet{2021PION}. 

To calculate the infrared emission maps we assume a gas-to-dust mass ratio of 160 \citep{2004ApJS..152..211Z}, which is comprised of 70\% silicates \citep{2003ApJ...598.1026D} and 30\% carbonaceous \citep{1996MNRAS.282.1321Z} grains. We assume that no polycyclic aromatic hydrocarbons survive within the H~\textsc{ii} region. For both the silicate and carbonaceous grains we assume minimum and maximum grain sizes of 0.005 and 0.1\,$\mu$m respectively. The size distribution between these limits is a power law $\textrm{d}n/\textrm{d}a \propto a^\text{-q}$ \citep{1977ApJ...217..425M}, where we take $q=3.3$. For the stellar spectrum we use a \citet{1993PhST...47..110K} spectral model with the same temperature and mass as $\zeta$~Ophiuchi. The dust is also removed wherever the temperature is more than 10$^6$\,K, this is to ensure a dust-free wind. 

An independent ray-tracing method to calculate synthetic images from {\sc pion} simulations was developed to allow us to compare simulations with real observational data extending the uniform-grid methods used in \citet{MacLanGva13}. For a 3D Cartesian grid, straight lines are projected through the 3D simulation at an angle $\theta$ to one of the coordinate axes, and emission from each cell is added along the ray according to the local quantities in that cell (i.e. density, temperature, etc.). The intensity along the rays are assigned to pixels on a 2D image to compare with observational data. Local absorption within the simulation domain is not included.

\begin{figure}
	\centering
	\includegraphics[height=.38\textwidth]{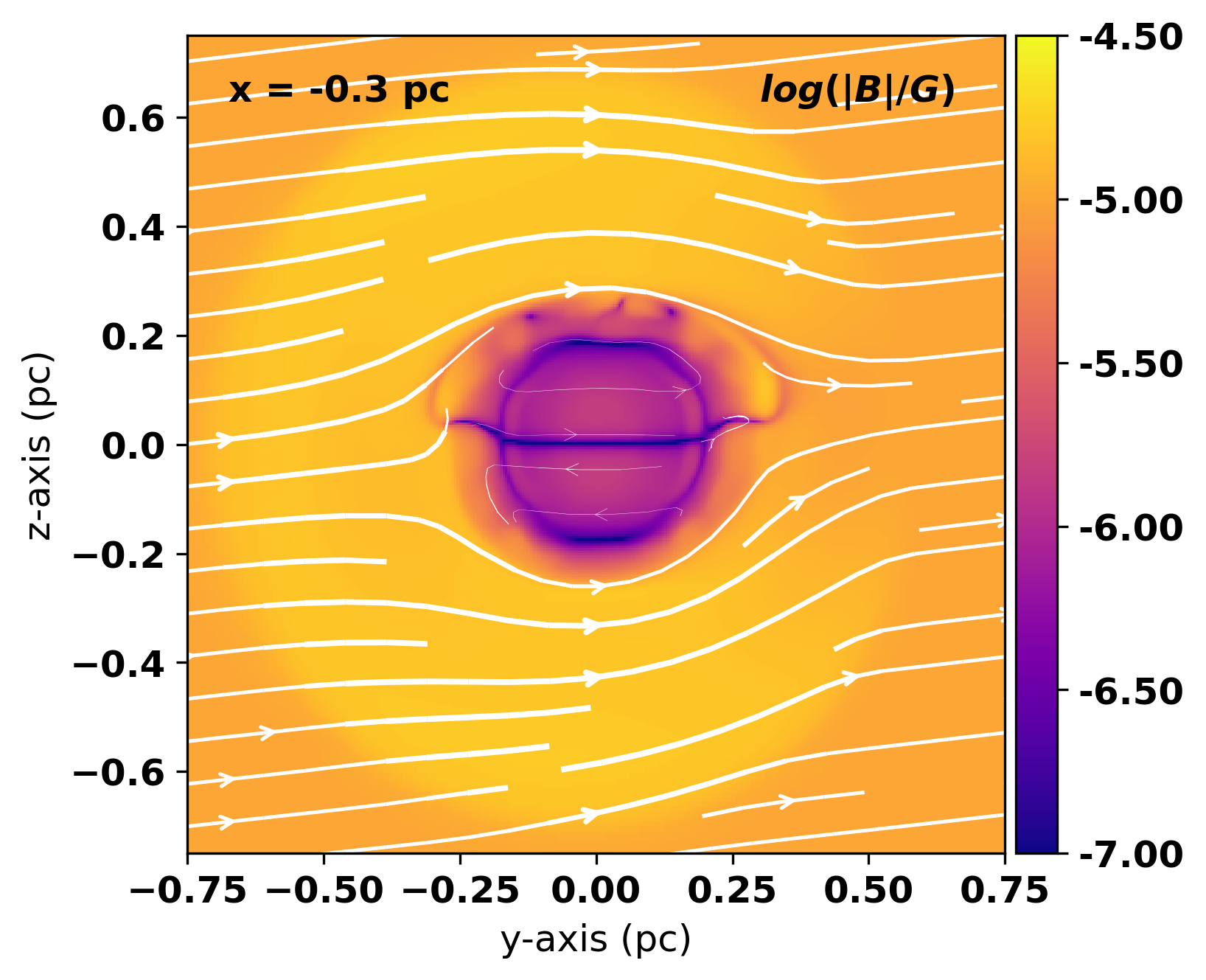}
	\includegraphics[height=.38\textwidth]{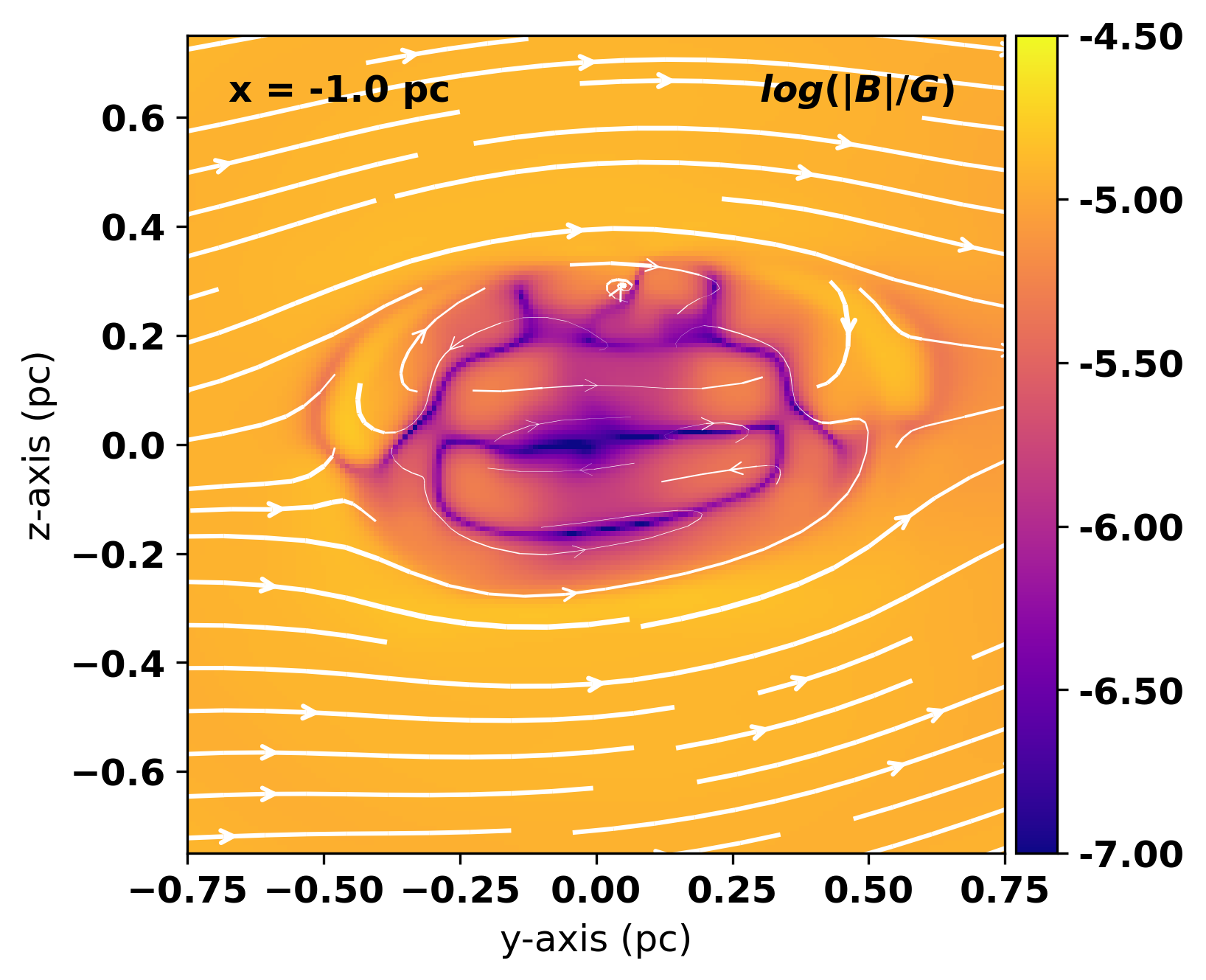}
	\caption{Plots of the Log$_{10}$ magnetic field magnitude (|$\bm{B}$|/G). Both plots are of Z01 in the ($y$-$z$) plane at x = -0.3 pc (top) and x = -1.0 pc (bottom). The streamlines overlaid in both panels show the direction of the magnetic field in this plane.}
	\label{fig:mag}    
\end{figure}

\subsection{Simulations Z01, Z02, and Z03}
\label{sec:z01}

We used features from the Infrared observations (Fig.~\ref{fig:obs_dust}) to 
constrain the simulation parameters; specifically the brightness, and the measured distances $R_\mathrm{CD}$ and $R_\mathrm{FS}$.
To create a model that best matched the size and infrared brightness of $\zeta$~Ophiuchi, several different 3D simulations were run, 
varying the uncertain parameters $v_\star$, $\dot{M}$, and $\rho_\text{ISM}$ while still producing a bow shock with approximately the right size and shape.
Table \ref{tab:sims_zeta} shows the parameters used to run the three simulations presented in the following sections.
All three simulations have a box length of $L_\mathrm{box}=7\times10^{18}$\, cm in each dimension (or $2.27^3$\,pc$^3$), resolved with $256^3$ grid cells per level, with three refinement levels as shown in Fig.~\ref{fig:den}. 
The focus of the mesh refinement is $(x,y,z)=(1.5\times10^{18},0,0)$\,cm.
On the finest level the cell size is $\Delta x=6.84\times10^{15}$\,cm, or $2.2\times10^{-3}$\,pc.

For the three simulations we impose a stellar surface field of 1\,G in a split-monopole configuration.
This is well below the non-detection of $\langle B_z \rangle = 118\pm61$\,G found by \citet[][tab.~5]{BagFosLan15}, and also low enough that the star does not have a magnetosphere.
For the ISM magnetic field we take $|\mathbf{B}_0|=10\,\mu$G, oriented partly in the direction of motion of the star and partly in the perpendicular plane: $\mathbf{B_0}=(5.0,8.66,1.0)\,\mu$G.
This is stronger than the typical magnetic field in the Galactic Plane, but the ISM density that we use is also a few times above the mean density in the Plane. The motivation for certain parameters in each of the three simulation are discussed further in Sect. 4.

A projected image of the gas density is shown in Fig.~\ref{fig:3D} with rotations of 45$^\circ$ to the $x$- and the $y$-axis.
The transfer function is chosen to highlight the contact discontinuity and the apex of the bow shock, and is plotted on the colour scale.
This is not intended to mimic an observable quantity such as spectral-line or continuum emission, only to show the 3D shape of the bow shock.

\subsubsection{Simulation Z01}
The Z01 simulation uses $v_\star=26.5$\,km\,s$^{-1}$, i.e., the case where the line-of-sight radial velocity is very small and the bow shock is seen almost edge-on.
We use an ISM number density of ions, $n_\mathrm{i}=8$\,cm$^{-3}$ and a mass-loss rate of $\dot{M}=1.46\times10^{-8}\,\mathrm{M}_\odot\,\mathrm{yr}^{-1}$.

The Z01 simulation was run for 0.13492 Myr which is just over 1.5 crossing times ($L_{\mathrm{box}}/v_\star$), and 50 dynamical timescales of the bow shock ($R_0/V_\star$). 
Numerical results from simulation Z01 are shown in Fig.~\ref{fig:den}, after 0.13492\,Myr, when the bow shock has reached equilibrium (i.e. the overall shape of the bow-shock/bubble doesn't change and has reached a stationary state).
The left panel shows $\log_{10}$ of the gas density (g cm$^{-3}$) in a slice at $y = 0$ in the ($x,z$) plane, and the right panel shows $\log_{10}$ of the gas temperature for a slice at $z = 0$ in the ($x,y$) plane.
The boundaries of the two refined grid-levels are shown in the right panel of the figure. Each level of refinement is a factor of 2 more refined than the previous level. The wind injection region is a sphere of radius 20 cells. 

At the apex of the bow shock, the peak density is $\rho = 5.6 \times 10^{-23}$ g\,cm$^{-3}$. The density remains at approximately this value throughout the simulation because the bow shock is dynamically stable.
The temperature at the apex of the bow shock is also in equilibrium with the rest of the ISM ($\sim 8.2 \times 10^{3}$\,K), i.e., the forward shock is approximately isothermal.
Some shock heating is evident from the faint outline of the forward shock visible in the temperature plot.

The contact discontinuity can clearly be seen in the density and temperature plots as both quantities change by a factor of $\sim10^3$ from the shocked ISM to the shocked wind. Inside the shocked wind region, the gas density is as low as $6.5 \times 10^{-27}$ g\,cm$^{-3}$ and as high as $3 \times 10^{-25}$ g\,cm$^{-3}$. From this simulation we measure $R_\mathrm{TS}=0.07$\,pc,  $R_\mathrm{CD}=0.11$\,pc, and $R_\mathrm{FS}=0.22$\,pc. 


\begin{figure}
	\centering
	\includegraphics[height=.40\textwidth]{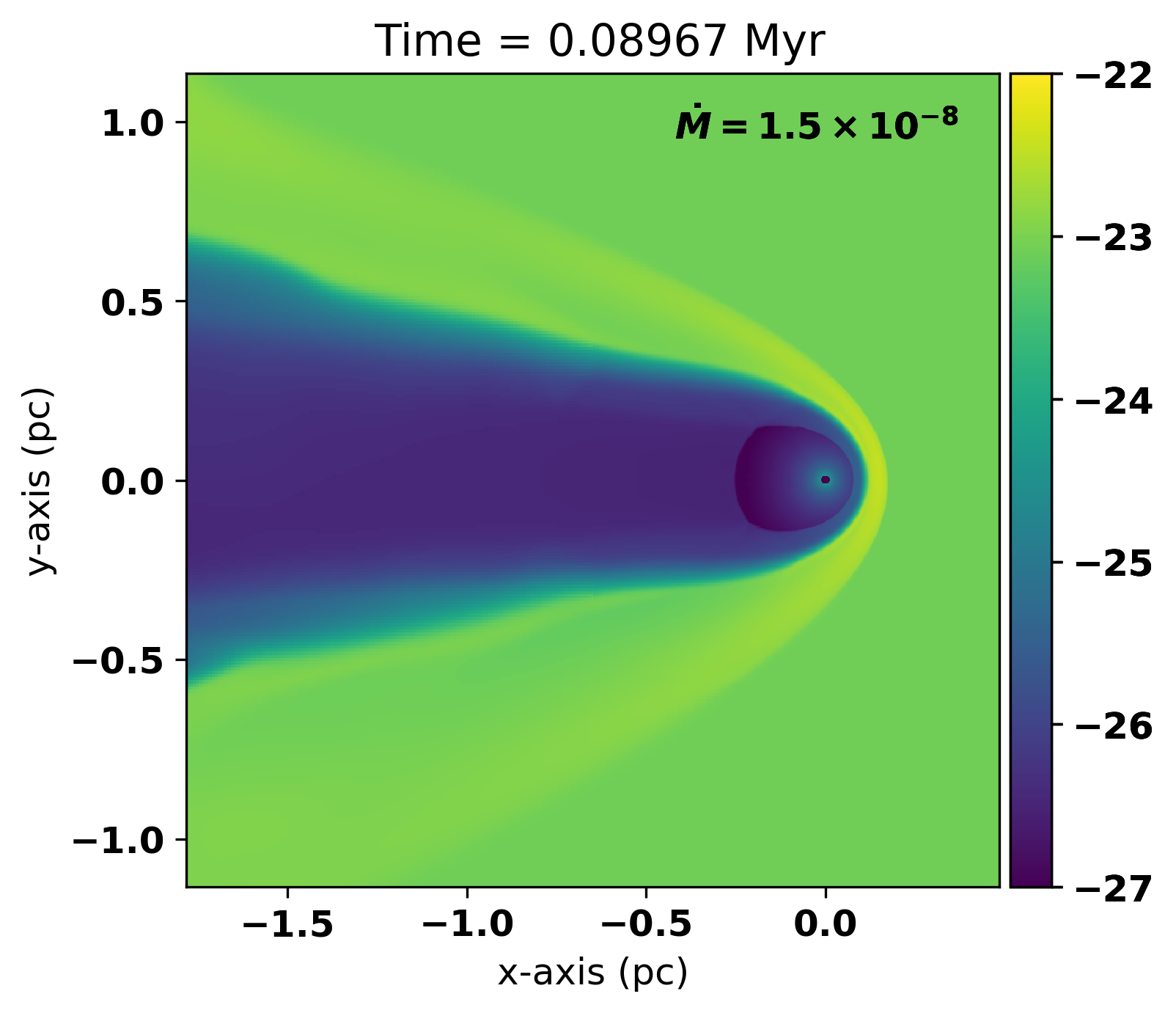} \\
	\includegraphics[height=.40\textwidth]{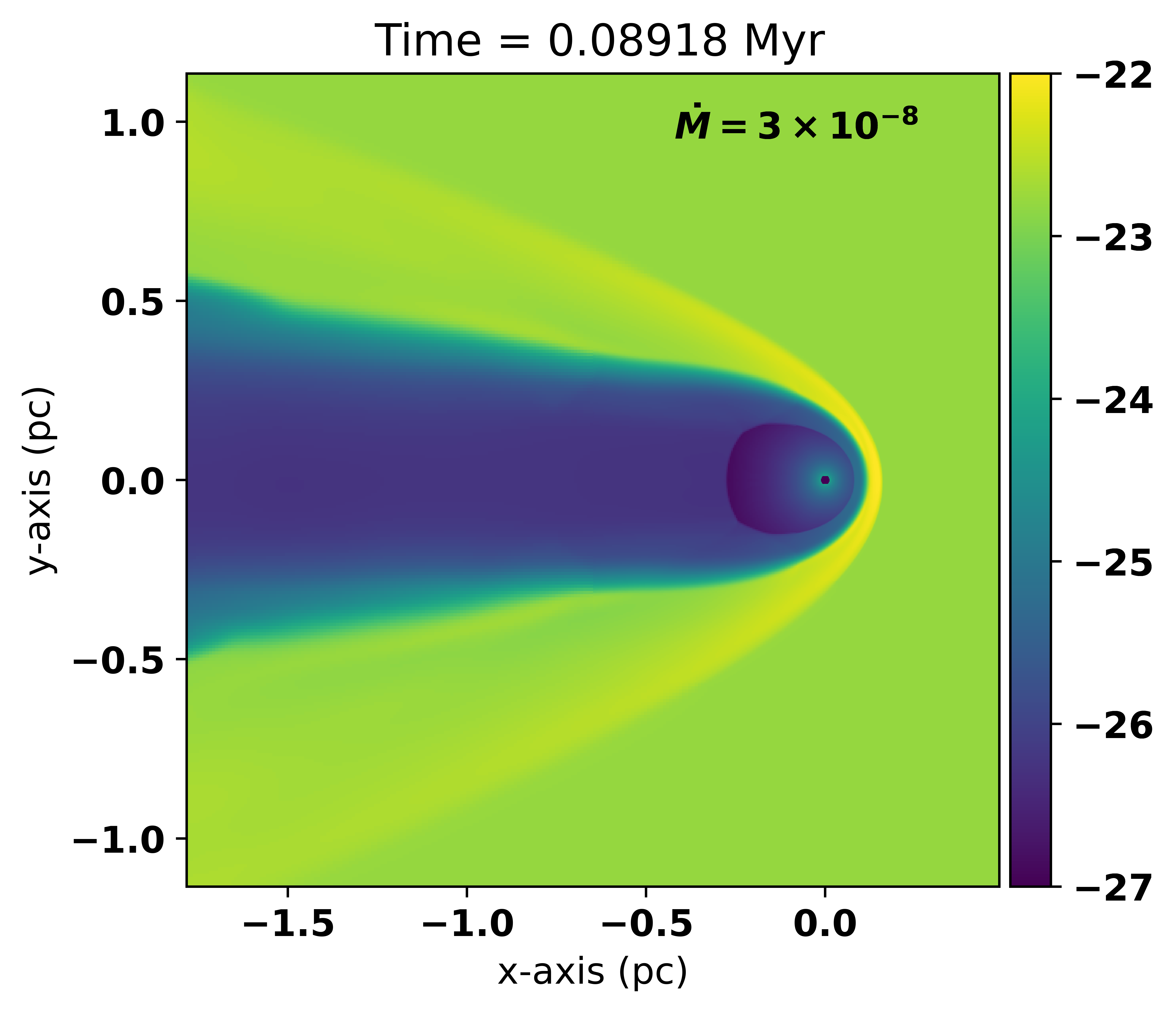}
	\caption{Top: Plots of the Log$_{10}$  gas density (g cm$^{-3}$) of simulation Z02. Bottom: Plots of the Log$_{10}$ gas density (g cm$^{-3}$) of simulation Z03. Slices through the ($x$-$y$) plane at z = 0. Star is at the origin.}
	\label{fig:z02}    
\end{figure}

The shape of the shocked wind region ("dorito" shape) is due to the orientation of the ISM's magnetic field which is primarily in the $\hat{y}$ direction (B$_0$ = [5, 8.66, 1] $\mu$G). In the ($x,z$) plane the magnetic field lines run perpendicular (not quite but close) to the outward pressure gradient of the bubble and the magnetic pressure compresses the shocked wind. On the other hand, in the ($x,y$) plane the field lines run parallel to the outward pressure of the bubble and allows the shocked wind to expand into a wider "dorito" shape. This is shown in Fig.~\ref{fig:mag} where slices in the $(y,z)$ plane at $x$=-0.3 pc (top panel) and $x$=-1 pc (bottom panel) of the $\log_{10}$ of magnitude of the magnetic field (|B|/G) are plotted. The streamlines overlaid in both panels show the direction of the magnetic field in this plane. The thickness of the streamlines represent the strength of the field with thin lines showing weak magnetic field and thicker lines showing stronger magnetic field. Both panels show how the pressure from the magnetic field is counteracting the pressure from the expansion of the bubble in the z-direction but doesn't restrict the bubble expansion as much in the y-direction. 


\subsubsection{Sim Z02}

The Z02 simulation
was run for 0.08967 Myr, again approximately 1.5 crossing times and 50 dynamical timescales of the bow shock.
The final snapshot is shown in Fig.~\ref{fig:z02}, showing the $(x,y)$ plane at $z=0$.
From this simulation we can measure $R_\mathrm{TS}=0.09$\,pc,  $R_\mathrm{CD}=0.13$\,pc, and $R_\mathrm{FS}=0.26$\,pc.
The simulation differs from Z01 in that we used $v_\star=40$\,km\,s$^{-1}$, corresponding to the case where there is a significant radial component to the star's space velocity.
We use a similar mass-loss rate to Z01, but a smaller ISM number density $n_\mathrm{i}=4$\,cm$^{-3}$ so that the ram pressure of the ISM is similar, and the bow shock has the correct size.
This simulation has a bow shock with approximately the same density as Z01 at the apex, and correspondingly similar infrared brightness (see section \ref{sec:infra}).

\subsubsection{Sim Z03}

The Z03 simulation, parameters used shown in Table~\ref{tab:zeta}, was run for 0.08918 Myr which is just over 1.5 crossing times and 50 dynamical timescales of the bow shock.
The final snapshot from the simulation Z03 is shown in Fig.~\ref{fig:z02}, for which we measure $R_\mathrm{TS}=0.09$\,pc,  $R_\mathrm{CD}=0.13$\,pc, and $R_\mathrm{FS}=0.26$\,pc.
Again we show the $(x,y)$ plane at $z=0$.
The bow shock is approximately the same size and shape as for Z02 because we again used $v_\star=40$\,km\,s$^{-1}$, but here the ISM density and mass-loss rate are both twice as high as for Z02.
This means that the total pressure throughout all parts of the bow shock is doubled, and the density at the apex of the bow shock is also twice that of Z02.
Because of this the infrared brightness (see section \ref{sec:infra}) is significantly higher than for Z01 and Z02, and the observational data.
The simulation was run because the higher pressure should give rise to higher X-ray emission in the shocked stellar wind (see section \ref{sec:xraysynth}).


\begin{figure}
	\centering
	\includegraphics[height=.31\textwidth]{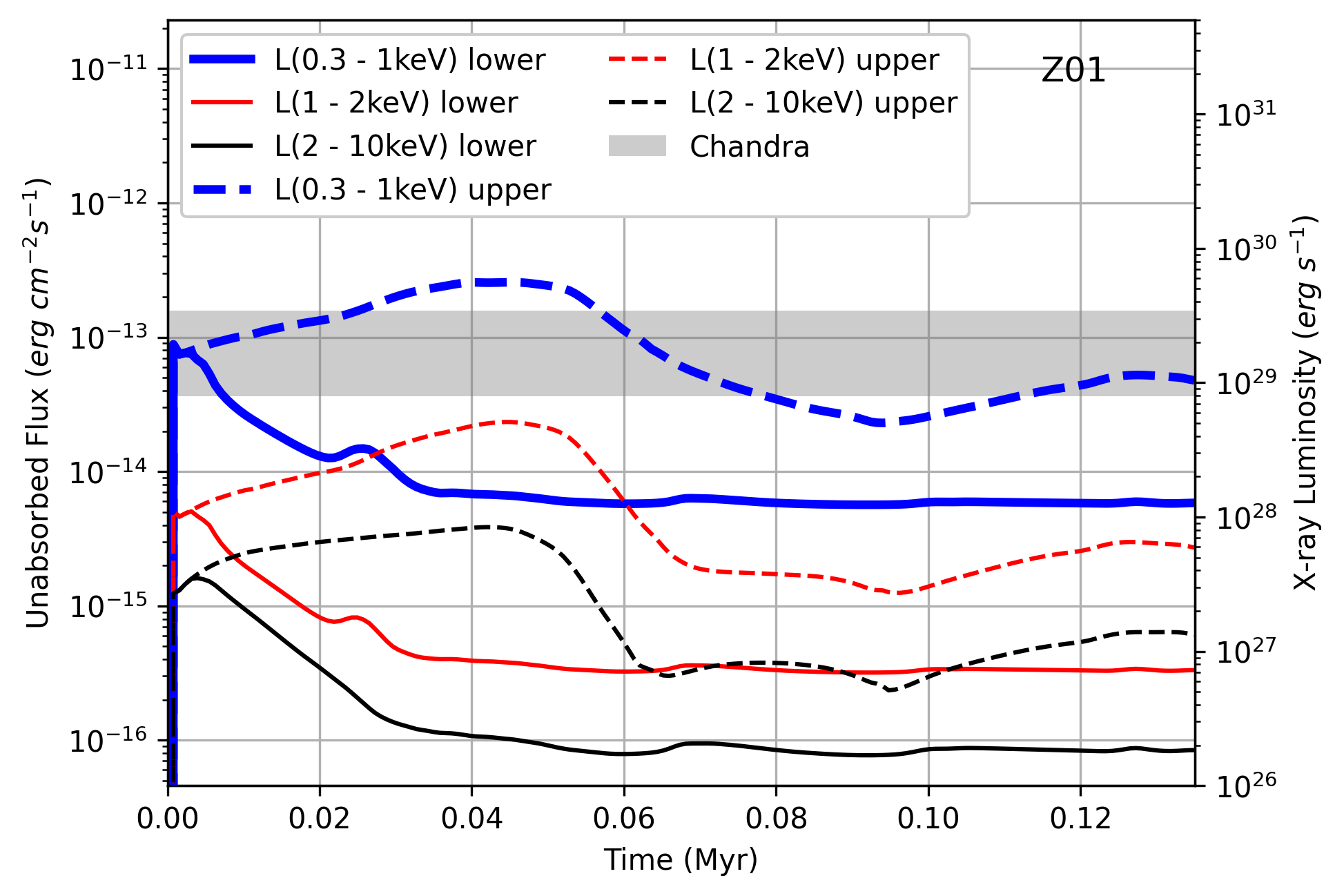} \\
	\includegraphics[height=.31\textwidth]{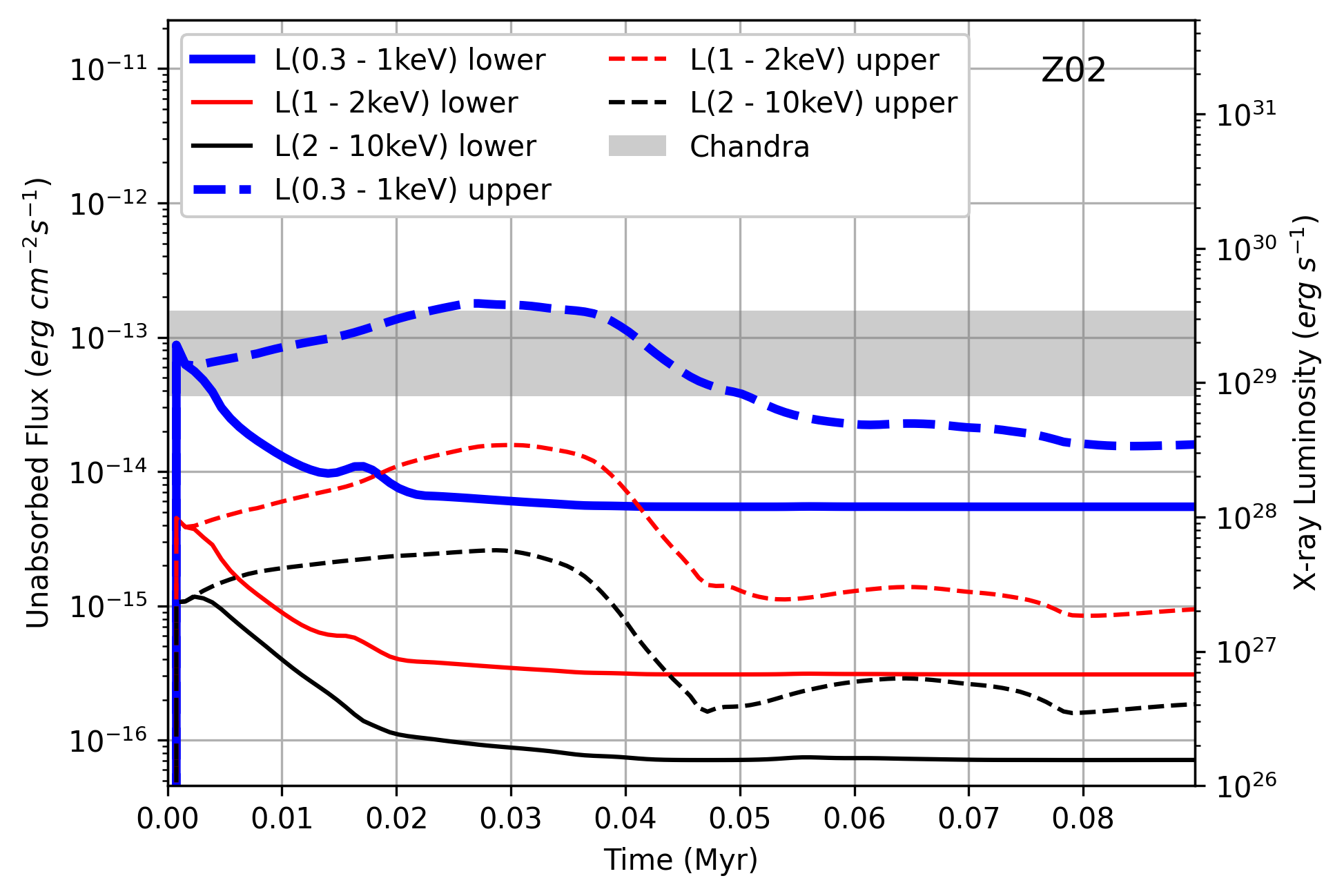} \\
	\includegraphics[height=.31\textwidth]{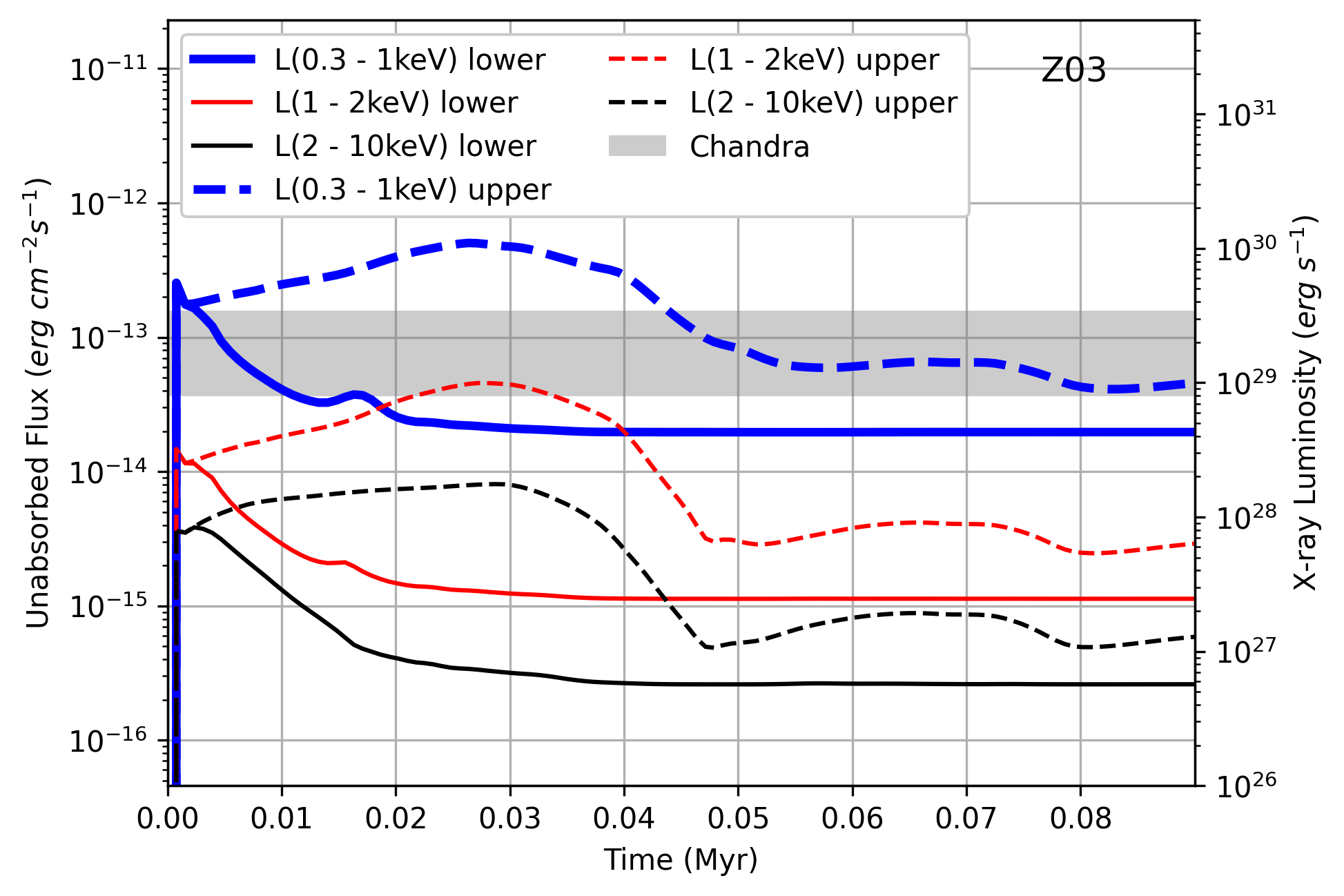} 
	\caption{Synthetic X-ray unabsorbed flux (erg\,cm$^{-2}$\,s$^{-1}$) and luminosity (erg\,s$^{-1}$) plot of $\zeta$~Ophiuchi as it evolves in time (Myr). Top: Z01. Middle: Z02. Bottom: Z03. Three X-ray bands: soft (0.3\,keV - 1\,keV), medium (1\,keV - 2\,keV), and hard (2\,keV - 10\,keV). Grey shaded box shows observed \textit{Chandra} luminosity/flux values. The solid lines represent the \textit{Chandra} field of view and the dashed lines are the whole simulation grid.} 
	\label{lum_zeta}
	
\end{figure}

\subsection{X-ray emission}
\label{sec:xraysynth}
The emissivity as a function of temperature for different X-ray bands was calculated using \textsc{xspec} v12.9.1 \citep{1996ASPC..101...17A} and tabulated. Solar abundances from \citet{2009ARA&A..47..481A} as implemented in \textsc{xspec} are used.
This was used in the same way as in \citet{Green} to calculate the X-ray luminosity from each grid cell in simulation snapshots.
Summing over all cells gives the total predicted X-ray luminosity, and raytracing through the domain gives synthetic X-ray surface-brightness maps.
Absorption within the simulation was neglected.

\begin{figure}
	\centering
	\includegraphics[height=.36\textwidth]{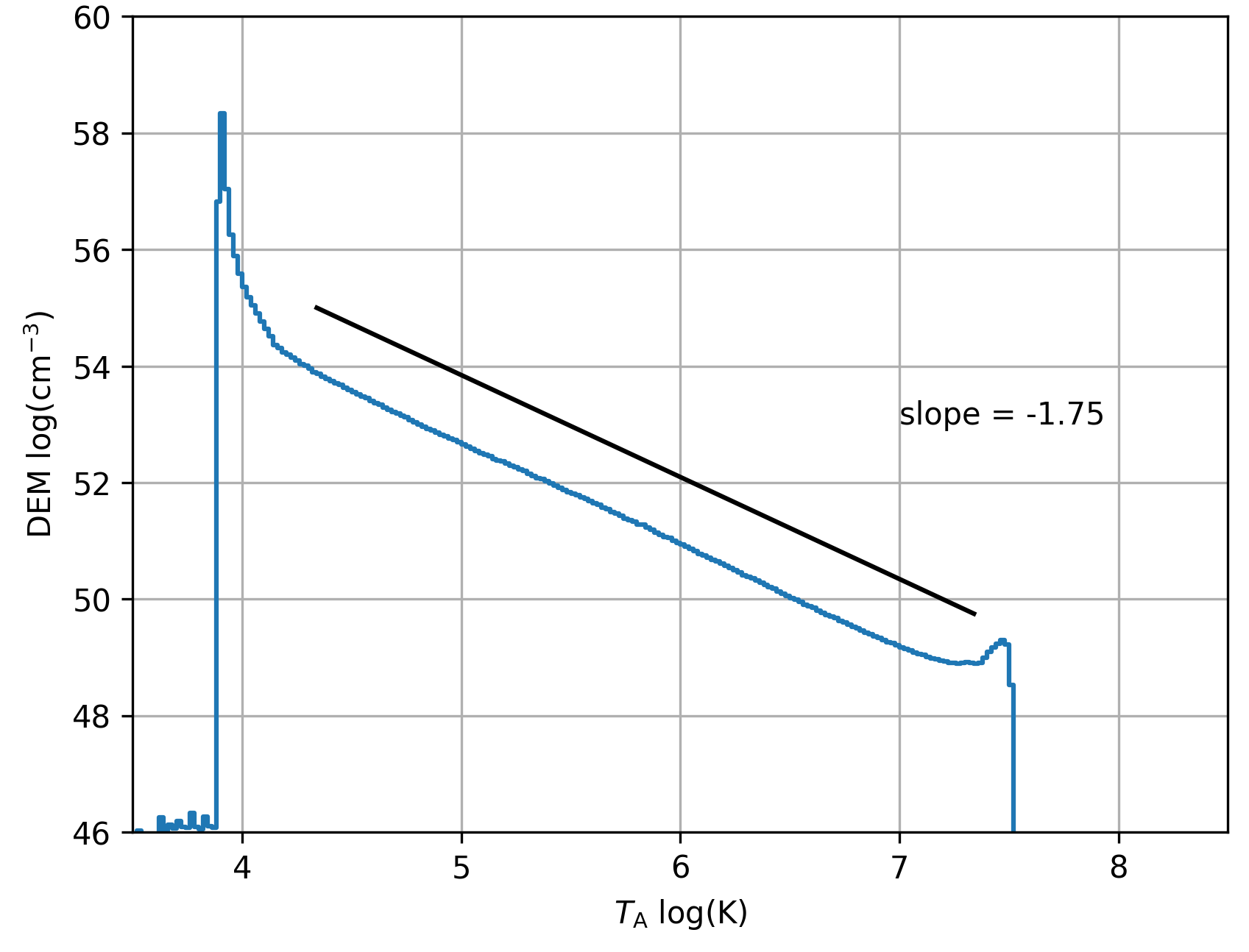} \\
	\caption{DEM profile of the simulated nebula (unabsorbed) from the Z01 simulation, after 0.13492 Myr of evolution. The black line represents the slope of the DEM profile between 10$^{4.2}$ K and 10$^{7.2}$ K.}
	\label{dem_zeta}
	
\end{figure}

\begin{figure}
	\centering
	\includegraphics[height=.34\textwidth]{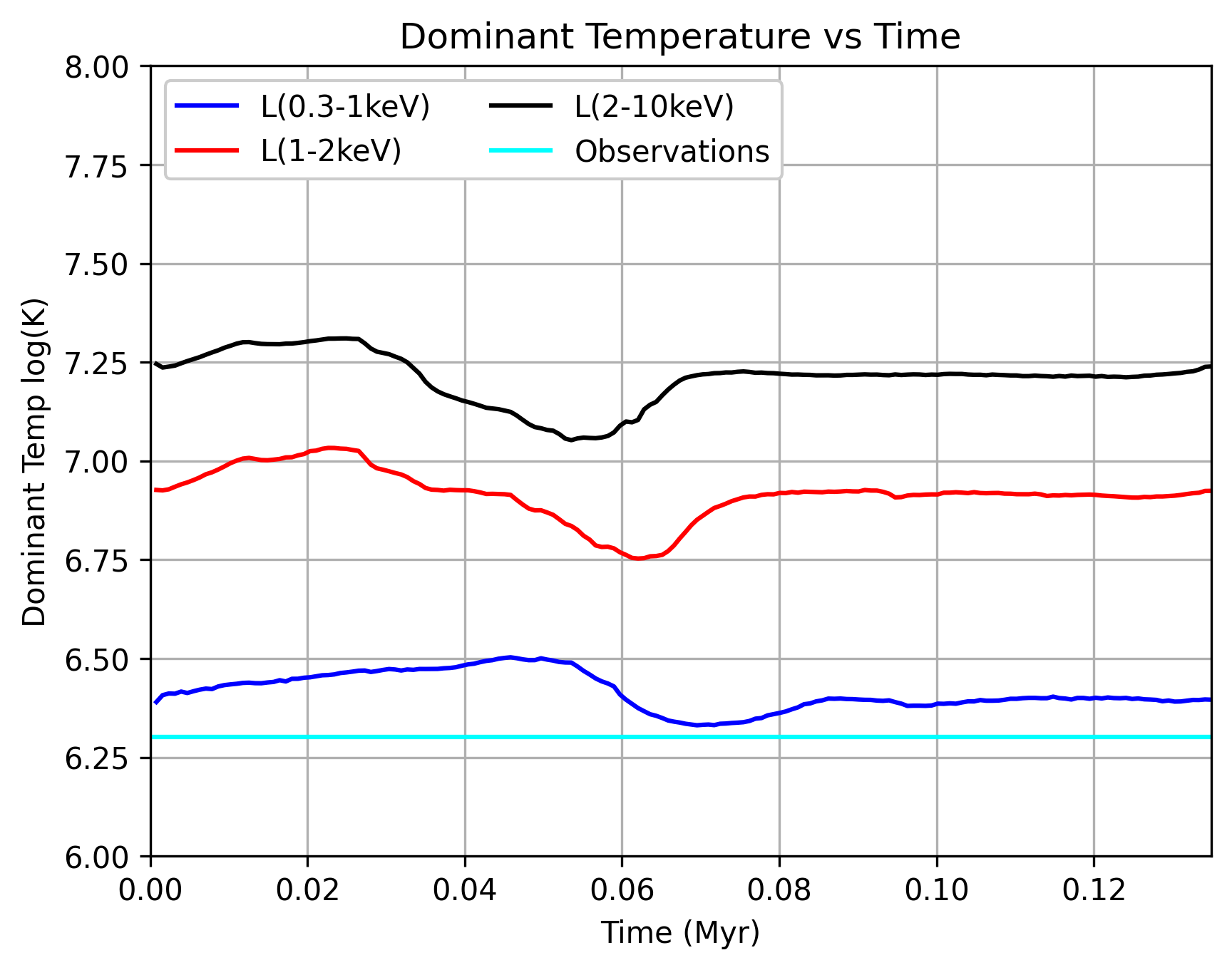} \\
	
	\caption{Log$_{10}$ Dominant temperature, T$_A$ (K), of the simulated nebula (unabsorbed) as it evolves in time (Myr).}
	\label{dom_zeta}
	
\end{figure}

Fig.~\ref{lum_zeta} shows the predicted soft (0.3 - 1 keV), medium (1 - 2 keV), and hard (2 - 10 keV) X-ray luminosity of the plasma as a function of time for the simulations Z01 (top), Z02 (middle), Z03 (bottom).
For each energy band we show an upper bound on the X-ray luminosity, corresponding to all of the X-ray emission from the whole simulation domain, and a lower bound corresponding to emission within a cube of side 0.5$\times$0.5$\times$0.5\,pc centred on the star. 
This smaller cube is similar to the region observed with \emph{Chandra} in Fig.~\ref{xray_obs}.
This allows us to best compare the values predicted from the simulation with observational values.
Fig.~\ref{lum_zeta} plots the unabsorbed flux (erg cm$^{-2}$ s$^{-1}$) for a distance of 135\,pc
and the luminosity (erg s$^{-1}$) versus time.
The observed luminosity with uncertainties is shown as the grey shaded region.

For comparison, the mechanical luminosity of the wind, $L_\mathrm{w}=0.5\dot{M}v_\infty^2$, is $L_\mathrm{w} \approx (1-2)\times10^{34}$\,erg\,s$^{-1}$ using the values in Table~\ref{tab:zeta}.
The measured X-ray flux corresponds to a luminosity $\approx10^5\times$ less than the mechanical luminosity of the stellar wind.
The Z01 simulation has a shocked-wind region that produces a soft X-ray luminosity (0.3-1\,keV) of $\sim 10^{28}$-$10^{29}$ erg\,s$^{-1}$ (flux $10^{-14}$-$10^{-13}$ erg\,cm$^{-2}$\,s$^{-1}$).
For this simulation the stellar motion is almost entirely in the plane of the sky, so the significant emission from the wake behind the star is not measured by the \emph{Chandra} observations.
The correct comparison is therefore to the lower flux limit, which is about $10\times$ lower than the observed flux.
For Z02 we find a similar soft X-ray luminosity, with lower limit $\sim 10^{28}$ erg\,s$^{-1}$ and upper limit $4\times 10^{28}$\,erg\,s$^{-1}$.
In this case the star has a significant line-of-sight velocity and some of the emission from the wake could be projected onto the \emph{Chandra} field of view, but still the flux is a factor of at least 3 too low.
The Z03 simulation has the highest pressure bubble and therefore the largest soft X-ray luminosity of the three simulations: $4\times 10^{28}$ to $10^{29}$\,erg\,s$^{-1}$.
Here again the star has a significant line-of-sight velocity and so much of the emission from behind the star is projected onto the \emph{Chandra} field of view.
The X-ray flux from this simulation is consistent with the flux measured by \emph{Chandra}, although still on the low side.
The luminosity/flux in the region around the apex of the bow shock stays constant over the whole life of the simulated nebula, because the flow is not subject to instability.
The total luminosity throughout the simulation does have some fluctuation on long timescales, similar to what was found by \citet{Green} although the level of the oscillations appears to be smaller.


The X-ray emission from the hot gas in the Z01 simulation is further analysed by calculating the differential emission measure (DEM) as a function of $T$, defined by

\begin{equation}
\text{DEM}(T_\mathrm{b}) = \sum_{k, T_\mathrm{k}\in{T_\mathrm{b}}} n_\mathrm{e}^2\Delta{V_\mathrm{k}},
\end{equation}

where $n_\mathrm{e}$ is the electron number density in cell $k$ and $\Delta{V_\mathrm{k}}$ is the volume of cell $k$ \citep{2018MNRAS.478.1218T}. Fig.~\ref{dem_zeta} shows the DEM of the simulated nebula (unabsorbed) from the Z01 simulation, after 0.1349 Myr of evolution. The DEM shows a profile strongly skewed towards lower temperatures, with a power-law behaviour similar to that shown by \citet{2018MNRAS.478.1218T} for stellar-wind bubbles with turbulent mixing layers and a power-law exponent of approximately -1.75 (the black line in the figure). 

We can also use the X-ray emissivity in a given energy band together with the DEM profile, to calculate $T_\mathrm{A}$ for the simulated wind bubble \citet{2018MNRAS.478.1218T}. Fig.~\ref{dom_zeta} shows the evolution of $T_\mathrm{A}$ as a function of time for different X-ray energy bands. The soft X-ray emission (0.3 - 2 keV) shown has a mean temperature of about $10^{6.4}$\,K. Whereas, the hard X-ray emission (2 - 10 keV) shown has a mean temperature of about $10^{7.25}$\,K. The figure also shows that the medium X-ray emission between $1$ and $2$\,keV has a mean temperature of $10^{6.8}$\,K. The values of $T_\mathrm{A}$ are almost constant for the duration of the simulation, and consistent with previous work using 2D simulations \citep{2018MNRAS.478.1218T, Green}.


\section{Synthetic Emission}
\label{sec:syn}

\subsection{Infrared synthetic data}
\label{sec:infra}

\begin{figure}
	\centering
    \includegraphics[height=.39\textwidth]{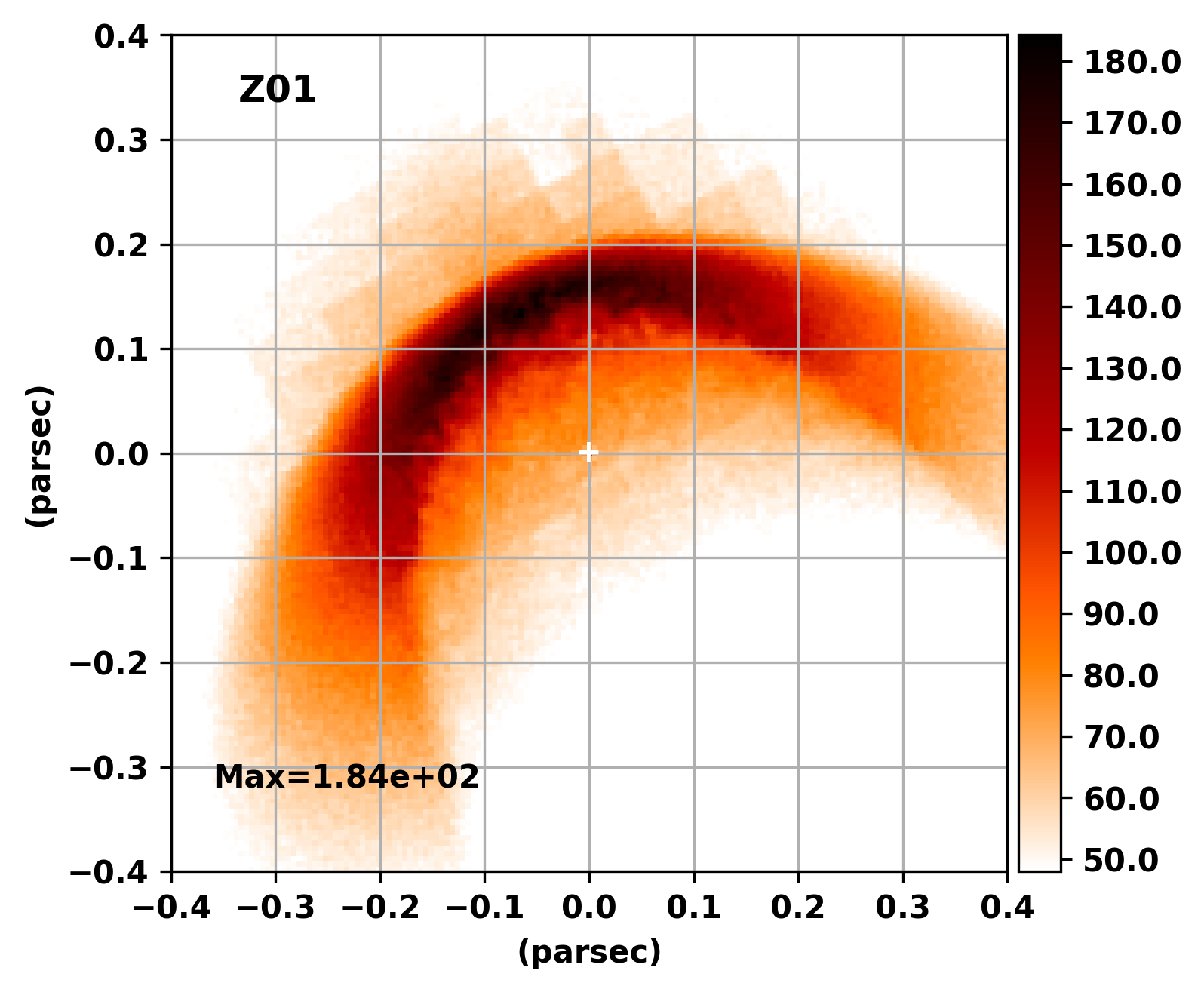} \\
    \includegraphics[height=.39\textwidth]{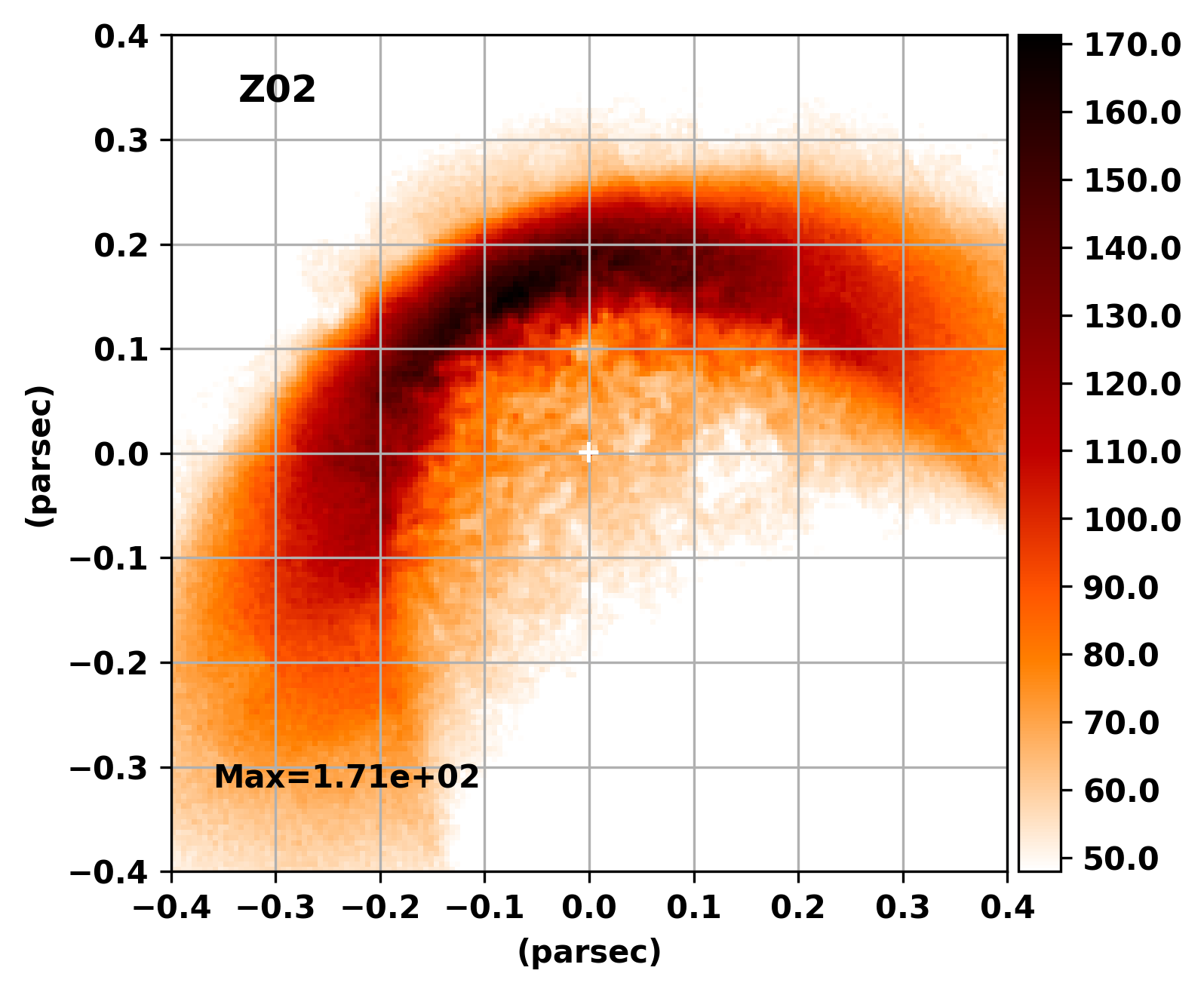} \\
    \includegraphics[height=.39\textwidth]{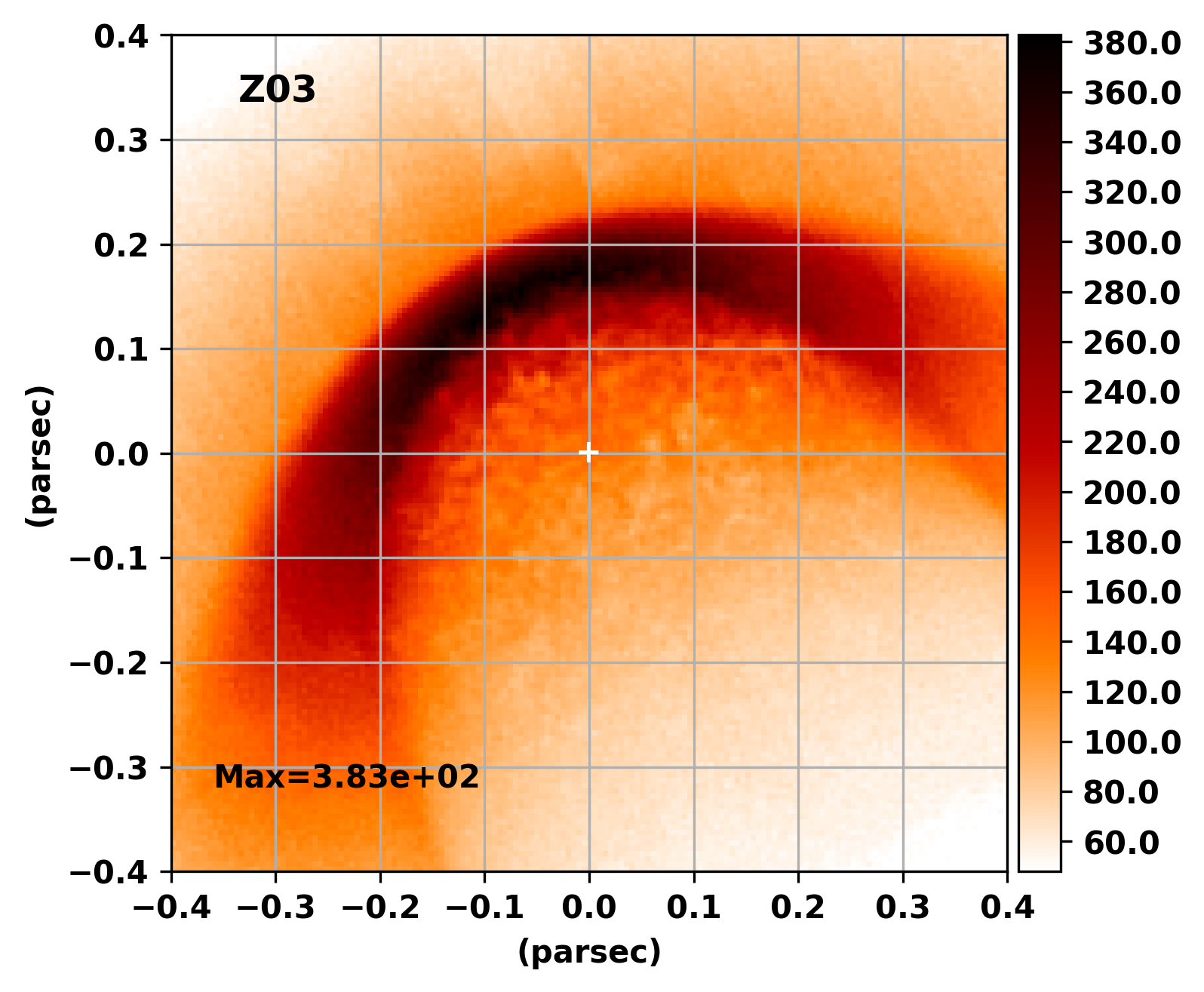}
	\caption{Synthetic infrared emission maps of the bow shock in units of MJy ster$^{-1}$ at 24$\mu$m. Top: Z01. Middle: Z02. Bottom: Z03. The star is at the origin with coordinates in parsecs relative to the position of the star. The image is rotated to the same orientation as the Spitzer image in Fig.~\ref{fig:obs_dust}.}
	\label{fig:dust}    
\end{figure}


Synthetic Infrared emission maps of the thermal dust emission at 24$\mu$m (MJy ster$^{-1}$), calculated from the Z01, Z02, and Z03 simulations, are shown in the top, middle, and bottom panels of Fig.~\ref{fig:dust} respectively. 
These were calculated using \textsc{torus} as described in section \ref{sec:meth}.
It was estimated that the ``background'' emission in the 24$\mu$m \textit{Spitzer} image was 47 MJy ster$^{-1}$ so this constant background value was added to the emission maps.
ecause of the high Galactic latitude of the source, it is likely that this is foreground emission from the Local Bubble.  Possible sources are thermal dust emission, or spectral lines of [O IV] and [Ne V] which are in the MIPS 24\,$\mu$m band \citep{2009AJ....138..691C}.
We made synthetic images at angles from 0$^\circ$ to 90$^\circ$ (with 15$^\circ$, step) between the line of sight and the velocity vector of the star. 

We show the 90$^\circ$ image for the Z01 simulation in the top panel Fig.~\ref{fig:dust} because if $v_r$ = -2.5 \kms then $v_\star$ is mostly in the plane of the sky. 
In the middle and bottom panel we show the 40$^\circ$ projection for the Z02 and Z03 simulations because if $v_r$ = +24.7 \kms then $v_\star$ is no longer in the plane of the sky. The other angles from the Z01 simulation can be seen in Fig.~\ref{angles_zeta}, where we show synthetic emission maps in 24$\mu$m, 70$\mu$m, H$\alpha$, Radio 6Ghz, Emission Measure, and soft X-rays at angles of $0^{\circ} - 90^{\circ}$. 

From the top panel of Fig.~\ref{fig:dust} it can be seen that the bulk of the infrared emission corresponds to the region between the contact discontinuity and the forward shock, at 0.11-0.22\,pc. This is consistent with the position and width of the infrared arc seen in \textit{Spitzer} data in Fig.~\ref{fig:obs_dust} for the previously used distance of $d = 112$\,pc \citep{gvaramadze2012zeta}, but for the new \emph{Gaia} distance of $d = 135$\,pc the bow shock is too small.
The opening angle of the bow shock ($\frac{R_{90}}{R_0}$) in the synthetic image is smaller than in the observational data also.
In the other two panels we can see that the infrared emission is at 0.13-0.26\,pc. This is consistent with the position and width of the infrared arc seen in \textit{Spitzer} data in Fig.~\ref{fig:obs_dust}.
The opening angle of the bow shock ($\frac{R_{90}}{R_0}$) in these synthetic images is similar to the observational data.
The bow shock in all panels, however, is also smooth whereas the observations show some density structure.
This is probably because we use a uniform ISM without any turbulent density structure.

The maximum brightness of the 24$\mu$m synthetic snapshot in the top (184 MJy ster$^{-1}$) and middle (171 MJy ster$^{-1}$) panels, matches the maximum brightness of the \textit{Spitzer} image: 170 MJy ster$^{-1}$. The bottom panel (383 MJy ster$^{-1}$) does not match the maximum brightness of the \textit{Spitzer} image. The overall agreement between the Z01 and Z02 synthetic Infrared emission maps and the observational Infrared data is good. Due to the simplifications in our model (i.e. uniform ISM density) it is not possible to match every aspect of the observations.

\begin{figure}
	\centering
	\includegraphics[height=.37\textwidth]{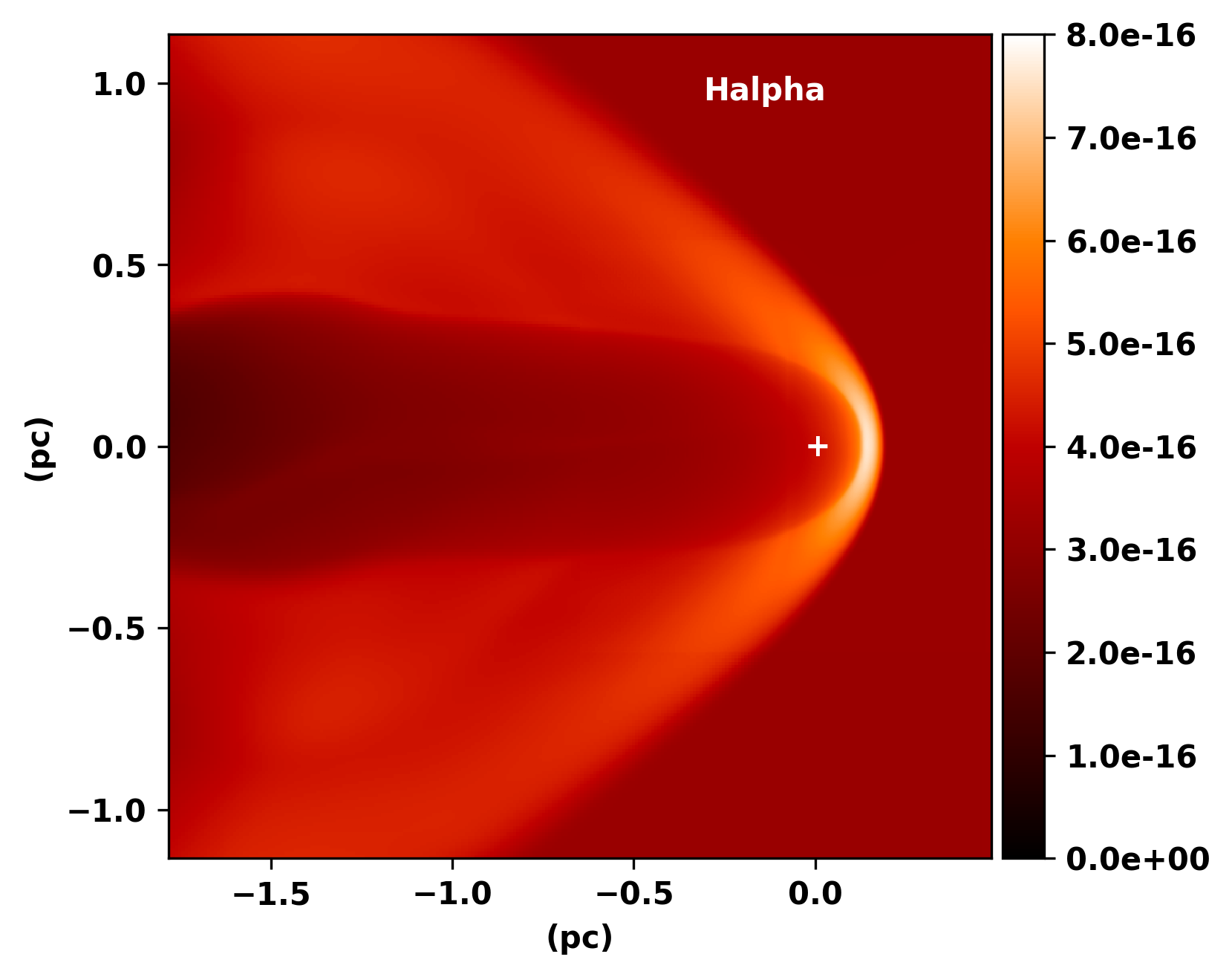} \\
	\includegraphics[height=.37\textwidth]{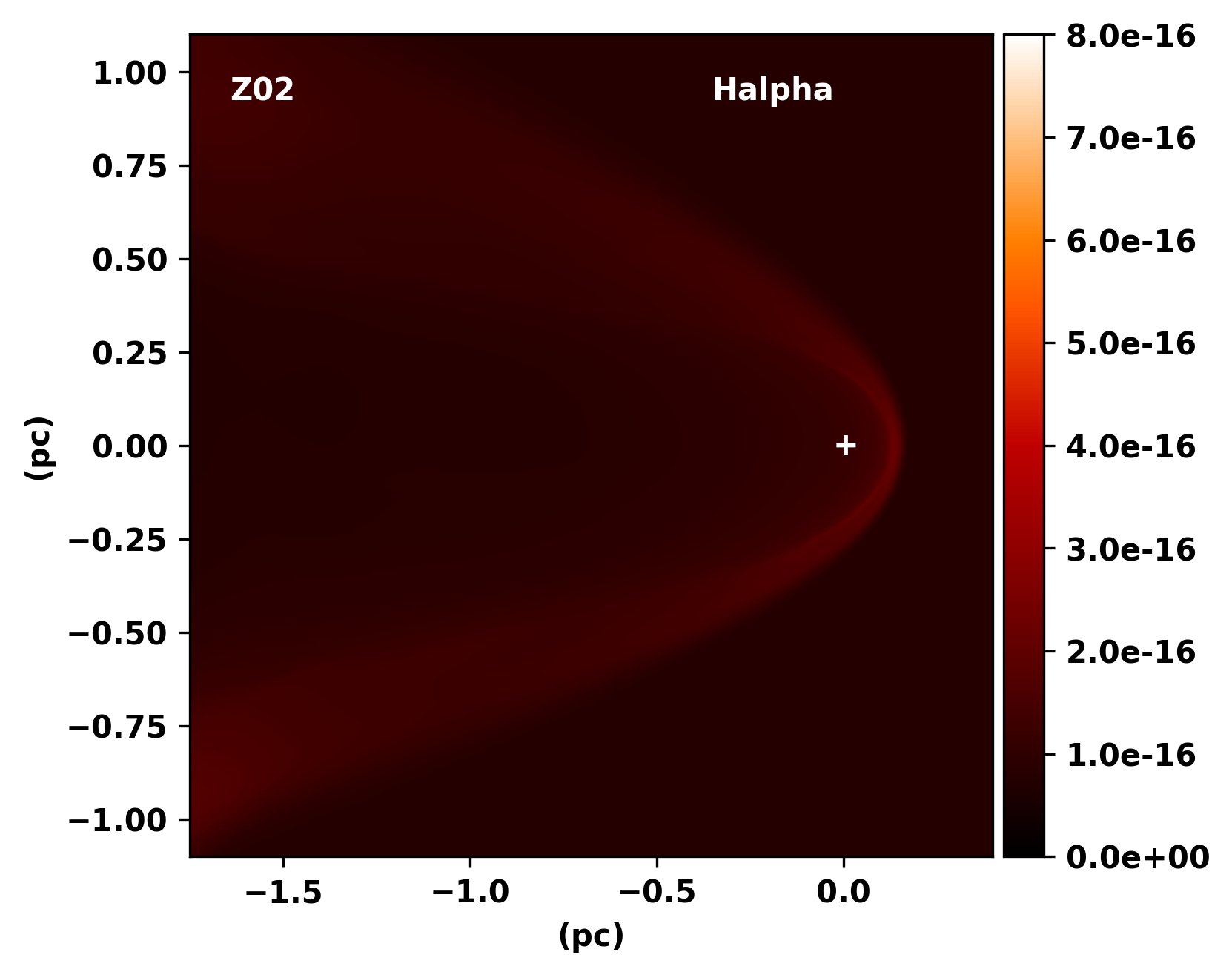} \\
	\includegraphics[height=.37\textwidth]{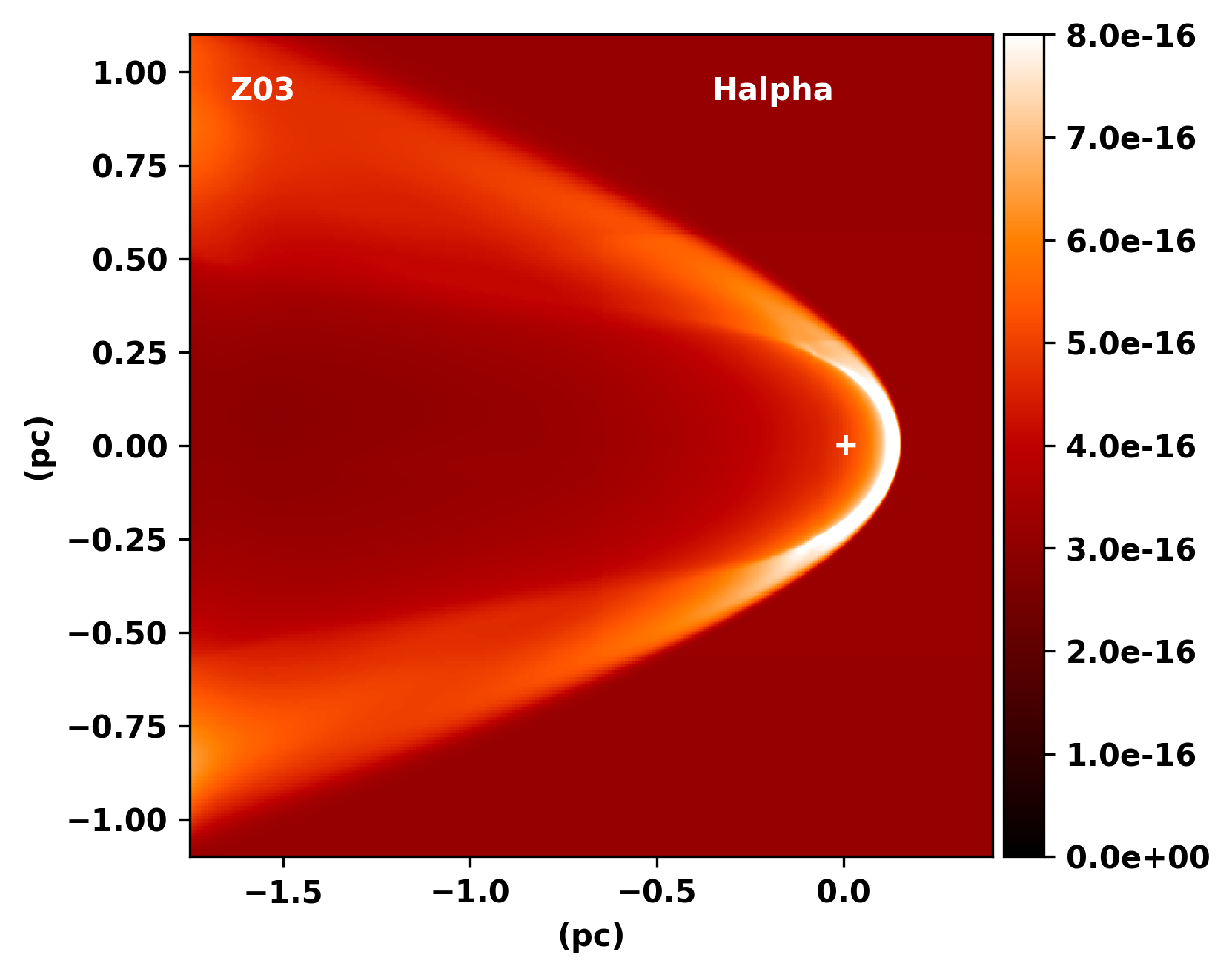} 
	\caption{Synthetic H$\alpha$ emission maps of the simulated nebula around $\zeta$~Ophiuchi on a linear scale with units erg\,cm$^{-2}$\,s$^{-1}$\,arcsec$^{-2}$. Top: Z01 with the line of sight being the y-axis. Middle: Z02. Bottom: Z03. Coordinates in parsecs relative to the position of the star (white cross).}
	\label{halpha_zeta}
\end{figure}

\subsection{H$\alpha$ synthetic data}
\label{sec:halphaemission}

Our 3D raytracing method described in \citet{MacLanGva13} and \citet{Green} has been used to produce synthetic H$\alpha$ emission maps of the Z01, Z02, and Z03 simulations, shown in Fig.~\ref{halpha_zeta} and ~\ref{angles_zeta}. 
The apex of the bow shock contains the brightest H$\alpha$ emission with intensity $8\times 10^{-16}$ erg\,cm$^{-2}$\,s$^{-1}$\,arcsec$^{-2}$. 

The H~\textsc{ii} region surrounding $\zeta$~Ophiuchi is less dense than the bow shock but is also significantly larger.
The H~\textsc{ii} region has diameter $\approx10$\,pc and mean density $n_\mathrm{i}\approx 3$\,cm$^{-3}$ \citep{gvaramadze2012zeta}, for an emission measure $\approx90$\,cm$^{-6}$\,pc.
This background emission has not been added to the synthetic images, and so the background emission level is determined by the ISM number density and the size of the domain (2.27\,pc) which is only 1/4 the true diameter of the H~\textsc{ii} region.  As such it is the brightness of the bow shock above the background emission that should be compared with observations.

The line-of-sight passing through the bow shock has length $\approx0.4$\,pc (Fig.~\ref{halpha_zeta}), and therefore the bow shock should have $n_\mathrm{i}\gtrsim\sqrt{90/0.4}=15$\,cm$^{-3}$ in order to have comparable emission measure to the H~\textsc{ii} region, and be easily visible above the background.
In all three simulations this condition is fulfilled and so the bow shock should be visible in H$\alpha$.
Indeed \citet{GulSof79} detected the bow shock in narrow-band images, more clearly in [O\textsc{iii}] than in H$\alpha$, but a digital version of their image is not available for more detailed comparison.
New narrow-band observations of this bow shock would be valuable to further constrain its physical properties.

\begin{figure}[!]
	\centering
	\includegraphics[height=.37\textwidth]{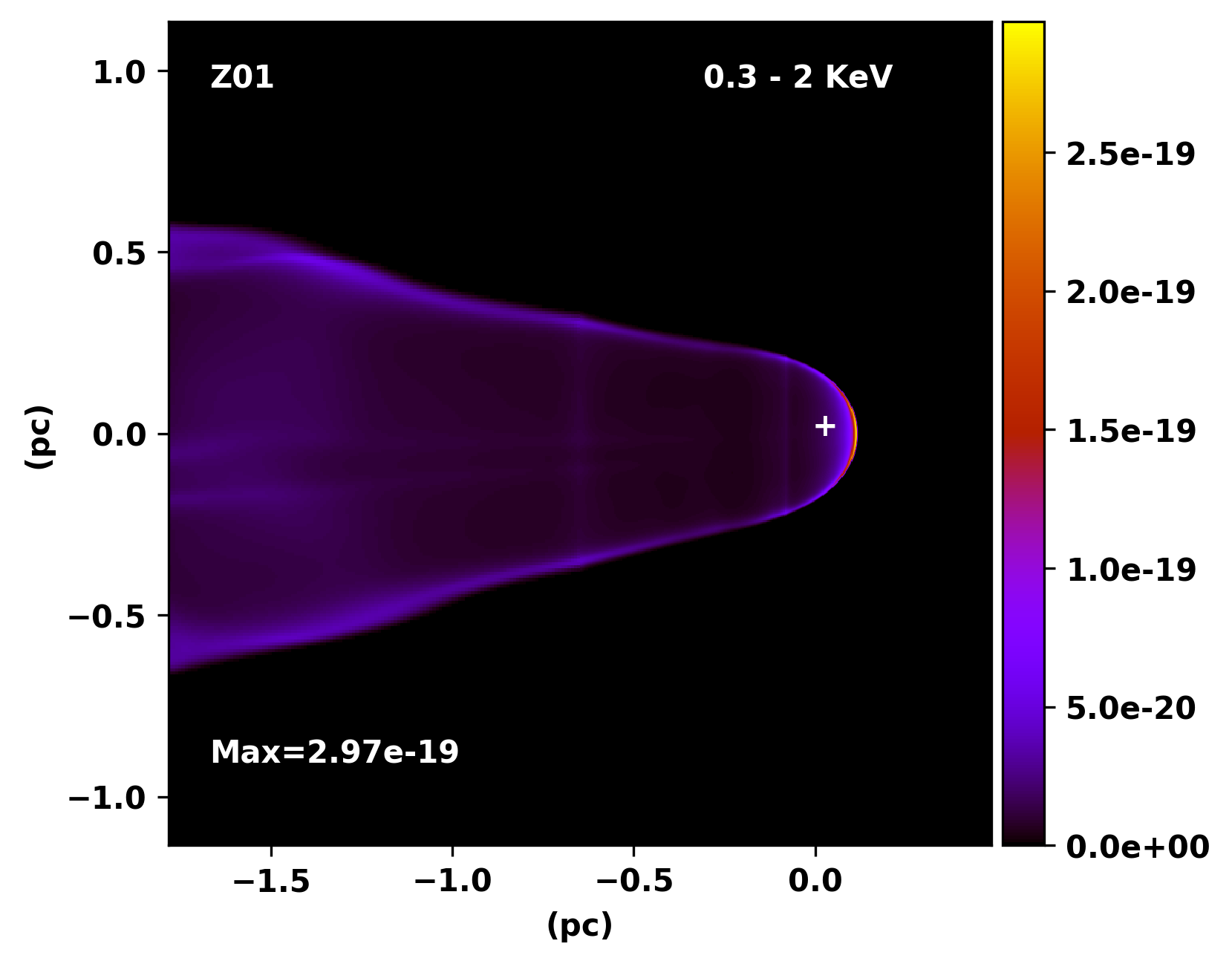} \\
	\includegraphics[height=.37\textwidth]{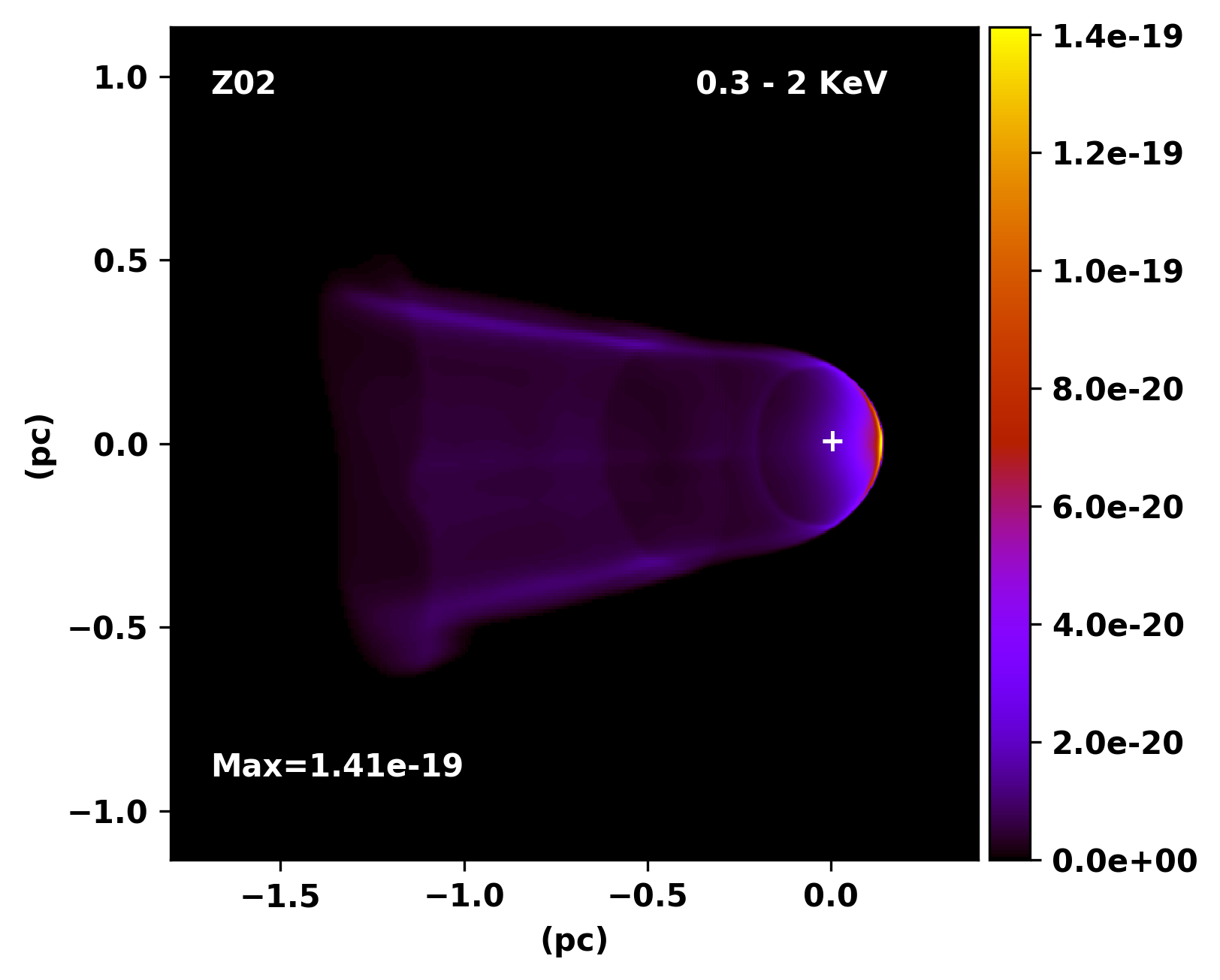} \\
	\includegraphics[height=.37\textwidth]{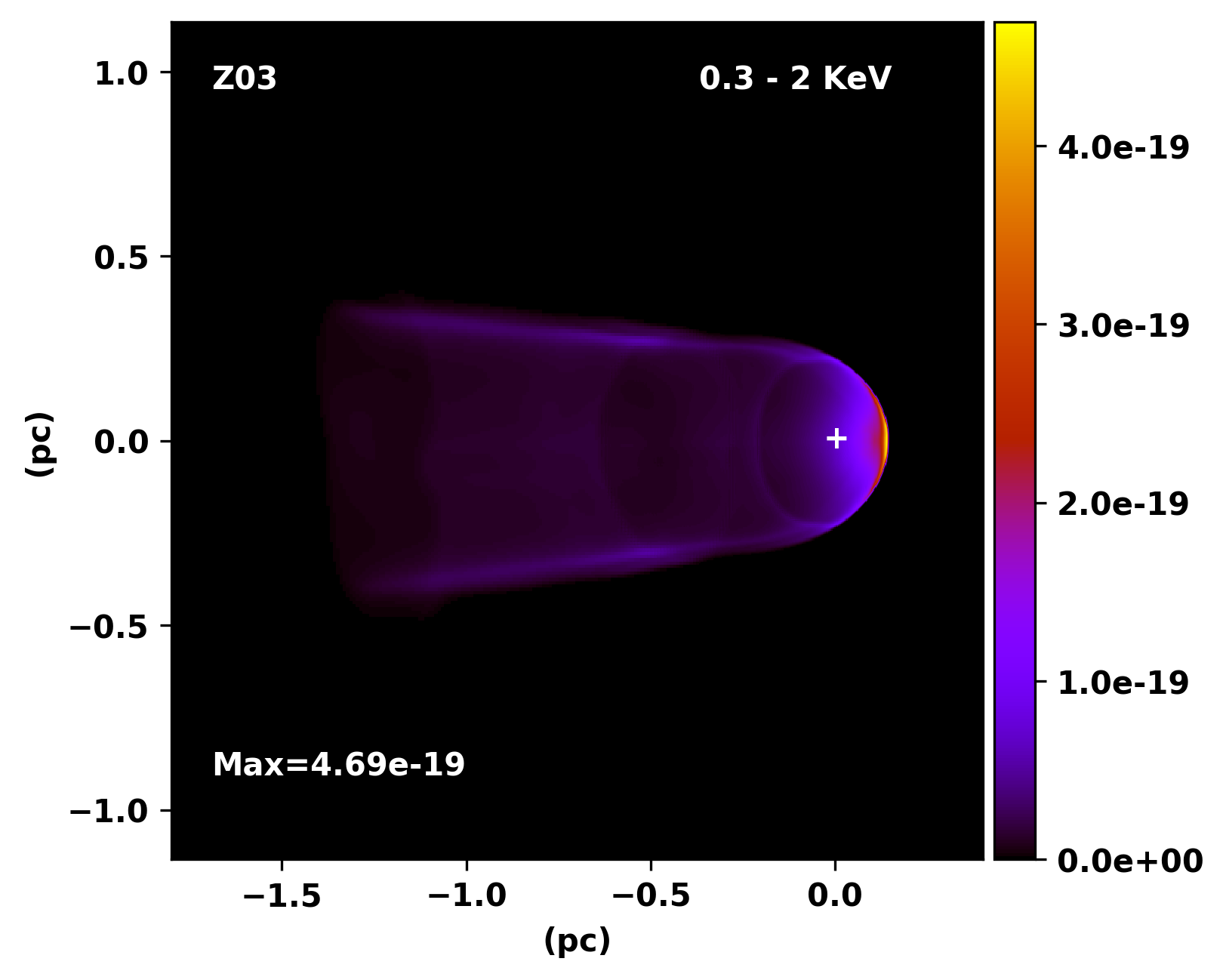}
	\caption{Synthetic soft X-ray (0.3 - 2 KeV) emission maps of the simulated nebula around $\zeta$~Ophiuchi (unabsorbed). Top: Z01 at an angle of $90^{\circ}$ with respect to the direction of stellar motion. Middle: Z02 at an angle of $45^{\circ}$. Bottom: Z03 at an angle of $45^{\circ}$. Coordinates are in parsecs relative to the position of the star (white cross) and the colour scale is from zero to maximum in erg\,cm$^{-2}$\,s$^{-1}$\,arcsec$^{-2}$. The image normal direction is the $z$-axis of the simulation.}
	\label{xray_syn}
\end{figure}

\begin{figure}[!]
	\centering
	\includegraphics[height=.32\textwidth]{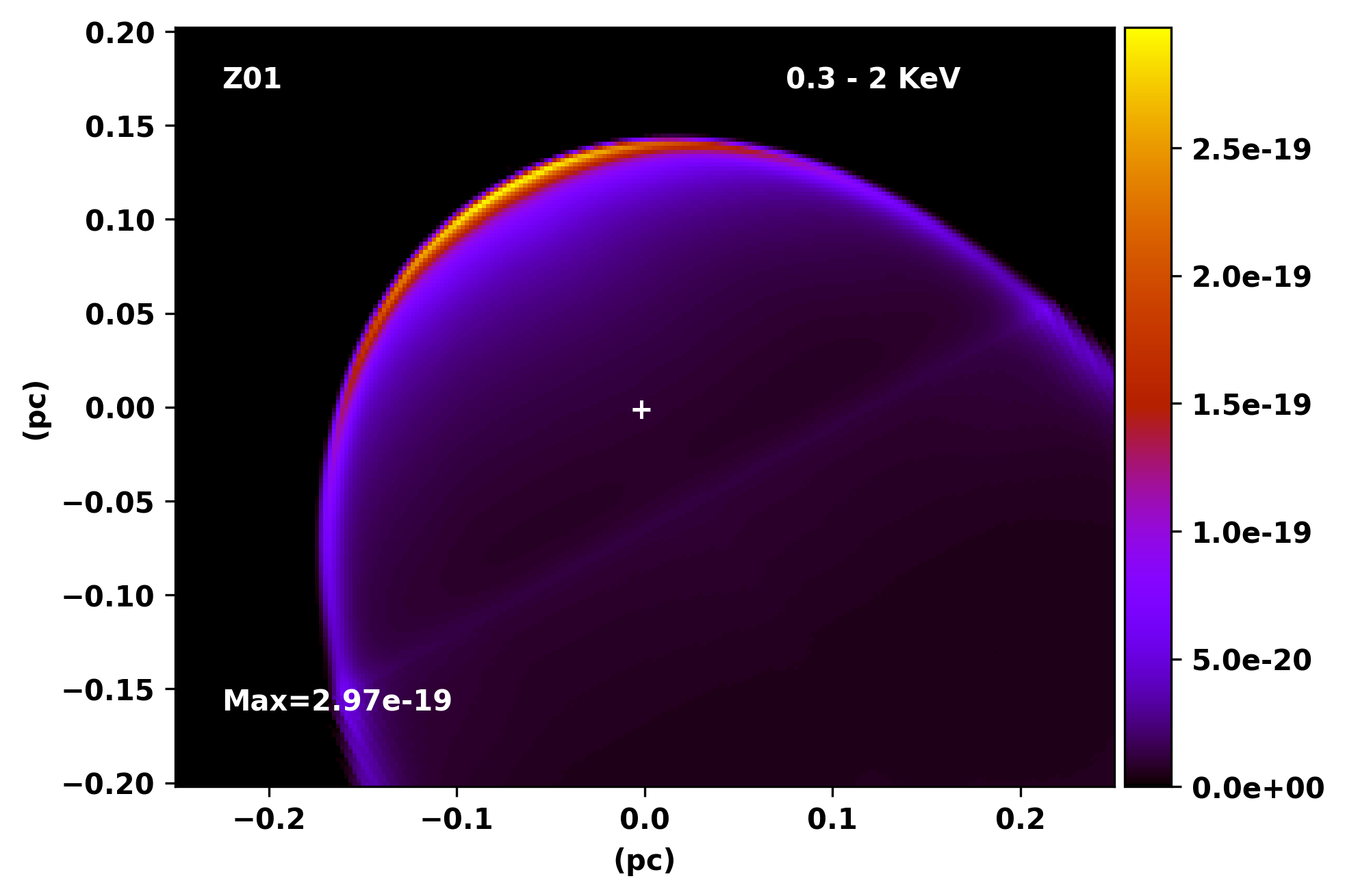} \\
	\includegraphics[height=.32\textwidth]{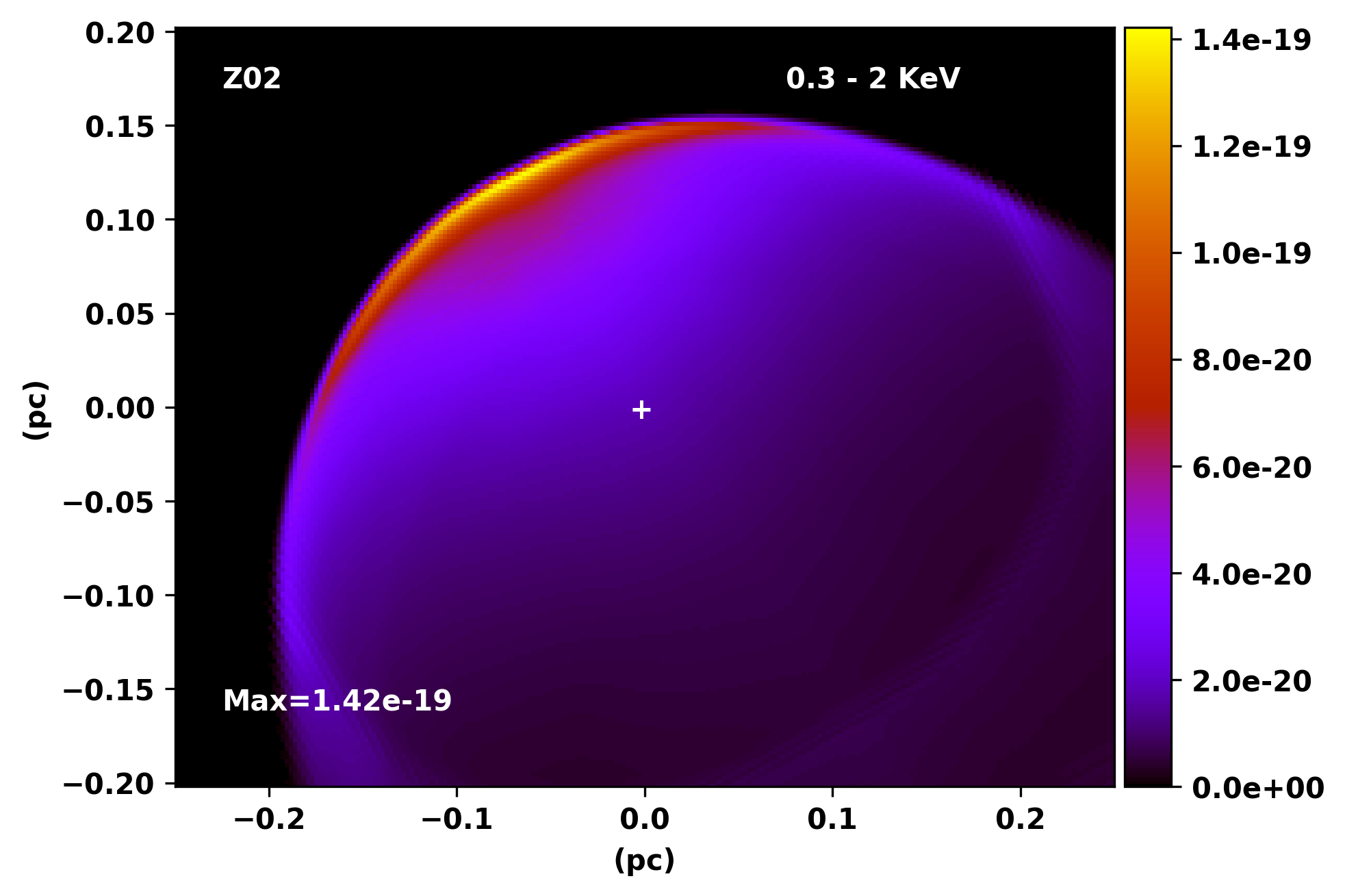} \\
	\includegraphics[height=.32\textwidth]{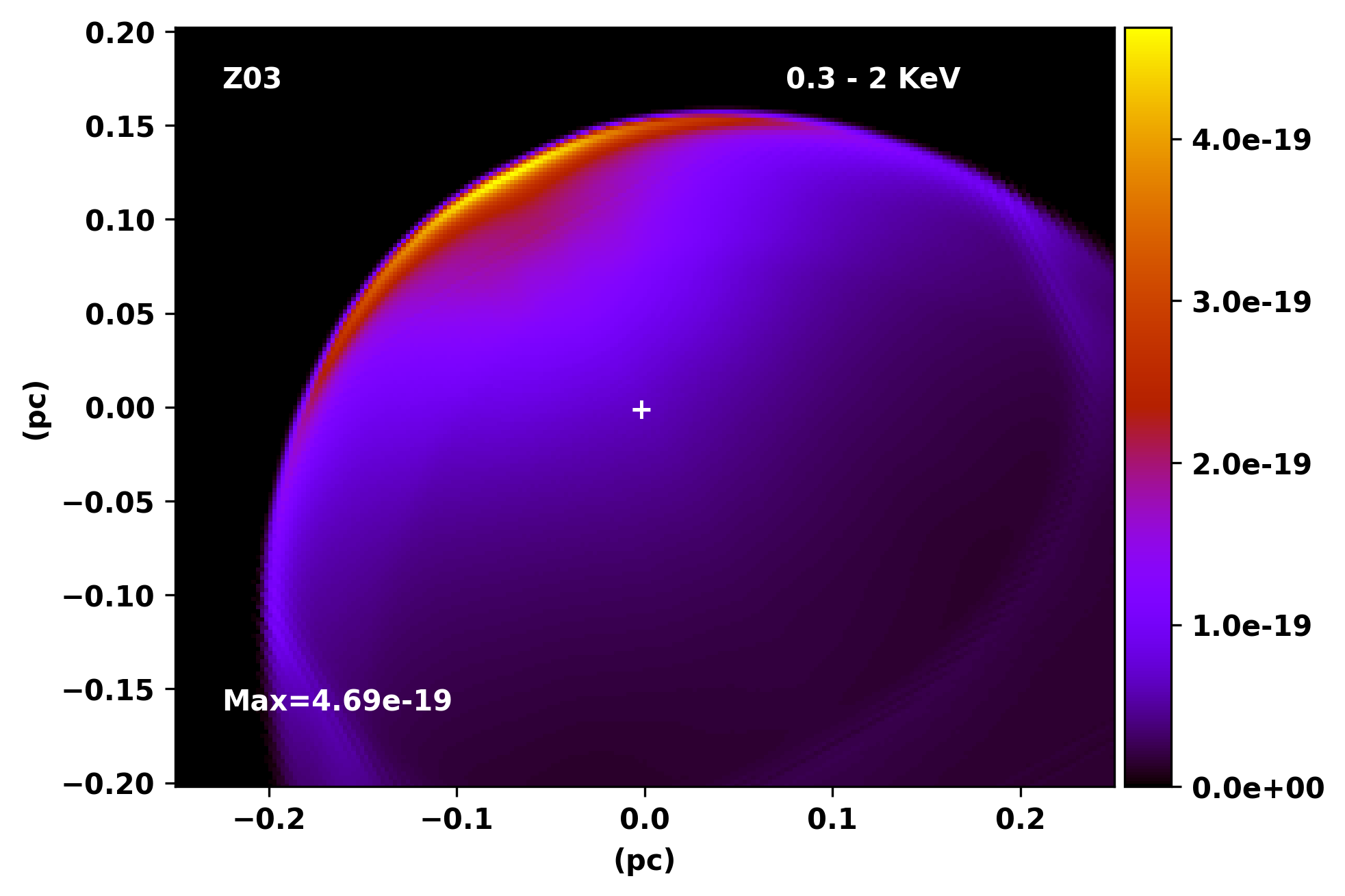}
	\caption{Synthetic soft X-ray (0.3 - 2 KeV) emission maps of the simulated nebula around $\zeta$~Ophiuchi (unabsorbed) rotated to the direction of motion of $\zeta$~Ophiuchi in Fig. 2 and set to the same scale. Top: Z01 at an angle of $90^{\circ}$ with respect to the direction of stellar motion. Middle: Z02 at an angle of $45^{\circ}$. Bottom: Z03 at an angle of $45^{\circ}$. Coordinates are in parsecs relative to the position of the star (white cross) and the colour scale is from zero to maximum in erg\,cm$^{-2}$\,s$^{-1}$\,arcsec$^{-2}$. The image normal direction is the $z$-axis of the simulation.}
	\label{xray}
\end{figure}

\subsection{X-ray synthetic data}

Using the 3D raytracing method we produced synthetic soft X-ray emission maps, shown in Fig.~\ref{xray_syn}, using the same snapshot as Figs.~\ref{fig:den} and \ref{fig:z02}.
The panels show the final snapshot from simulations Z01, Z02, and Z03 from top to bottom respectively.
For Z01 we plot a projection along the $z$-axis so that the image $x$- and $y$-axes are also the simulation $x$- and $y$-axes.
For Z02 and Z03, because the model includes a significant radial velocity along the line of sight, we plot a projection along a line 45$^\circ$ between the $z$- and $x$-axes.

The surface brightness seen in Fig.~\ref{xray_syn} should be compared with the the average surface brightness of the \textit{Chandra} observations in Fig.~\ref{xray_obs}, of $2.04^{+0.82}_{-1.35}\times10^{-18}$\,erg\,cm$^{-2}$\,s$^{-1}$\,arcsec$^{-2}$.
The X-ray emission in the top panel (Z01) near the apex of the bow shock is about 10 times too faint, in agreement with the global luminosity calculation above.
Simulation Z02 (middle panel) has somewhat brighter emission, but still much fainter than the observed flux.
As expected, simulation Z03 (bottom panel) is much brighter in soft X-rays than the other two simulations (by design), but still is is a factor of three fainter than the observed emission.

Fig.~\ref{xray} shows the soft X-ray emission presented in Fig.~\ref{xray_syn} set to the same axis scales and rotated to the same direction as the {\it Chandra} image in Fig.~\ref{xray_obs}.
The top panel is from the Z01 simulation, the middle panel in the Z02 simulation, and the bottom panel is the Z03 simulation.
The majority of the X-ray emission is coming from the apex of the bow shock at the contact discontinuity where the shocked ISM and shocked wind meet, and also following this boundary downstream from the star.
This is the region of mixed wind and ISM gas, containing intermediate temperatures that cool strongly and radiate efficiently in soft X-rays.
The emission from the synthetic map at the apex of the bow shock can be compared to the outer contour from the observational image, where the outline of the emitting region is also an arc that traces the inner edge of the infrared bow shock.
However, the observational data shows some of the emission may also be occurring closer to the star which is not seen in the synthetic image.
From the uncertainty due to pixel pile-up from stellar emission within the white circle in the observational image, this fainter emission observed closer to the star could also have a contribution from stellar emission, although we have gone to great lengths to remove the stellar contamination as much as possible.

\begin{figure}
	\centering
	\includegraphics[height=.37\textwidth]{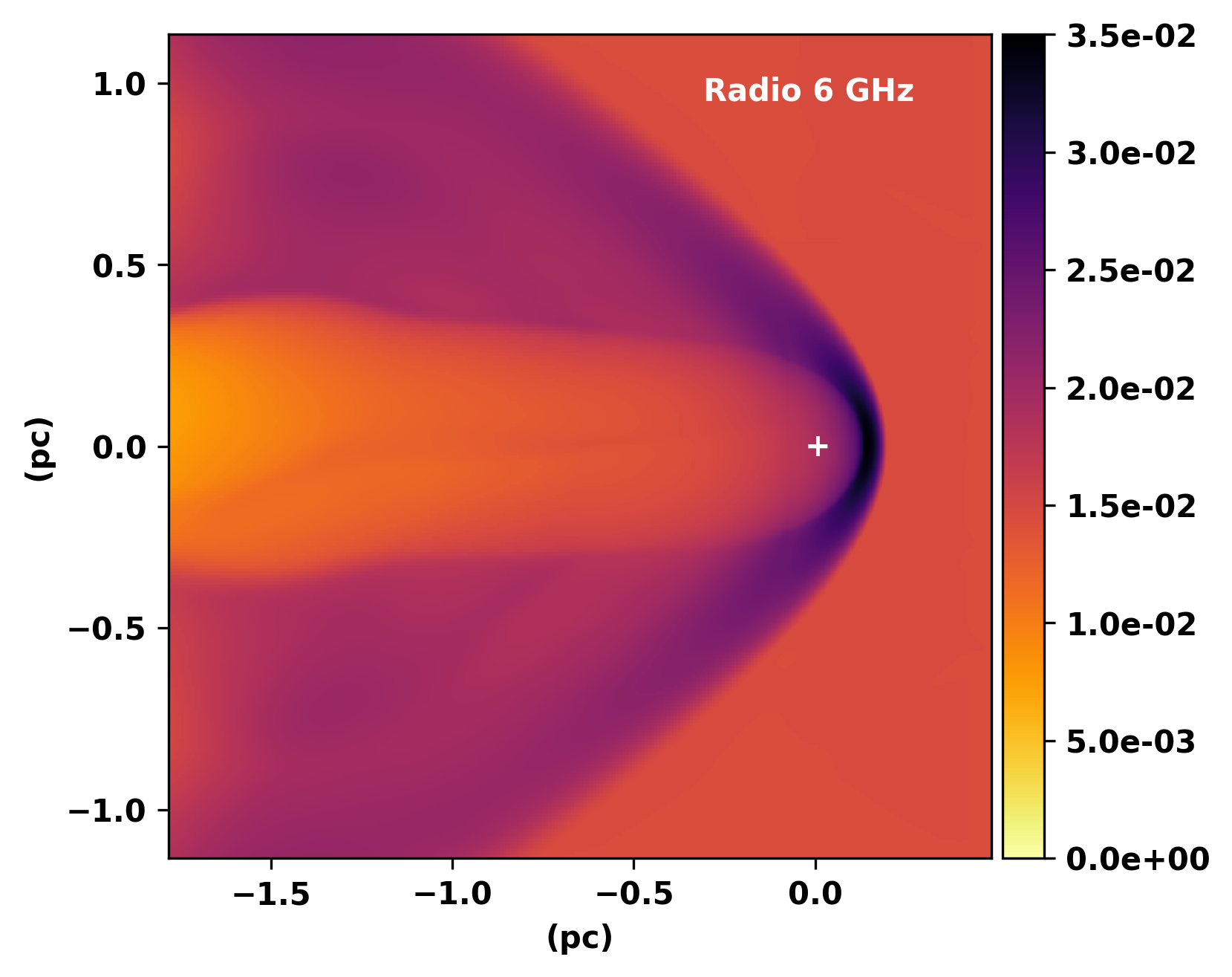} \\
	\includegraphics[height=.37\textwidth]{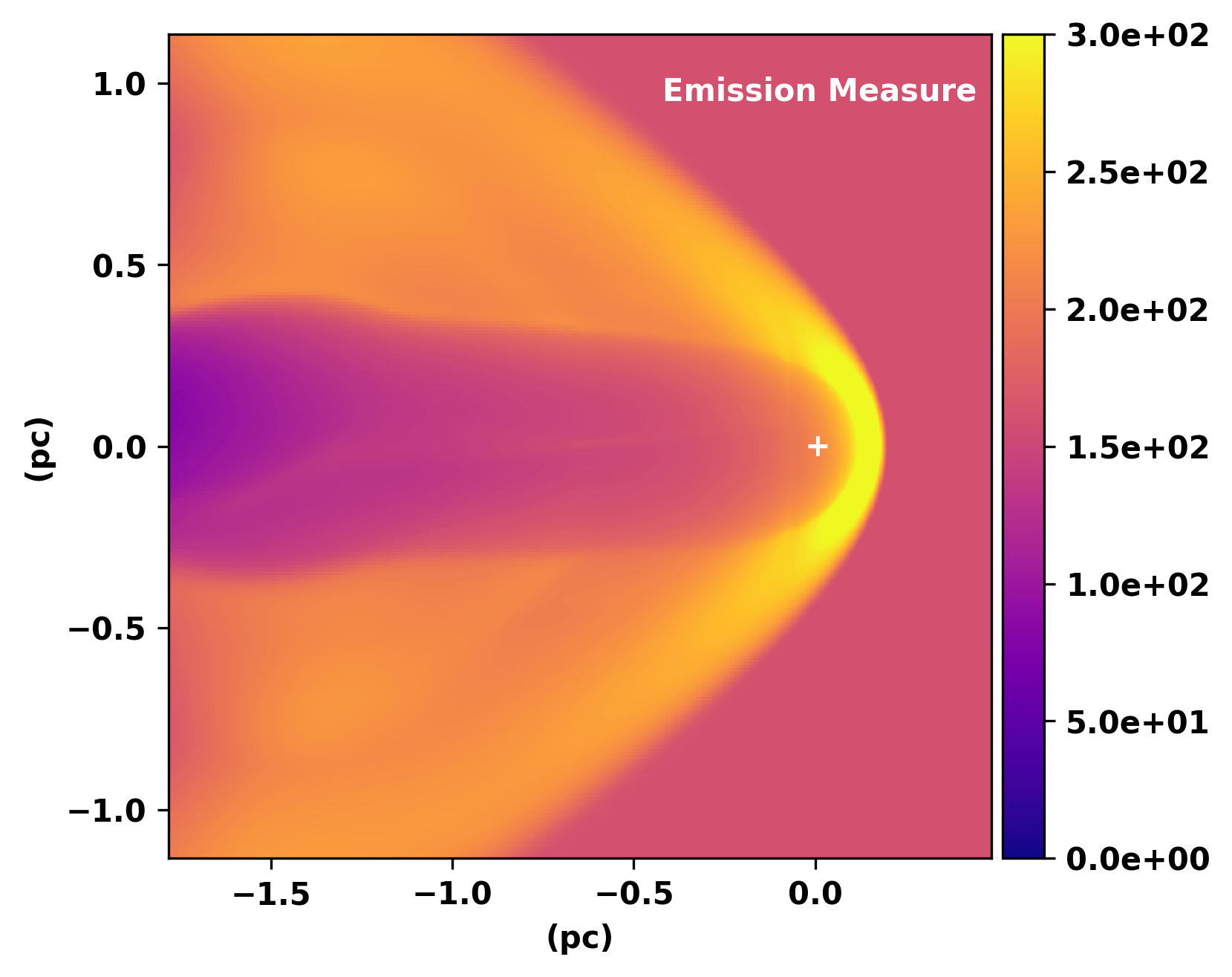}
	\caption{Above: Synthetic Radio 6 GHz emission maps of the simulated nebula around $\zeta$~Ophiuchi on a linear scale with units MJy ster$^{-1}$. Below: Synthetic Emission Measure maps of the simulated nebula around $\zeta$~Ophiuchi on a linear scale with units cm$^{-6}$\,pc. Both are the Z01 simulation. Coordinates are in parsecs relative to the position of the star (white cross).}
	\label{bremm}
\end{figure}

\subsection{Radio and Emission Measure synthetic data}
Using our 3D raytracing method again, we produced synthetic 6\,GHz radio emission (assuming thermal  bremsstrahlung) and synthetic Emission Measure maps of the Z01 simulation, shown in Fig.~\ref{bremm}. The Emission Measure is defined by

\begin{equation}
\text{EM} \equiv \int n_\mathrm{e}^2 d\ell,
\end{equation}

where $n_\mathrm{e}$ is the electron number density as a function of position along a line of sight $\ell$, and EM is traditionally measured in units cm$^{-6}$\,pc.
Both panels show a projection where the line of sight is the $y$-axis and the image coordinate axes are $x$ and $z$.
The maximum brightness of the radio emission is 0.035 MJy\,ster$^{-1}$, and 300 cm$^{-6}$\,pc for the Emission Measure.
In both instances the max brightness occurs at the apex of the bow shock.

As discussed in section \ref{sec:halphaemission}, $\zeta$~Ophiuchi is located within a large H~\textsc{ii} region that itself has significant emission measure (and also thermal radio emission) at the level of $\sim 100$\,cm$^{-6}$\,pc.
Fig.~\ref{bremm} shows that the background-subtracted emission measure of the bow shock is EM$\approx 150$\,cm$^{-6}$\,pc, which should be detectable.
The fact that the background emission is almost as bright as the bow shock, combined with the large size of the bow shock on the sky, makes this a challenging observation, unless the bow shock emits bright non-thermal emission as well as thermal.
Downstream from the apex, the background-subtracted emission measure is only $\sim (50-100)$\,cm$^{-6}$\,pc, and would be much more challenging to detect against the non-uniform background emission from the H~\textsc{ii} region.

Until recently only one bow shock (of BD+43$^\circ$3654) had been detected at radio frequencies \citep{Benaglia2010}, but in the past year a number of others have been detected: Vela X-1 \citep{2022MNRAS.510..515V}, bow shocks in the NGC\,6357 and RCW\,49 regions \citep{2022MNRAS.512.5374V} and also non-thermal emission from the Bubble Nebula, NGC\,7635 \citep{2022arXiv220411913M}.
These detections, together with our simulation results, suggest it could be worthwhile attempting to detect the bow shock of $\zeta$~Ophiuchi at radio frequencies.

\section{Discussion}
\label{sec:discusion}

\subsection{Simulation motivation}
Three simulations, Z01, Z02, and Z03, were run to create synthetic observations that are comparable to $\zeta$~Ophiuchi's features. Here we describe the motivation for running each simulation, and summarise the results obtained:

\textbf{Z01} - The Z01 simulation was modelled using parameters based on the system parameters calculated by \citet{gvaramadze2012zeta}, where the distance (d) to $\zeta$~Ophiuchi was estimated to be 112\,pc and the stellar space velocity was 26.5\,\kms.
As mentioned in section \ref{sec:infra}, the synthetic 24\,$\mu$m infrared emission map from Z01 had a comparable peak intensity and bow shock stand-off distance to observations, but the opening angle of the bow shock was too small.
The bow shock is now too small when compared to the updated distance, $d = 135$\,pc, from GAIA EDR3 \citep{Gaia2021}.
Fig.~\ref{xray_obs} shows diffuse X-ray emission with luminosity $2.33^{+1.12}_{-1.54}\times10^{29}$~erg~s$^{-1}$ from \textit{Chandra} observations.
This value is at least 8$\times$ more than what is predicted with our Z01 simulation ($\approx 10^{28}$ erg s$^{-1}$).
Also, the majority of X-ray emission was seen to occur around the apex of the bow shock, whereas the emission is seen to be closer to the star in observations. 

We discuss in section~\ref{sec:zeta_obs} that the radial velocity of $\zeta$~Ophiuchi could be significantly larger than previously estimated \citep{2018AN....339...46Z}: v$_r= 24.7\, \kms$ as oppose to $v_r = -2.5\,\kms$.
This implies a larger space velocity $v_\star$ and different viewing angle of the bow shock, which requires simulations with different parameters, motivating simulations Z02 and Z03.

\textbf{Z02} - The Z02 simulation was set up using a new distance to the star of 135 pc from \citet{Gaia2021}, a new peculiar transverse velocity (29\,\kms) and peculiar radial velocity (24.7\,\kms) from \citet{2018AN....339...46Z}.
We approximate the total space velocity as $v_\star=40\,\kms$.
To compensate for the larger space velocity, we reduced the ISM number density so that the standoff distance would be correct.
Z02 was found to have a comparable infrared intensity to the \textit{Spitzer} observations. The bow shock stand-off distance and opening angle were also comparable to observations.
We obtain $R_\mathrm{TS}=0.09$\,pc,  $R_\mathrm{CD}=0.13$\,pc, and $R_\mathrm{FS}=0.26$\,pc which are consistent with the position and width of the infrared arc seen in \textit{Spitzer} and \textit{WISE} data.
The synthetic X-ray emission maps from this simulation reduced the amount of emission from the bow-shock, potentially bringing the emission closer to the star.
The calculated thermal X-ray emission from the simulated wind bubble (luminosity $\approx (1$-$4) \times 10^{28}$ erg\,s$^{-1}$) is fainter than the the \textit{Chandra} diffuse X-ray observations ($2.33^{+1.12}_{-1.54}\times10^{29}$~erg~s$^{-1}$), although the simulation upper limit is only a factor of 2 below the observational lower limit.

\textbf{Z03} -
Because the X-ray emission of Z02 was still too faint, we set up Z03 with a twice larger mass-loss rate from the star and twice larger ISM number density.
We expected that the twice higher density within the wind bubble would produce X-ray luminosity approximately 4 times larger (because it scales with $n_e^2$), but that the size and shape of the bow shock would be unchanged.
These expectations were proven correct: Z03 was found to have a comparable a bow shock stand-off distance and opening angle to observations but its infrared intensity was $\sim 2$ higher than \textit{Spitzer} observations because of the higher gas density.
The calculated thermal X-ray emission from the simulated wind bubble does show a more comparable luminosity ($4 \times 10^{28}$ to $1 \times 10^{29}$ erg\,s$^{-1}$) to the \textit{Chandra} diffuse X-ray observations ($2.33^{+1.12}_{-1.54}\times10^{29}$~erg~s$^{-1}$).


\begin{figure*}
	\centering
	\includegraphics[height=.40\textwidth]{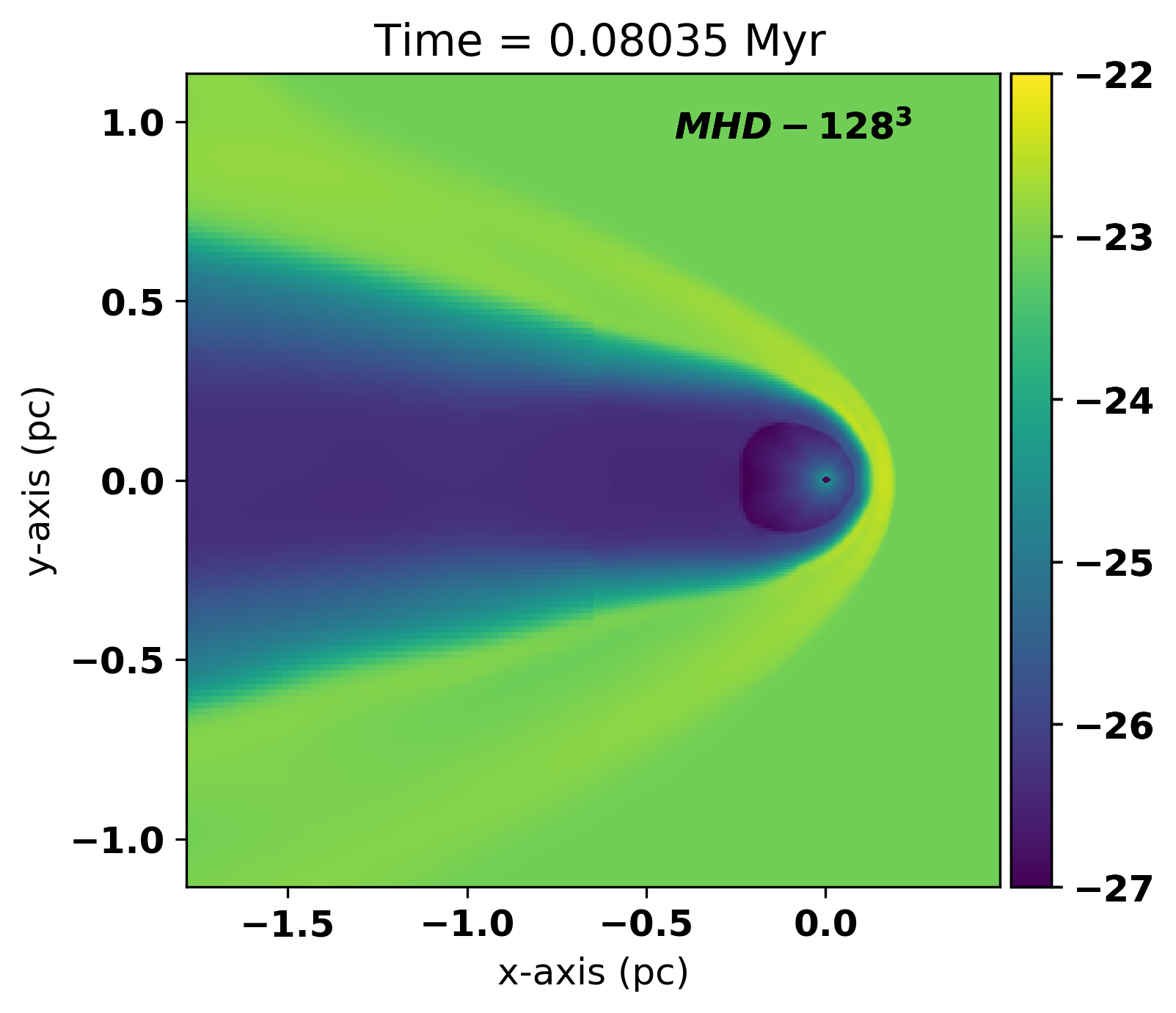} 
	\includegraphics[height=.40\textwidth]{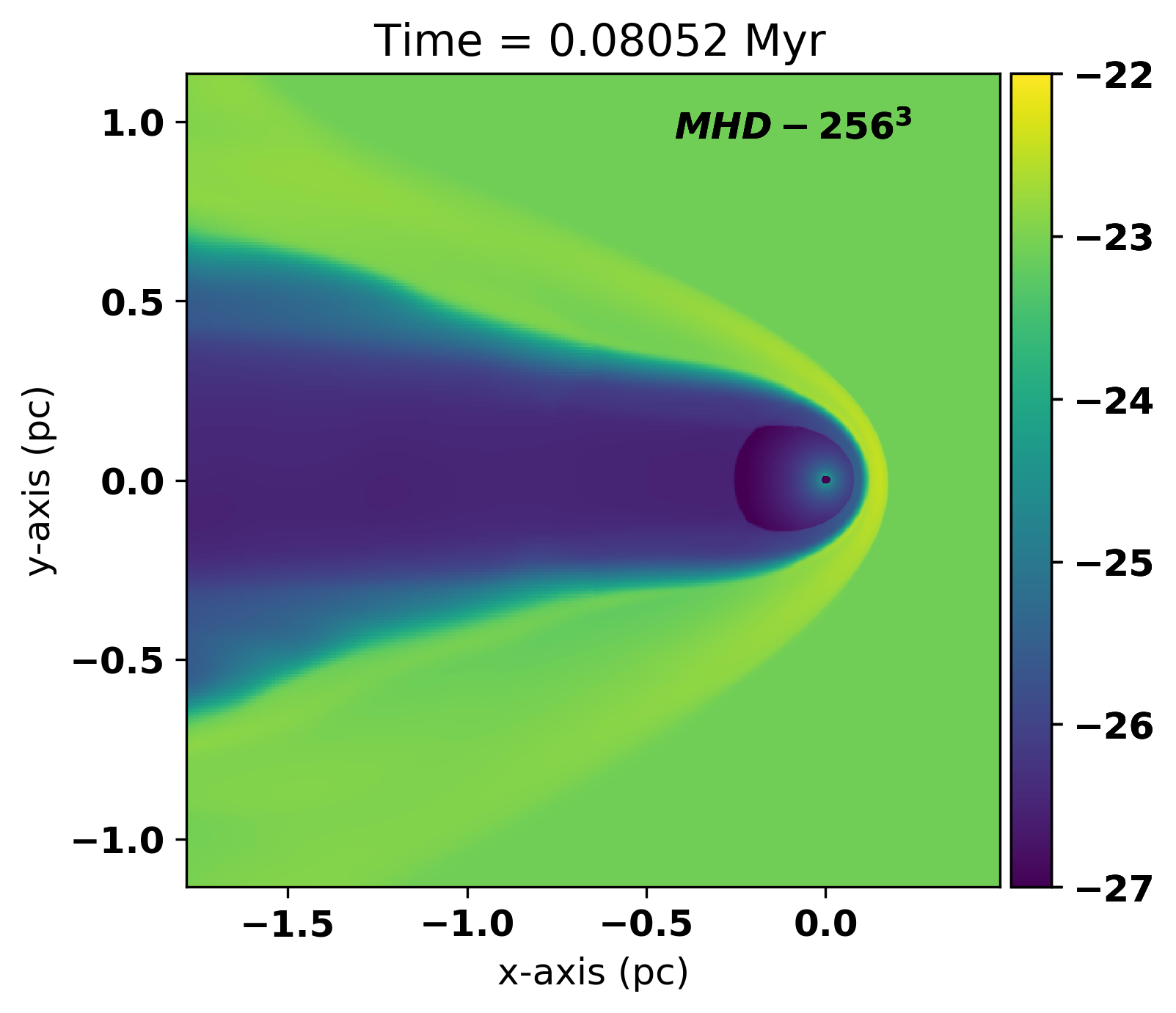} \\
	\includegraphics[height=.40\textwidth]{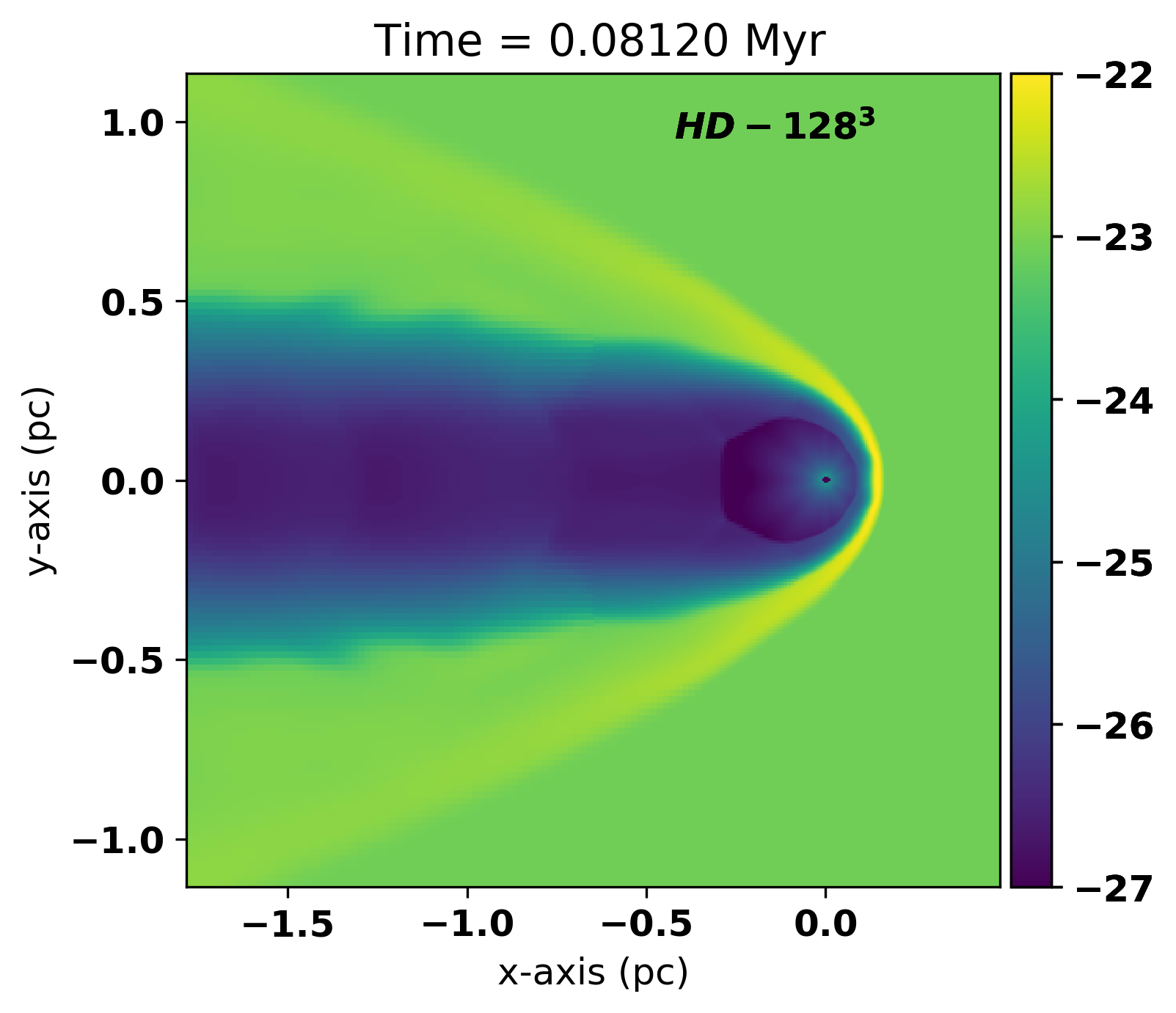} 
	\includegraphics[height=.40\textwidth]{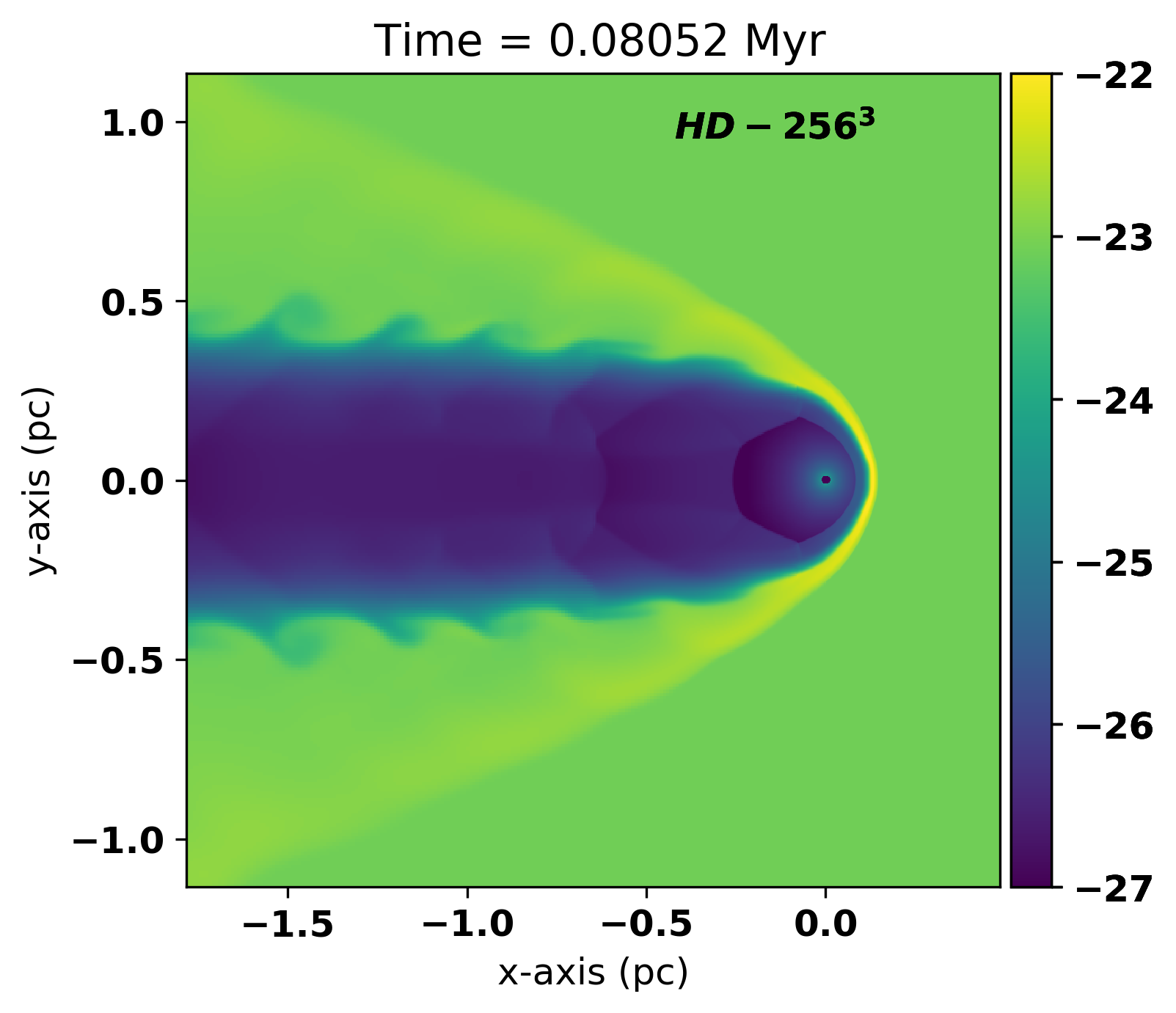} \\
	\caption{Plot of $\log_{10}$ of gas density (g cm$^{-3}$) of simulation Z02 for four different cases: top-left with MHD and resolution 128$^3$ cells per level, top-right with MHD and 256$^3$, bottom-left with hydrodynamics and 128$^3$, and bottom-right with hydrodynamics and 256$^3$. Slices through the ($x$-$y$) plane at $z = 0$ are shown, with the star at the origin.}
	\label{fig:z02_res}    
\end{figure*}

Overall none of the simulations provides a perfect match to the available observational data.
In particular the morphology of the observed X-ray emission is not matched by the simulation results, which always show a bright arc at the location of the contact discontinuity, and the overall X-ray luminosity of our simulations is a factor of a few too low except for Z03.
\emph{A priori}, we expected the opposite to be the case, because similar models over-predicted the thermal X-ray emission from NGC\,7635 \citep{Green, 2020MNRAS.495.3041T}.
It is tempting to argue that the inclusion of thermal conduction in our simulations would increase the X-ray emission, and perhaps move the peak of the emission closer to the wind-driving star \citep[cf.][]{2014meyer}, but this would exacerbate the problems with NGC\,7635.
Nevertheless it would be valuable to run similar simulations to those presented here, but including anisotropic thermal conduction as implemented in, e.g., \textsc{pluto} \citep{2012ApJS..198....7M, Meyer2017}, to compare the predictions for synthetic X-ray emission.
From our simulations we found an emission-weighted temperature of $\log_{10} (T_\mathrm{A}/\mathrm{K})\approx6.4$, or 0.22\,keV, comparable to the temperature derived from observations (0.2\,keV) by \citet{Toala2016}.
Thermal conduction would be expected to lower the emission-weighted temperature even further \citep{2011ApJ...737..100T}, and so we do not expect it to have a strong effect on our results.

Another explanation for the discrepancy between our simulations and observations could be that the separation of stellar and diffuse X-ray photons in the \textit{Chandra} data reduction is not perfect, and that the true diffuse emission could be fainter than what we measured. While we have taken every measure to quantify the contribution of the stellar photons to the spectrum from the extended emission region, it is challenging to disentangle from this from the intrinsically diffuse emission.
Further observations would be very valuable with e.g.\ \textit{XMM-Newton} \citep{Jansen2001}. We used the spectral parameters and flux of the diffuse emission determined in this work with the WebPIMMs tool\footnote{\url{https://heasarc.gsfc.nasa.gov/cgi-bin/Tools/w3pimms/w3pimms.pl}} to determine that only a modest exposure time is required with the EPIC \citep{Struder2001,Turner2001} to better constrain the extent and parameters of the diffuse emission. However, contamination from the PSF wings of the $\zeta$~Ophiuchi emission will still make the analysis demanding.
In the longer term, \textit{Athena} \citep{Nandra2013} and {\it Lynx} \citep{Gaskin2019} should easily allow us to trace and characterise the diffuse emission to even lower flux levels, with the high-sensitivity, sub-arcsecond imaging of {\it Lynx} being particularly suited to the case of $\zeta$~Ophiuchi \citep[see][for a comparison of the capabilities of future X-ray observatories]{Kavanagh2020b}.

\begin{figure}
	\centering
	\includegraphics[height=.32\textwidth]{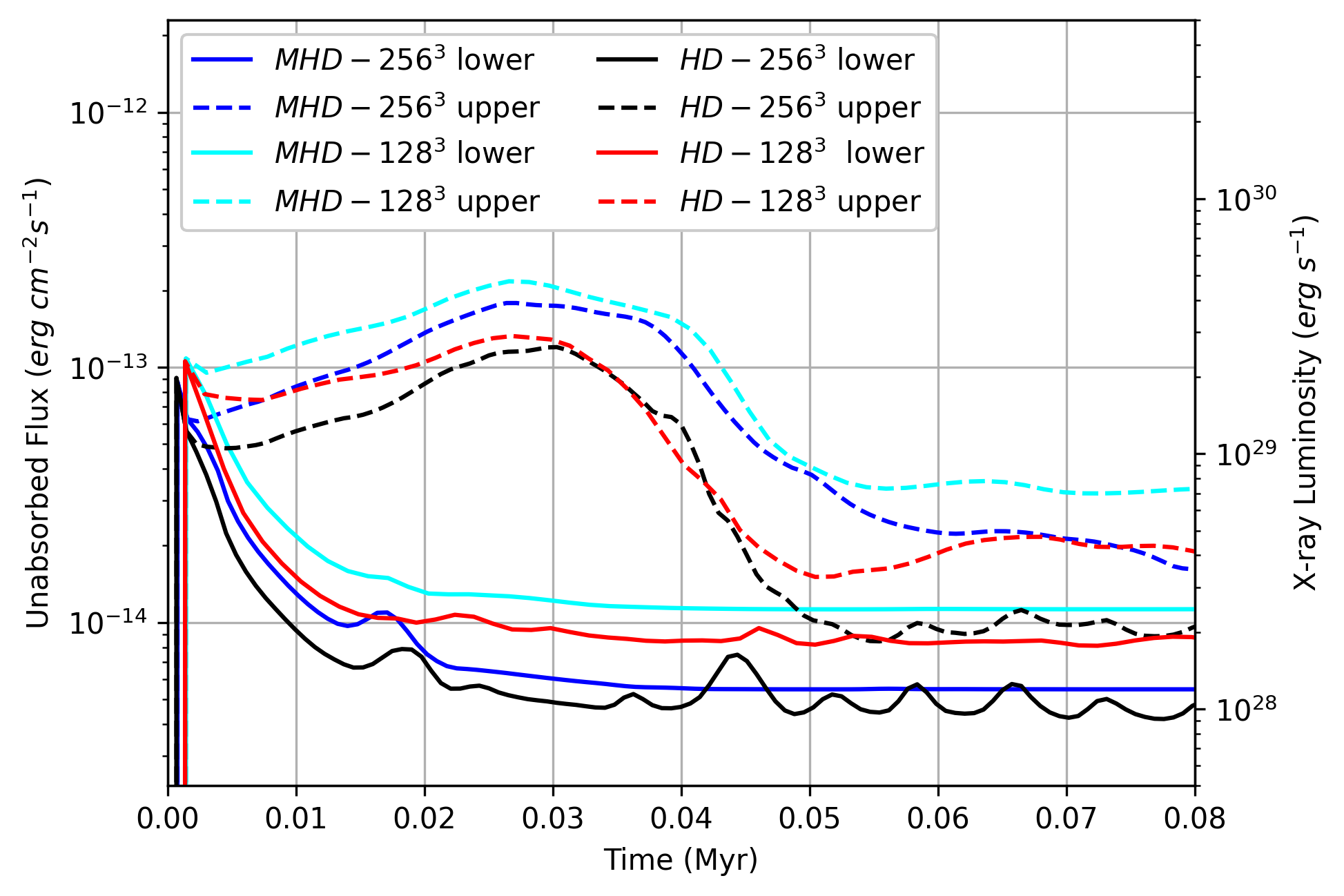}
	\caption{Synthetic soft (0.3\,keV - 1\,keV) X-ray unabsorbed flux (erg\,cm$^{-2}$\,s$^{-1}$) and luminosity (erg\,s$^{-1}$) plot of emission from the bow shocks shown in Fig.~\ref{fig:z02_res} as they evolve in time (Myr). The solid lines represent the \textit{Chandra} field of view and the dashed lines are the whole simulation grid, for the simulations as indicated in the legend.}
	\label{fig:lum_res}
\end{figure}

\begin{figure*}
	\centering
	\includegraphics[height=.29\textwidth]{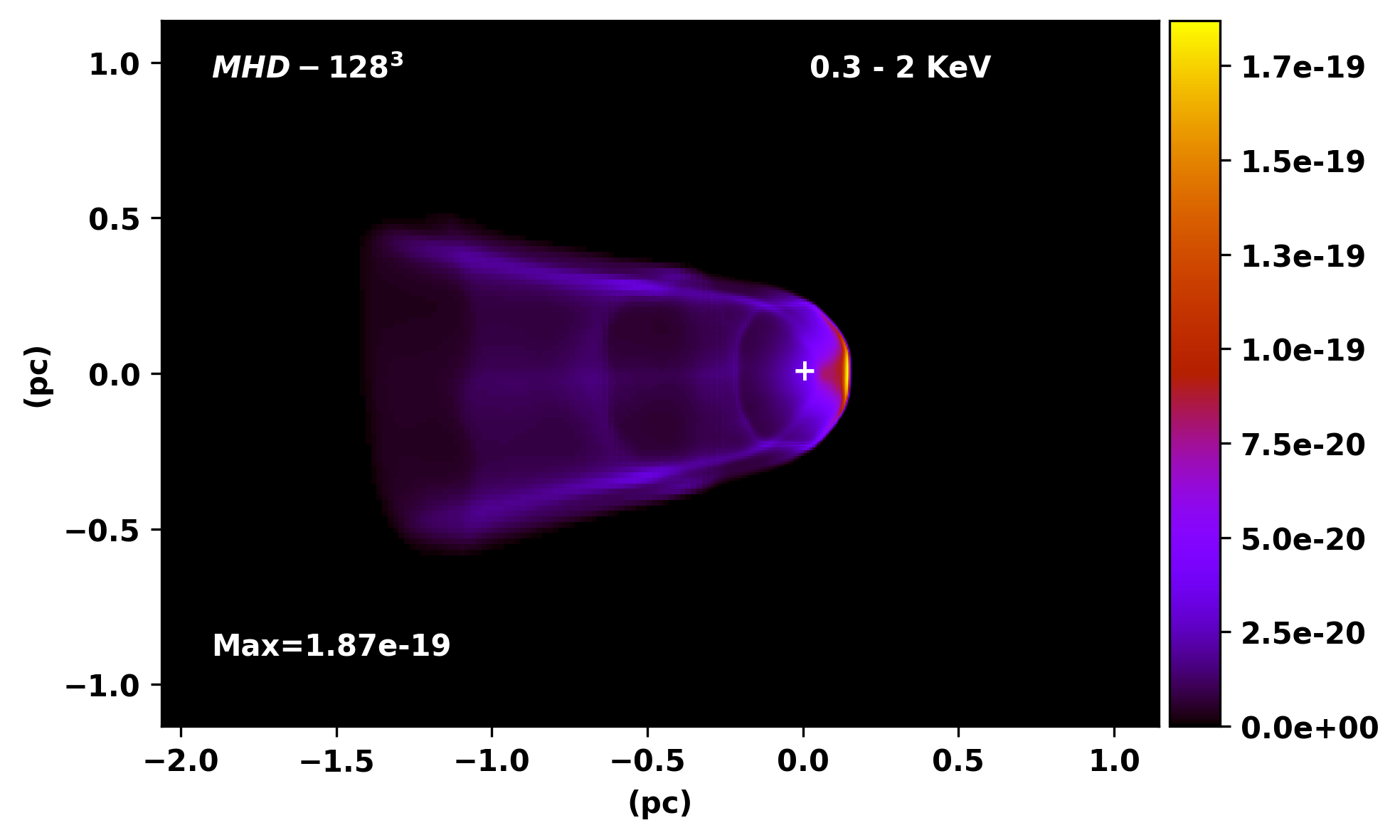} 
	\includegraphics[height=.29\textwidth]{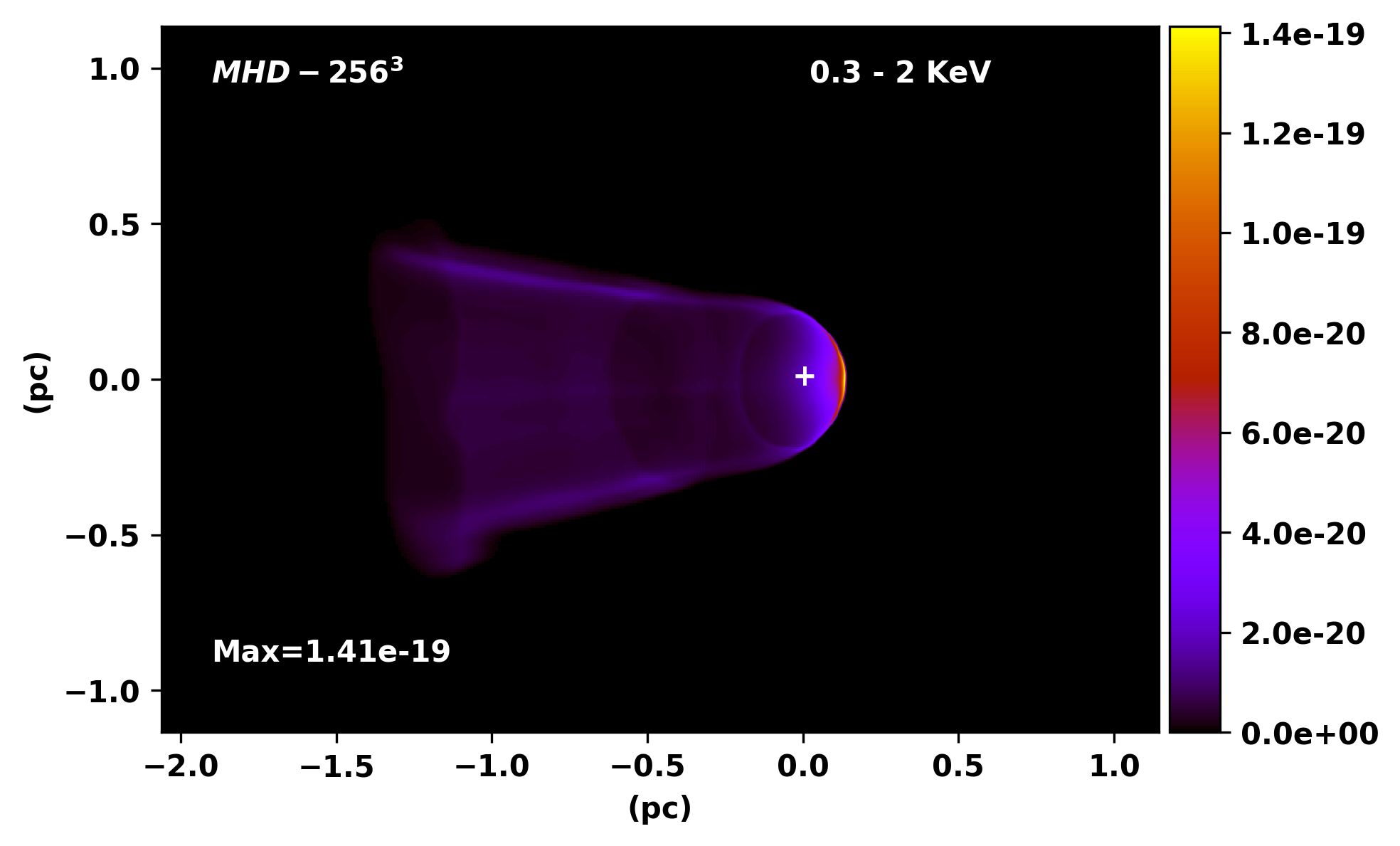} \\
	\includegraphics[height=.29\textwidth]{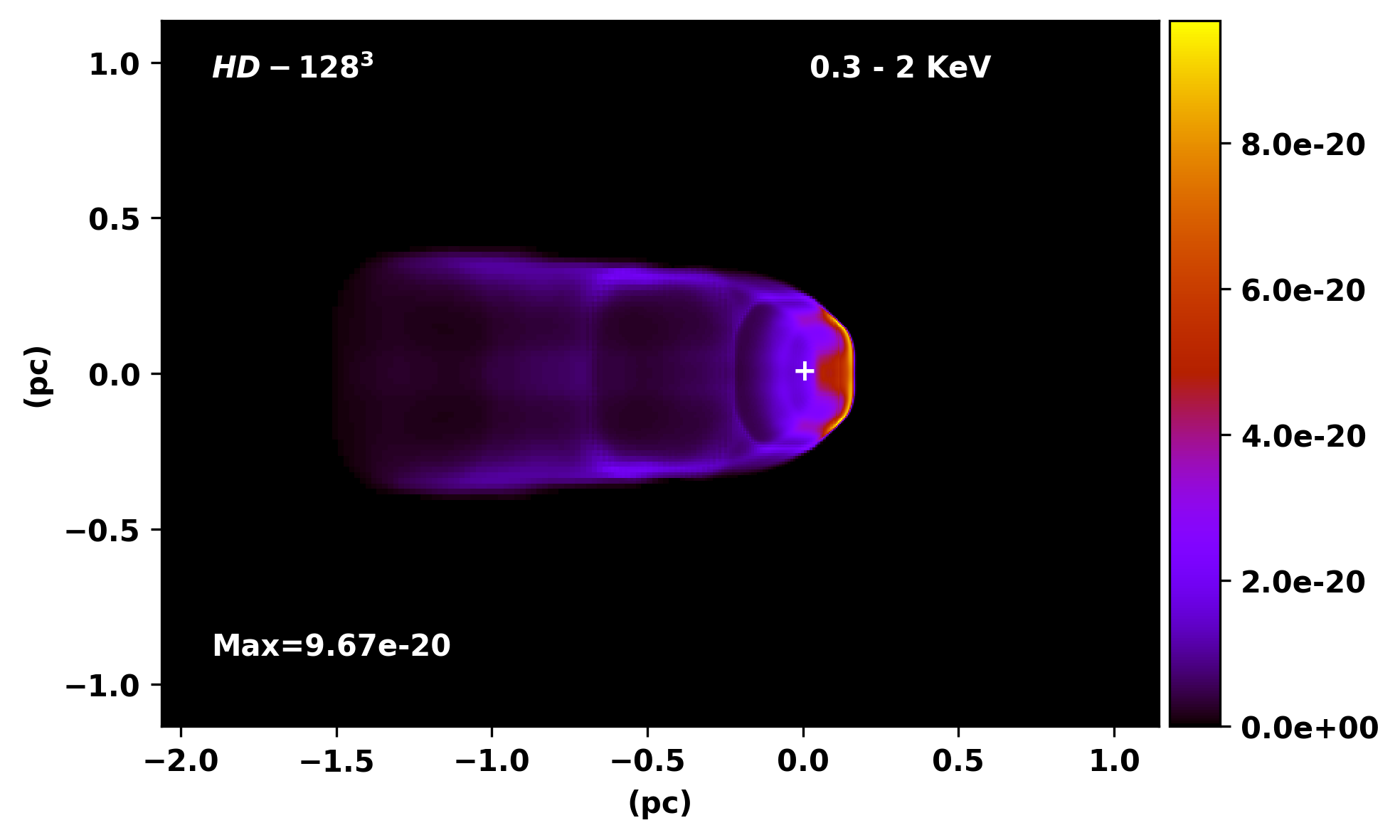}
	\includegraphics[height=.29\textwidth]{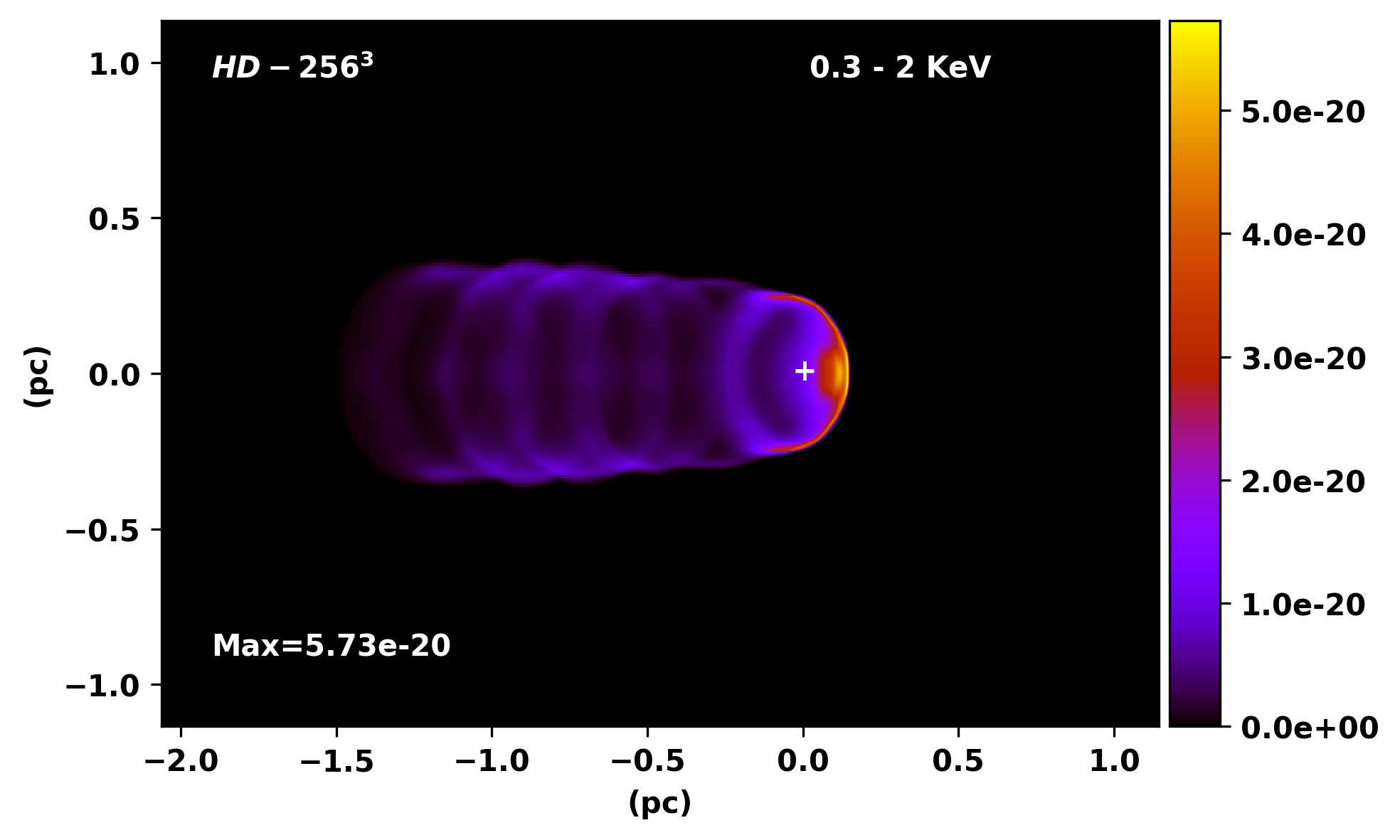}
	\caption{Synthetic soft X-ray (0.3 - 2 KeV) emission maps of the simulated nebula around $\zeta$~Ophiuchi (unabsorbed) for the snapshots shown in Fig.~\ref{fig:z02_res}. Top-left: simulation Z02 with MHD and at 128$^3$. Top-right: simulation Z02 with MHD and at 256$^3$. Bottom-left: simulation Z02 with HD and at 128$^3$. Bottom-right: simulation Z02 with HD and at 256$^3$. All images are projected at an angle of $45^{\circ}$ to the $x$-axis. Coordinates are in parsecs relative to the position of the star (white cross) and the colour scale is from zero to maximum in erg\,cm$^{-2}$\,s$^{-1}$\,arcsec$^{-2}$.}
	\label{fig:xray_res}
\end{figure*}

\subsection{Effects of resolution and magnetic fields}
In the 2D simulations of \citet{Green}, the X-ray emission was mainly produced by the mixing of wind and ISM gas in the wake downstream from the star, but here this is not the case and the brightest X-ray emission is from the apex of the bow shock.
There are three differences between the simulations presented here and those in \citet{Green}: 3D instead of 2D with somewhat lower resolution, MHD instead of hydrodynamics, and the ISM density is significantly lower in this work than in the previous paper.
Any of these three effects could be playing a role in the qualitatively different X-ray emission maps that we obtain here.

To investigate this we ran the simulation Z02 with lower resolution of $128^3$ cells and 3 refinement levels, and also without a magnetic field, i.e., hydrodynamics (at both resolutions).
Results of the four simulations are shown in Fig.~\ref{fig:z02_res} at approximately the same evolutionary stage, for a slice through the simulation at $z=0$.
The hydrodynamic simulations are prone to Kelvin-Helmholtz (KH) instability at the contact discontinuity; these can barely develop at $128^3$ resolution but are much clearer at $256^3$ resolution.
This instability is suppressed in the MHD simulations; this appears not to be a resolution effect and is expected on physical grounds \citep{1996ApJ...460..777F}.

Fig.~\ref{fig:lum_res} shows the time evolution of X-ray luminosity from the four simulations from the $(0.5\,\mathrm{pc})^3$ region around the star (solid lines) and the full simulation domain (dashed lines).
For the full domain both MHD simulations are more luminous than the two hydrodynamic simulations, despite the lack of KH instability-driven mixing.
This may be because the HLL MHD solver is diffusive and generates a broad mixing region at the wind-ISM interface.
For both hydrodynamics and MHD, the higher-resolution simulations have lower X-ray emission than the lower-resolution simulations, the difference being about a factor of 2 in luminosity.
This indicates that the results have not yet converged, a sign that the numerical diffusivity of the contact discontinuity is impacting the derived emission.
For the emission from the $(0.5\,\mathrm{pc})^3$ region around the star, the hydrodynamical and MHD simulations give very similar results at the same resolution, apart from the modulation of the hydrodynamical results by KH instability.
Fig.~\ref{fig:xray_res} shows synthetic X-ray emission maps from the four simulation snapshots; in all cases the brightest emission is from the apex of the bow shock, although emission from the wake shows some differences.

None of the simulations show the strong mixing of dense clumps into the shocked wind that was found for simulations of NGC\,7635 \citep{Green}.
We may speculate that this may be because here the ISM is lower density, with less inertia, easily pushed aside and not able to penetrate into the low-density wake.
This would need to be tested with a systematic study of bow shocks in different density environments.
Alternatively it could simply be a lack of resolution, but we did not have the computational resources for this project to run higher resolution simulations.
The issue of energy dissipation in stellar-wind bubbles and bow shocks, through turbulent mixing or thermal conduction, is currently actively investigated from both theoretical \citep{Green, 2021MNRAS.501.1352G, 2021ApJ...914...89L, 2021ApJ...914...90L} and observational \citep{2014MNRAS.442.2701R, 2021ApJ...908...68O} perspectives.
It is important to constrain this because the level of energy dissipation in wind bubbles crucially determines whether or not stellar-wind feedback is important for the dynamics of the ISM in galaxies.

\section{Conclusions}
\label{sec:conclusions}

This paper is the continuation of a project to investigate thermal emission from stellar wind bubbles surrounding runaway O stars.
We followed up our 2D study of NGC\,7635 with a detailed 3D study of the bow shock of $\zeta$~Ophiuchi.
To date, this has the only detection of diffuse thermal X-ray emission from a wind bubble of an isolated massive star, and ours is the first investigation of this emission with multi-dimensional simulations.
3D MHD simulations have been run to model the interaction of the star's wind with the interstellar medium, using a range of stellar and ISM parameters appropriate for comparison with $\zeta$~Ophiuchi.

We re-analysed \emph{Chandra} archival observations to produce a diffuse emission map and estimate the total diffuse, thermal, X-ray flux from the shocked wind region of the bow shock.
Overall our results are consistent with the published analysis by \citet{Toala2016}, and in addition we present a new map plotting the diffuse emission and a bounding contour enclosing the emission above 3$\sigma$.
We found a total unabsorbed X-ray flux in the 0.3-2\,keV band corresponding to diffuse X-ray luminosity of $L_\mathrm{X}=2.33^{+1.12}_{-1.54}\times10^{29}$~erg~s$^{-1}$.

Three simulations (Z01, Z02, and Z03) were presented and compared with observational data.
The Monte-Carlo radiative-transfer code \textsc{TORUS} was used to post-process the simulations to generate synthetic 24$\mu$m emission-map predictions to compare with observational \textit{Spitzer} MIPS data.
A ray-tracing projection code was also used to produce synthetic Emission Measure, Radio Bremsstrahlung, H$\alpha$ and soft ($0.3-2$ keV) X-ray emission maps to compare with the relevant observational data.

The Z01 simulation, for the case where the star's space velocity is entirely in the plane of the sky, was found to have a comparable mid-IR intensity and bow shock stand-off distance to observations, although the opening angle of the bow shock was too small. The maximum brightness of our 24$\mu$m synthetic emission maps from Z01 are also comparable with the corresponding observational data. The synthetic X-ray emission maps show that the majority of X-ray emission occurs at the apex of the bow shock at the contact discontinuity.
This arc-shaped emission region is not apparent in the observational data.
Calculated thermal X-ray emission from the simulated wind bubble is significantly fainter (luminosity $\sim 10^{28}$ - $ 10^{29}$ erg\,s$^{-1}$) than the \textit{Chandra} diffuse X-ray observations ($2.33^{+1.12}_{-1.54}\times10^{29}$~erg~s$^{-1}$).

Recent spectroscopic observations suggest that $\zeta$~Ophiuchi could have a significant radial velocity, however, comparable to its plane-of-sky velocity.
This motivated simulations Z02 and Z03, with a faster moving star (40\,\kms{} compared with 26.5\,\kms{} for Z01).
Simulation Z02 has many similar properties to Z01, and fits the observational data somewhat better: the opening angle of the bow shock is closer to that inferred from mid-IR observations, and the morphology of the X-ray emission is more of a filled bubble rather than an arc because of the different angle to the line of sight.
The total diffuse X-ray flux remains well below observations.

Simulation Z03, with a larger mass-loss rate and ISM density, has a higher total pressure and density in all parts of the bow shock, and hence more intense emission at all wavelengths.
The predicted X-ray emission is much closer to the observational values (luminosity $\sim [0.4-1] \times 10^{29}$ erg\,s$^{-1}$), although the predicted mid-IR emission is 2$\times$ brighter than observed.
We have made no attempt to fine-tune the dust properties for this calculation, and so the factor of 2 disagreement in mid-IR intensity is perhaps not so significant.

Comparison between hydrodynamic and MHD simulations showed that the Kelvin-Helmholtz instability is apparent at the contact discontinuity in the former case but absent (or too weak to see at these resolutions) when magnetic fields are included.
Our results for X-ray luminosity are also somewhat resolution-dependent and not converged, in the sense that higher resolution has lower luminosity.
Further work is required to assess whether instabilities would arise in even higher-resolution MHD simulations.

The shocked-wind region around $\zeta$~Ophiuchi is the closest object to Earth where bubble energetics and dissipative processes for the wind of a massive star can be investigated, and as such it is an ideal laboratory for constraining the relevant physical processes.
This first numerical study of the bow shock and wind bubble around $\zeta$~Ophiuchi does not give simple answers to the important questions, but our work can be used as a basis for building more complicated models including inhomogeneous and turbulent ISM, anisotropic thermal conduction, particle acceleration and transport, and more detailed wind models.
Better observational data would also be very helpful, because the existing X-ray dataset has significant contamination of the diffuse emission by stellar emission.


\section*{Acknowledgements}
\addcontentsline{toc}{section}{Acknowledgements}
SG acknowledges funding from the Dublin Institute for Advanced Studies. 
We acknowledge the SFI/HEA Irish Centre for High-End Computing (ICHEC) for the provision of computational facilities and support (project dsast023b). 
JM acknowledges funding from a Royal Society-Science Foundation Ireland University Research Fellowship (UF140375, 14/RS-URF/3219) and a University Research Fellowship Renewal (URFR201015, 20/RS-URF-R/3712). 
TJH is funded by a Royal Society Dorothy Hodgkin Fellowship. 
MM acknowledges funding from a Royal Society Research Fellows Enhancement Award (RGF\textbackslash EA\textbackslash 180214, 17/RS-EA/3468).
We are grateful to the referee for a detailed report that substantially improved the presentation of our work.
This work has made use of the NASA/IPAC Infrared Science Archive, which is operated by the Jet Propulsion Laboratory, California Institute of Technology, under contract with the National Aeronautics and Space Administration, and the  SIMBAD database, operated at CDS, Strasbourg, France. 
This work also has made use of  data from the European Space Agency (ESA) mission {\it Gaia} (\url{https://www.cosmos.esa.int/gaia}), processed by the 
{\it Gaia} Data Processing and Analysis Consortium (DPAC, \url{https://www.cosmos.esa.int/web/gaia/dpac/consortium}).
Funding for the DPAC has been provided by national institutions, in particular the institutions participating in the {\it Gaia} Multilateral Agreement.
This research made use of Astropy, a community-developed core Python package for Astronomy \citep{astropy:2013, astropy:2018},  Numpy \citep{harris2020array}, matplotlib \citep{Hunter:2007} and yt \citep{2011ApJS..192....9T}.


 


\bibliographystyle{aa}
\bibliography{refs}

\appendix




\section{Projected emission at all angles}
\begin{figure*}
	\includegraphics[width=.138\textwidth]{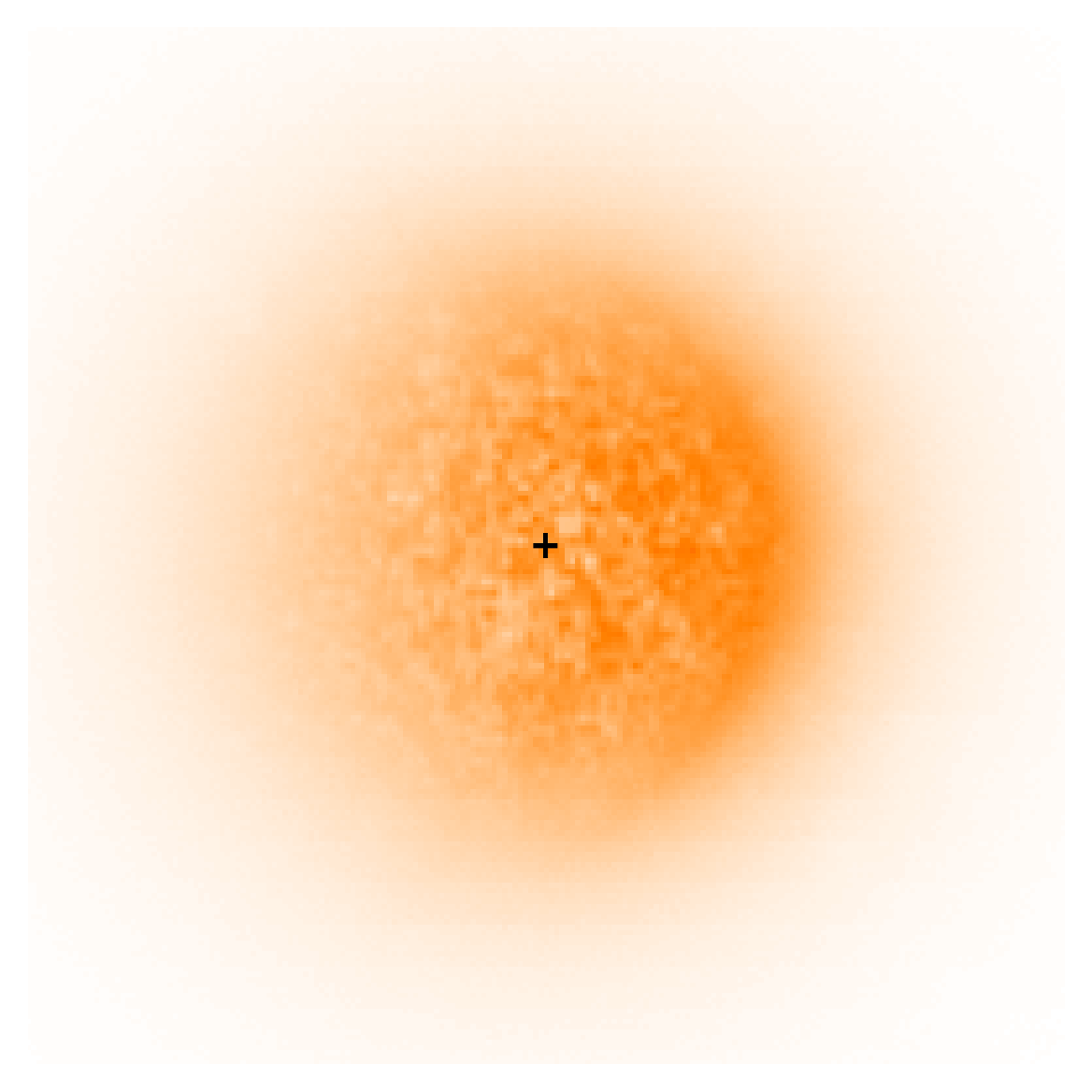}
	\includegraphics[width=.138\textwidth]{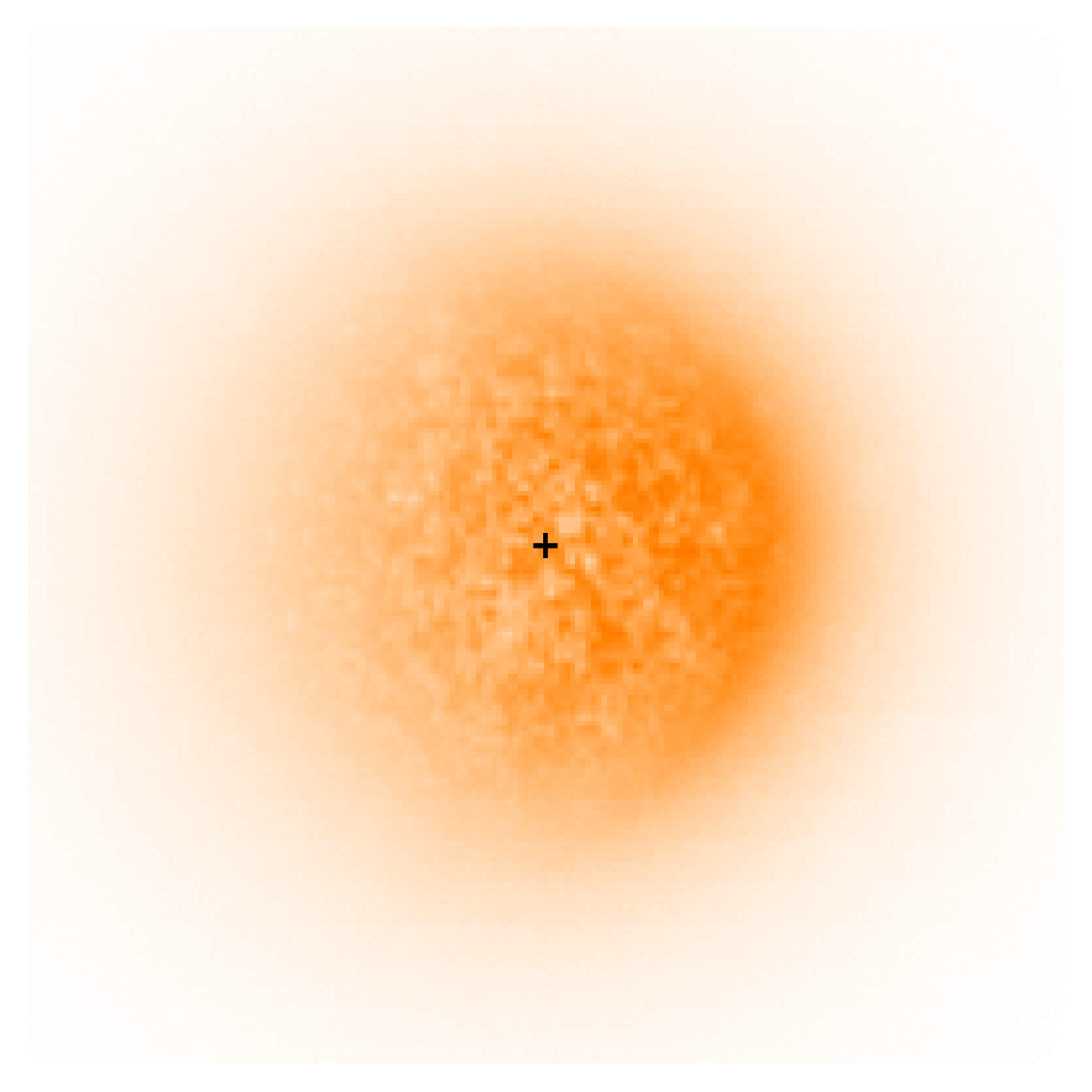}
	\includegraphics[width=.138\textwidth]{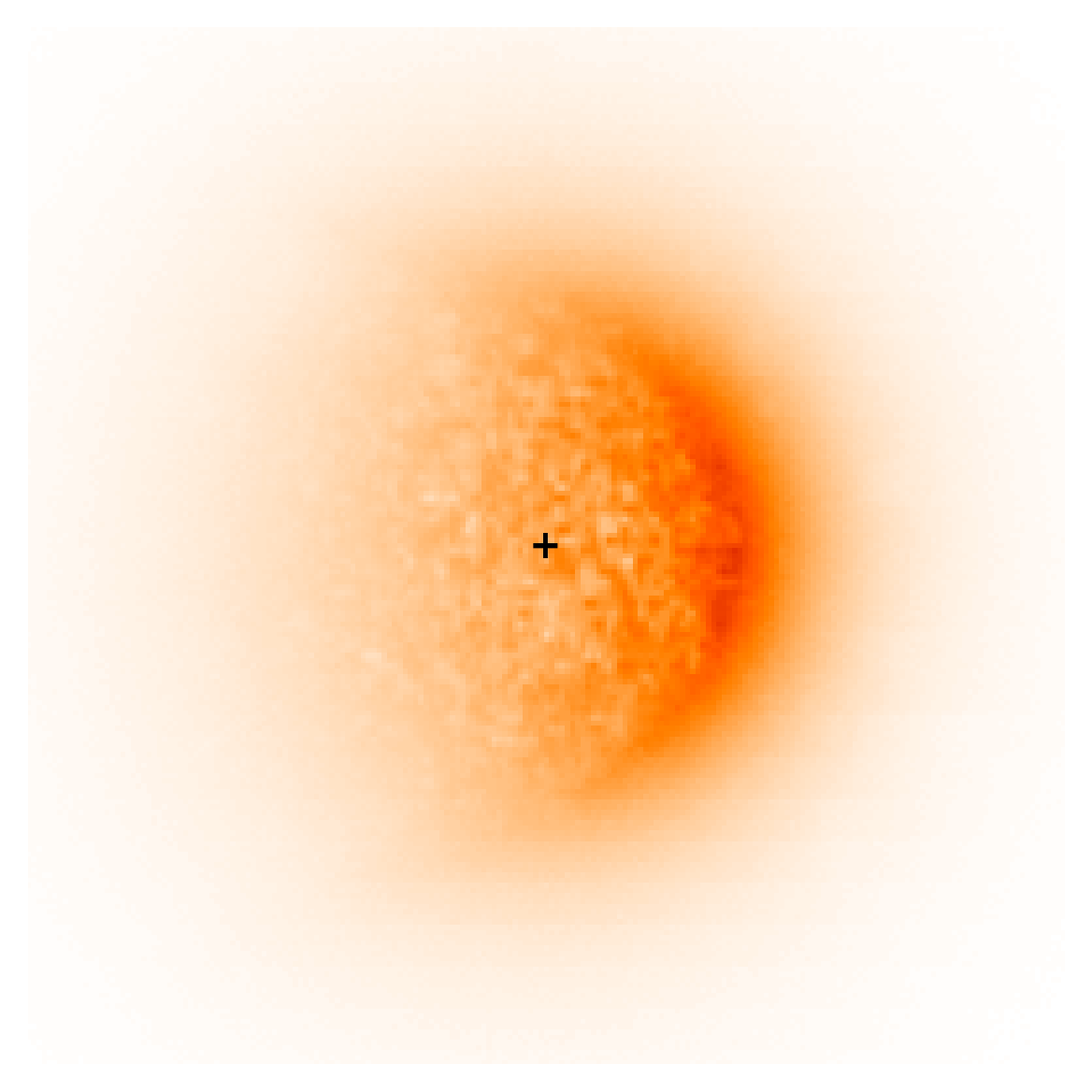}
	\includegraphics[width=.138\textwidth]{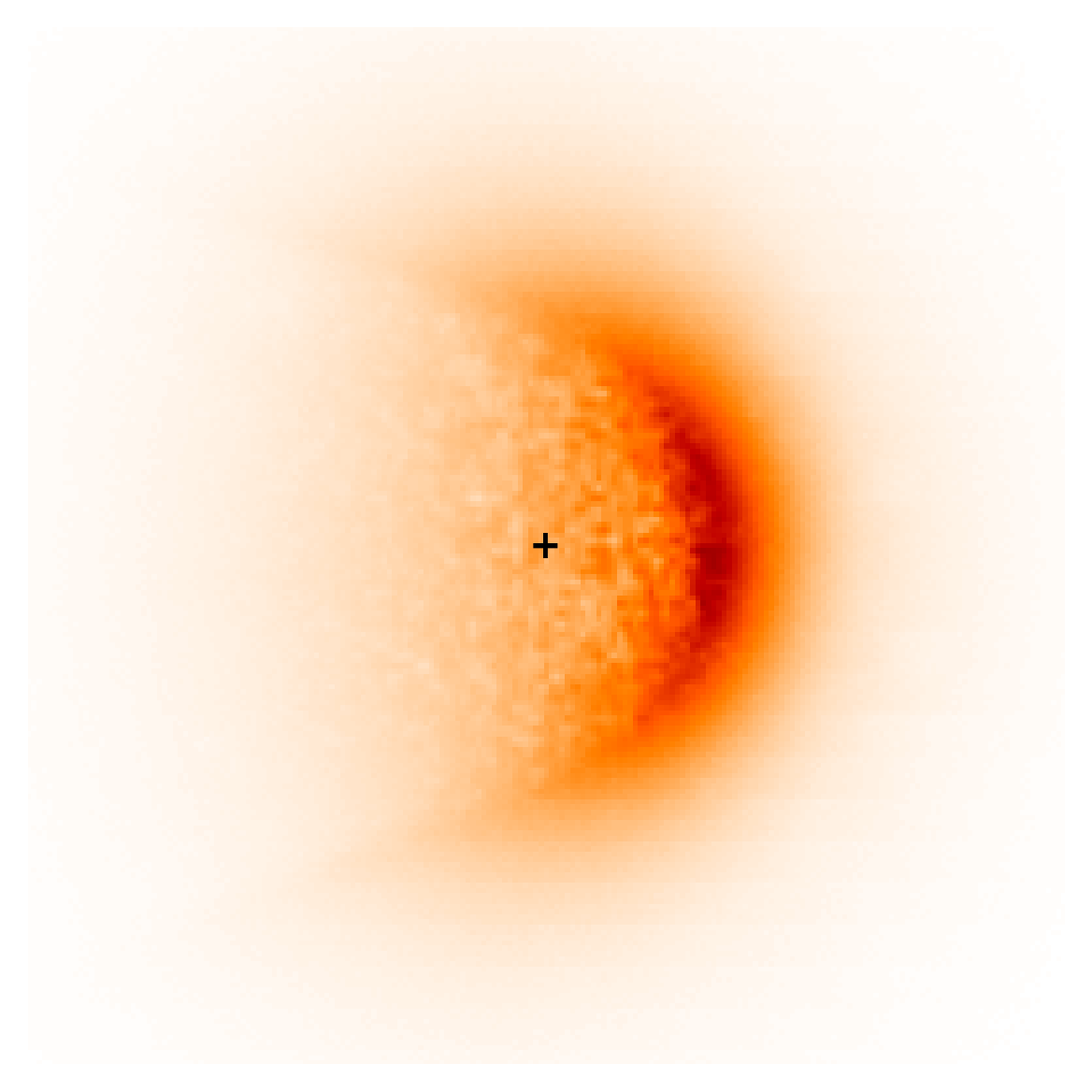}
	\includegraphics[width=.138\textwidth]{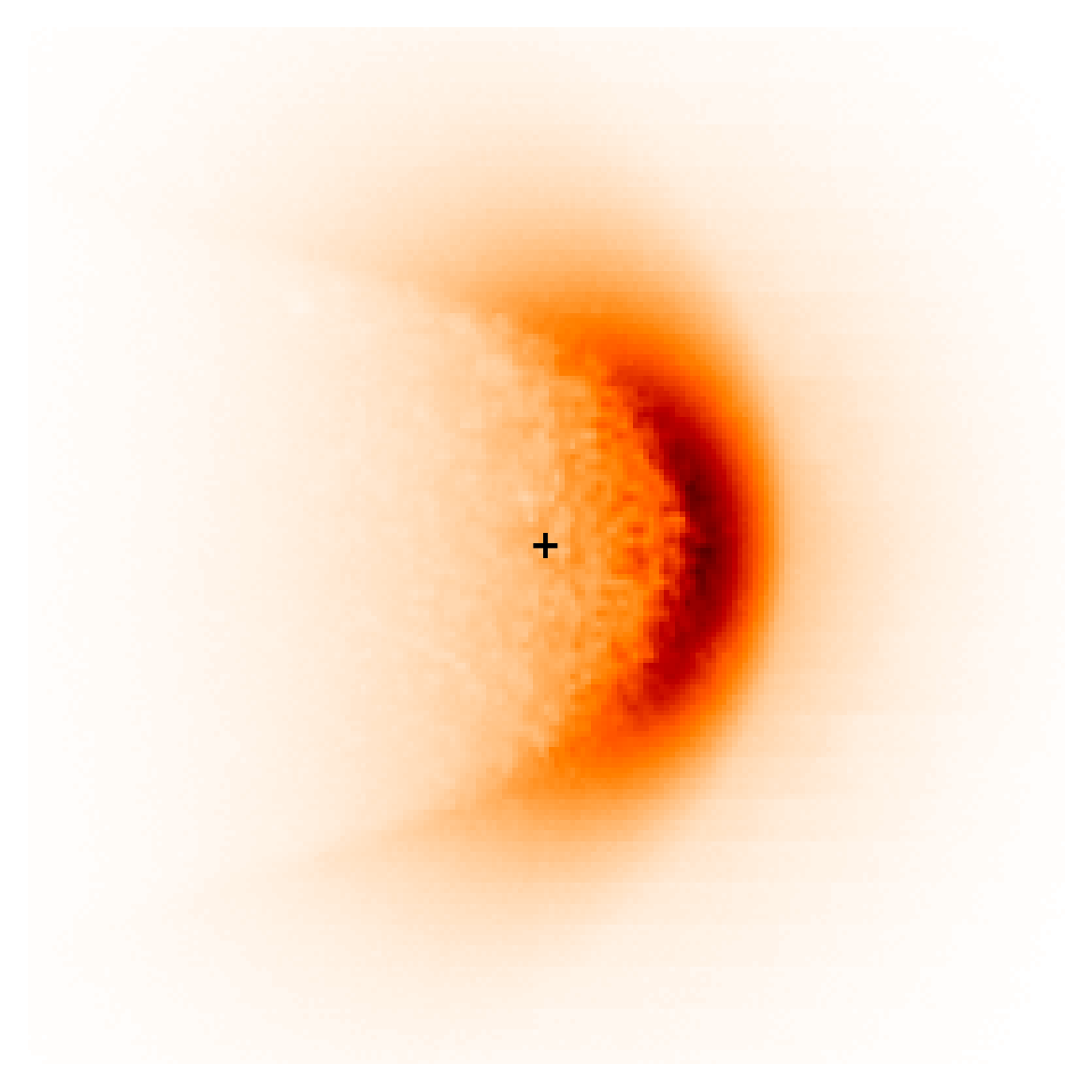}
	\includegraphics[width=.138\textwidth]{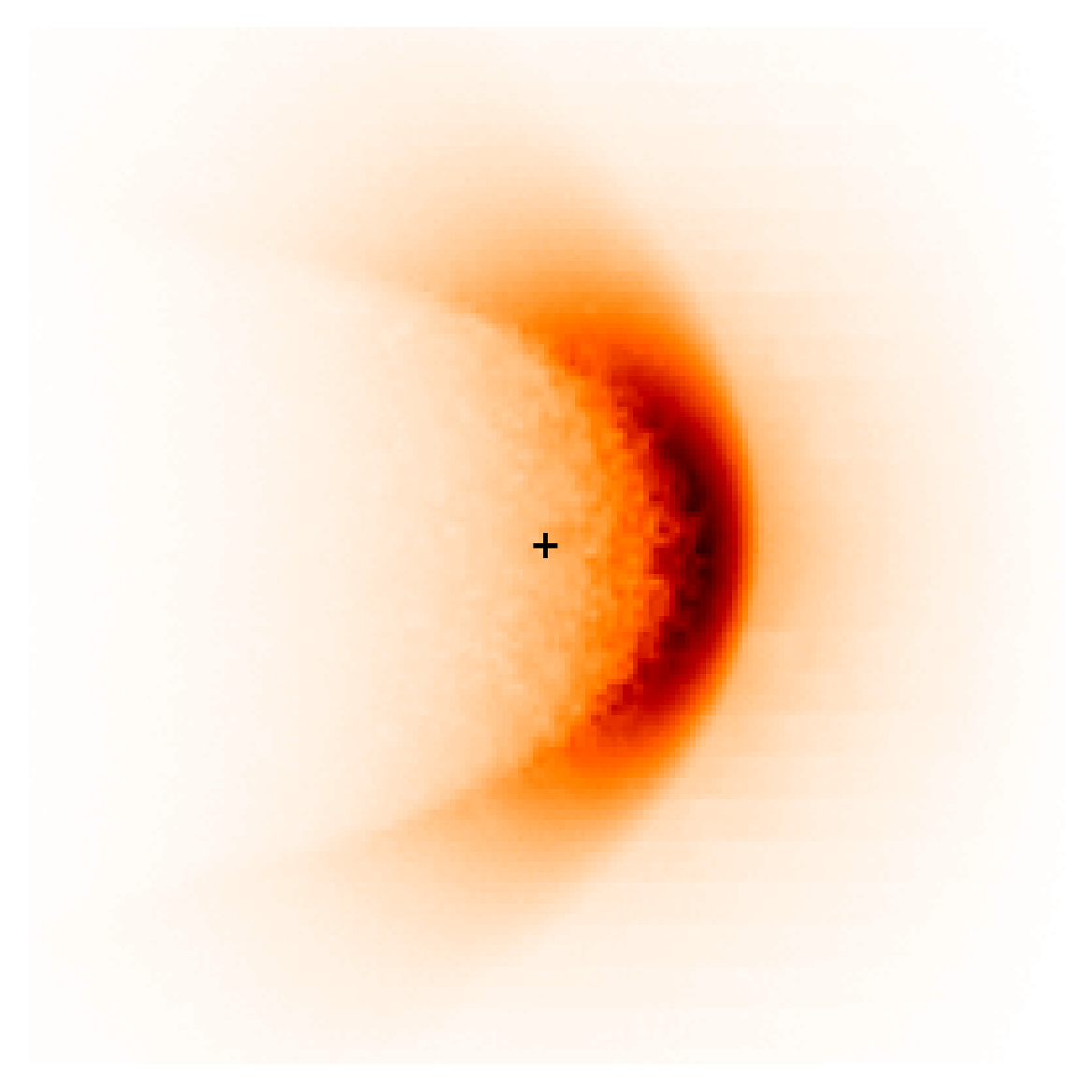}
	\includegraphics[width=.138\textwidth]{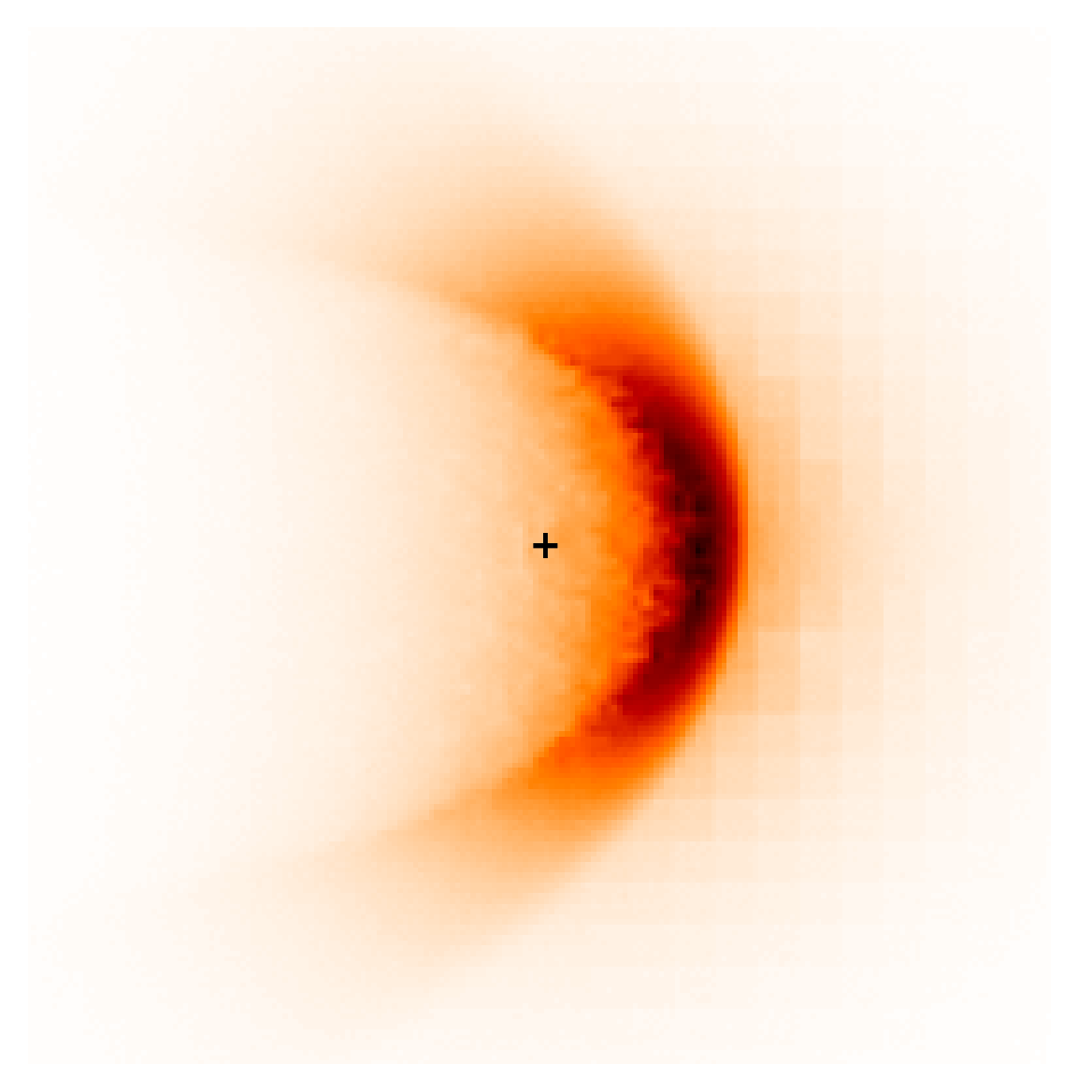} \\
	\includegraphics[width=.138\textwidth]{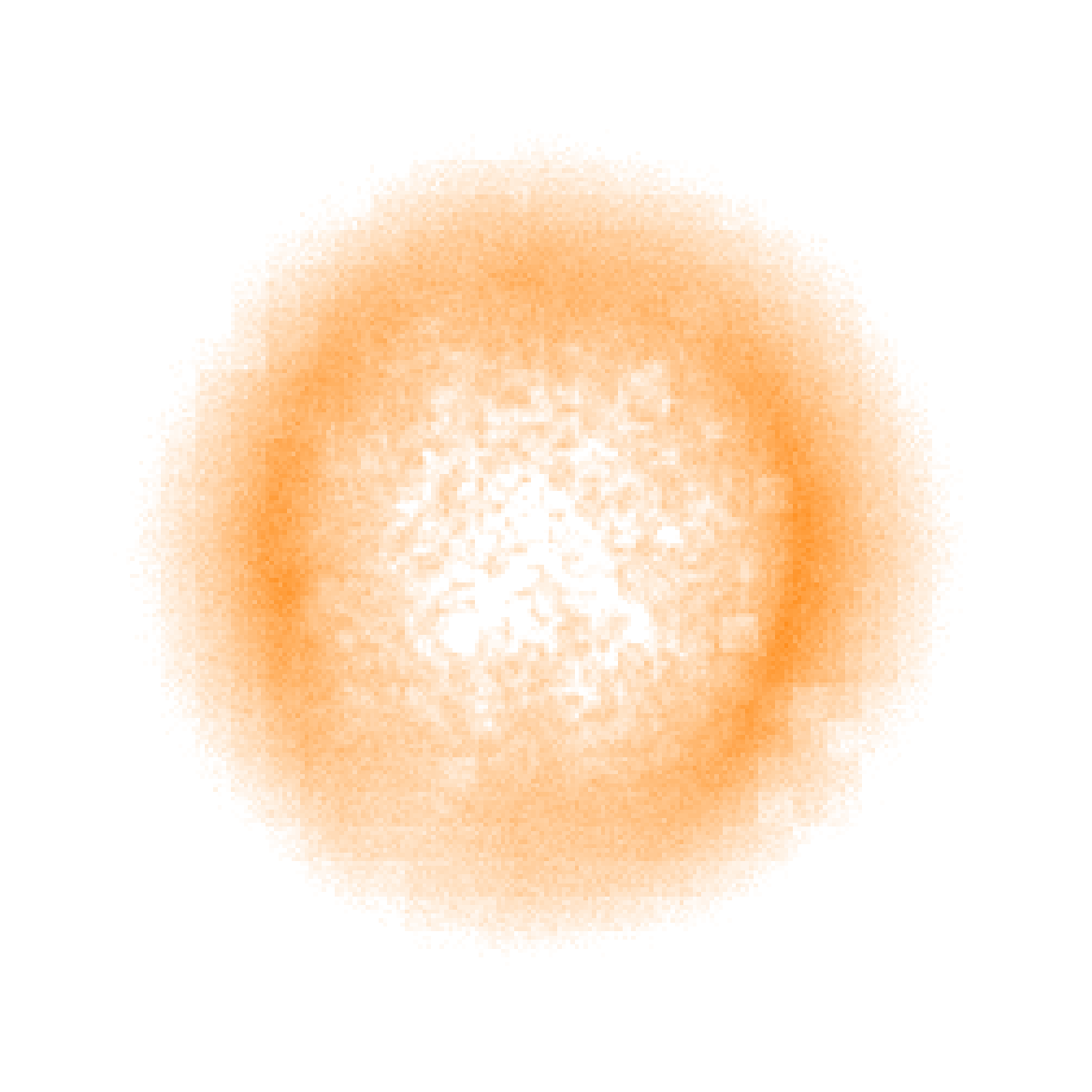}
	\includegraphics[width=.138\textwidth]{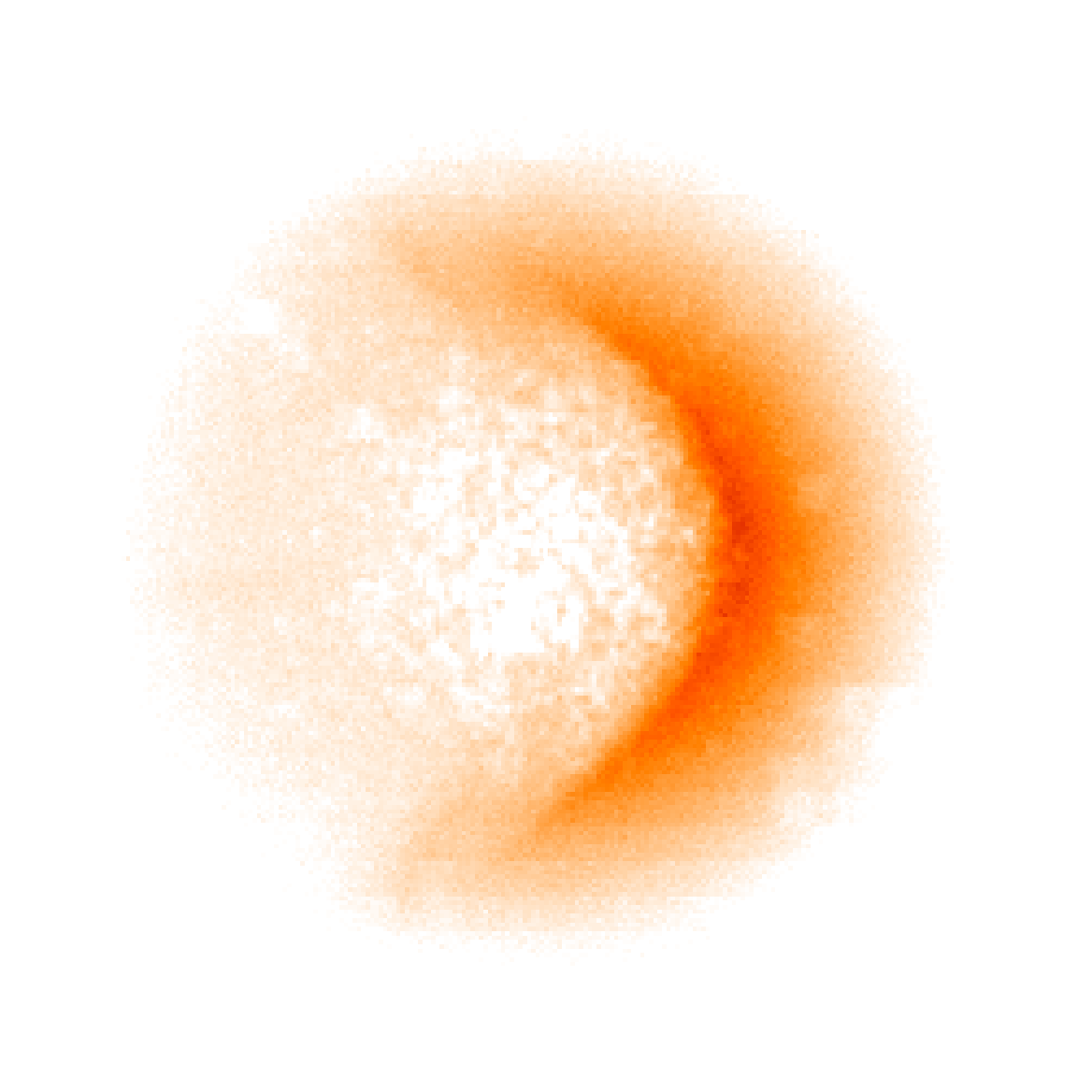}
	\includegraphics[width=.138\textwidth]{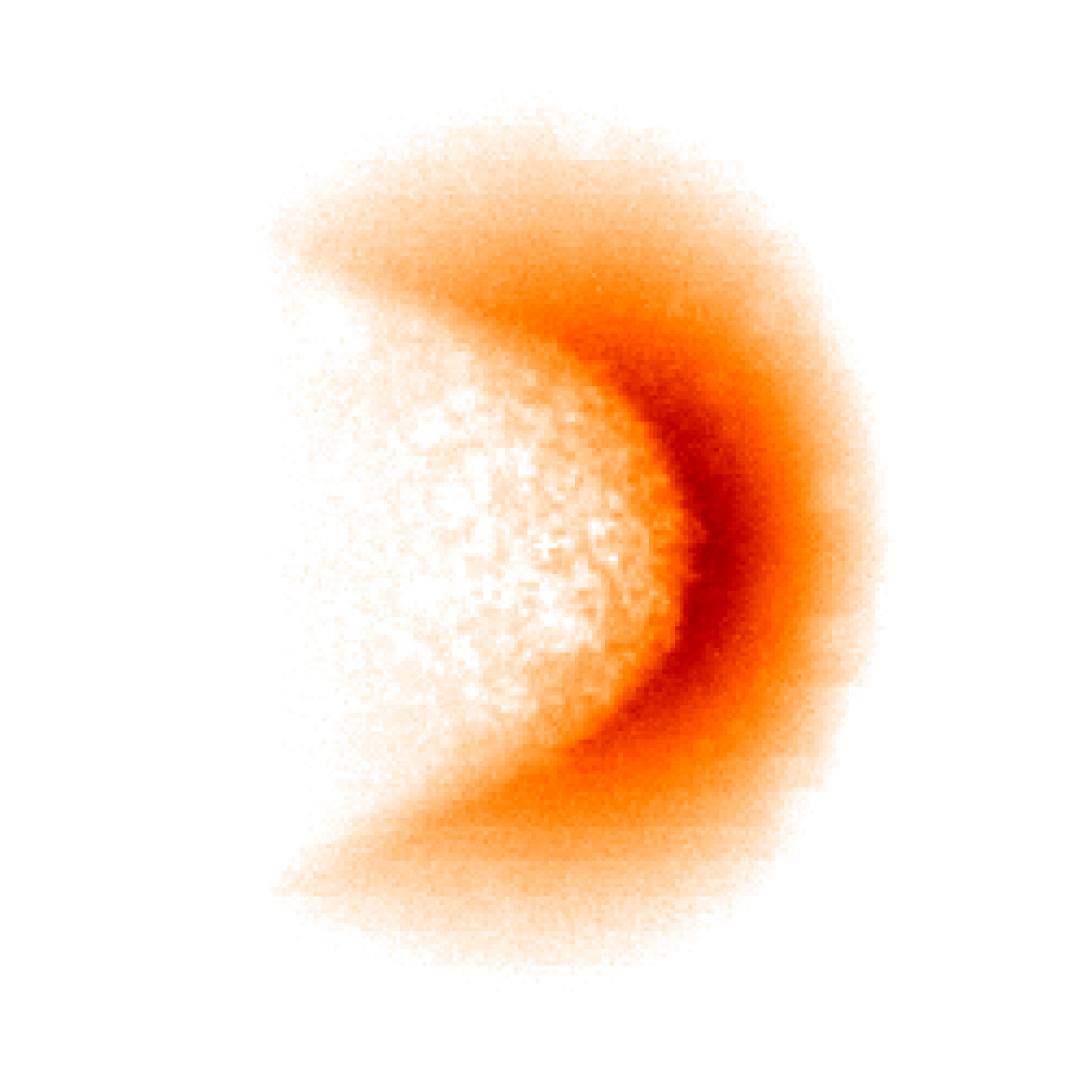}
	\includegraphics[width=.138\textwidth]{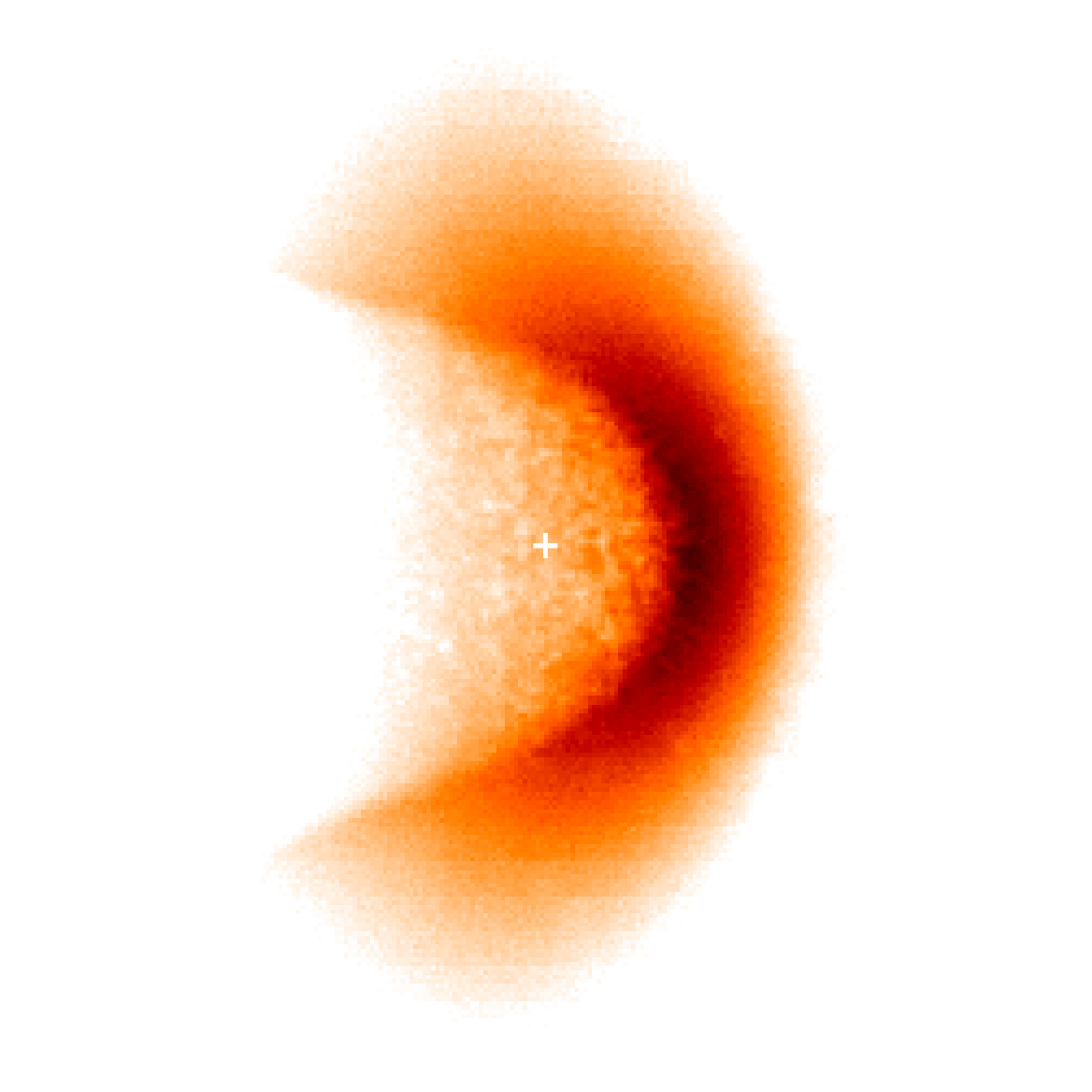}
	\includegraphics[width=.138\textwidth]{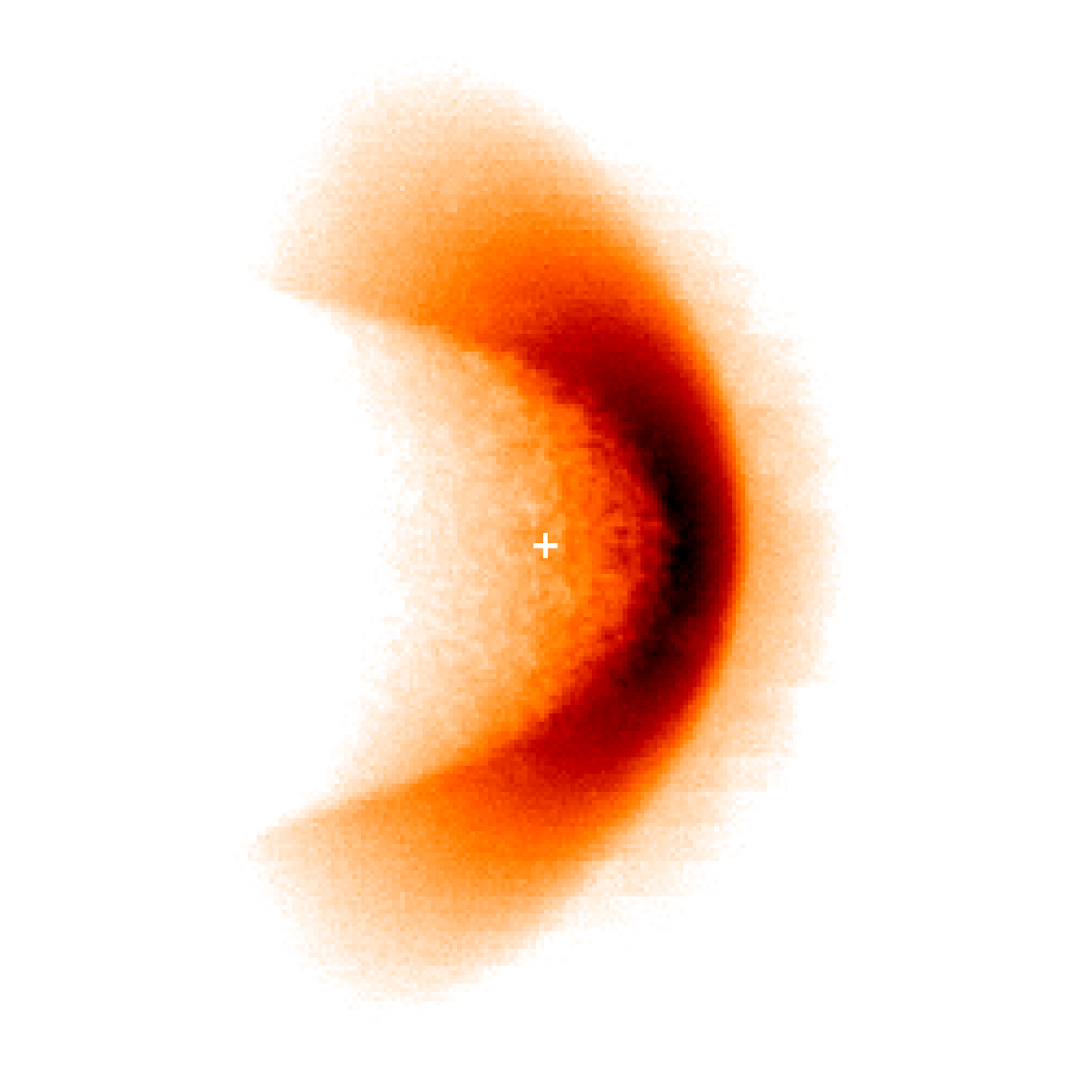}
	\includegraphics[width=.138\textwidth]{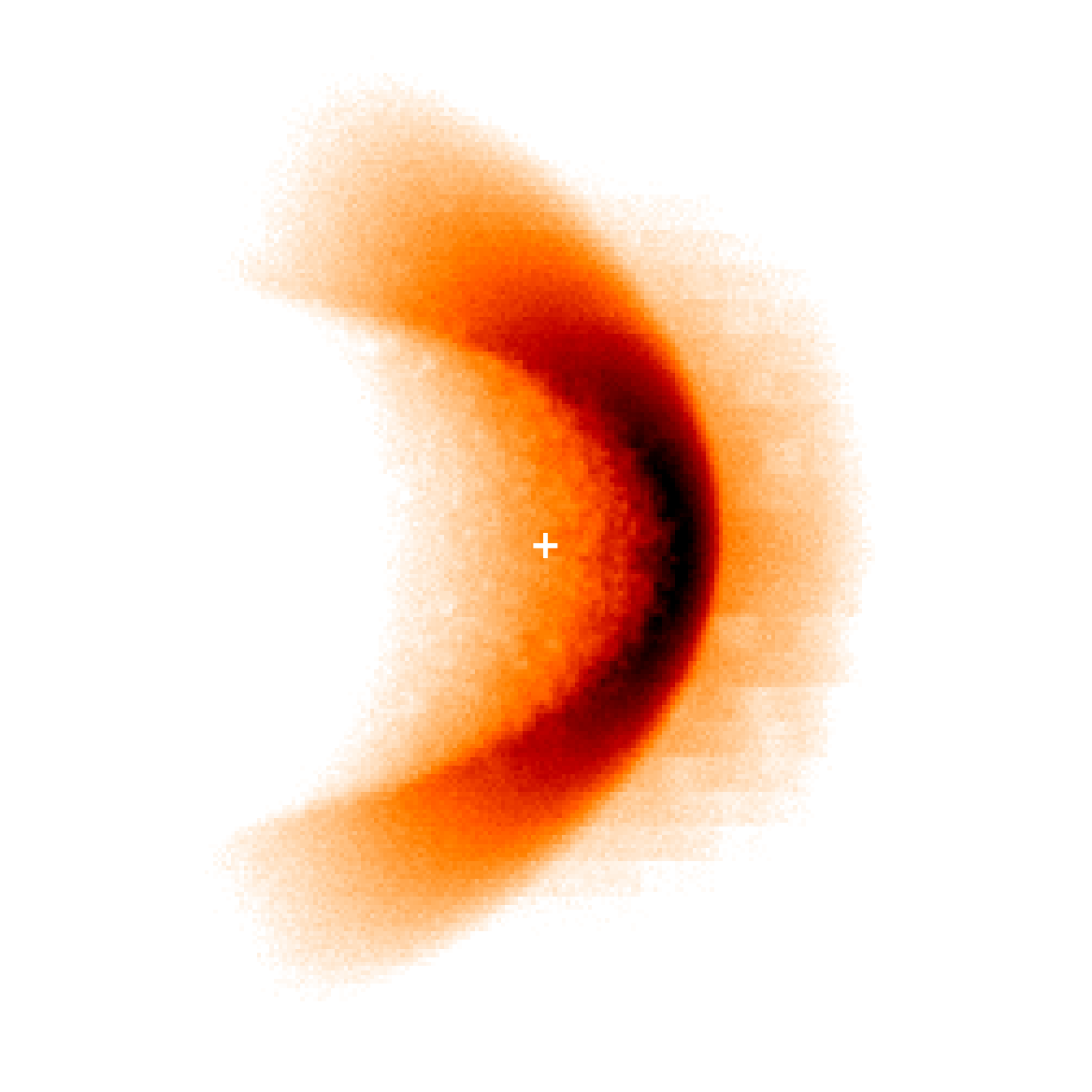}
	\includegraphics[width=.138\textwidth]{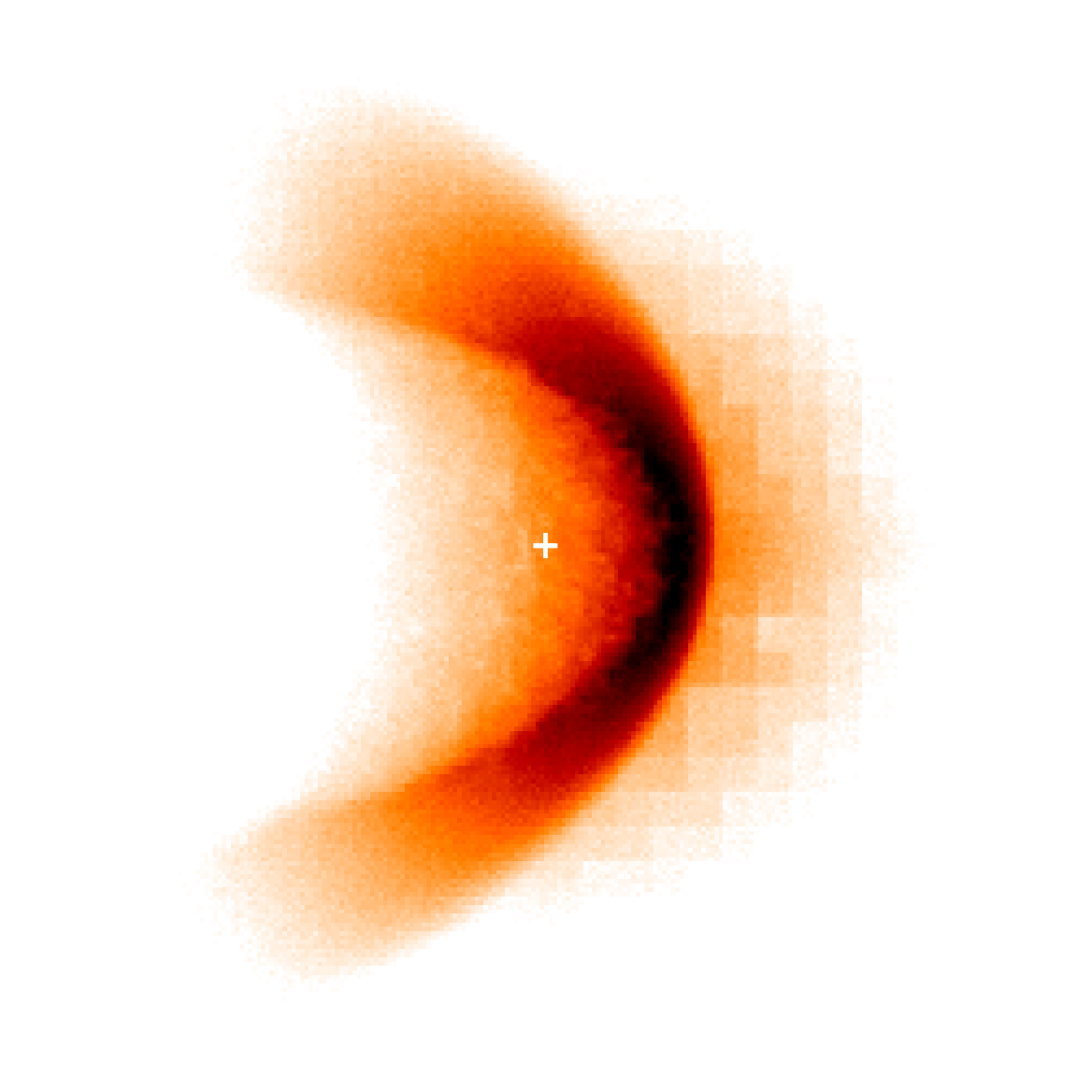} \\
	\includegraphics[width=.138\textwidth]{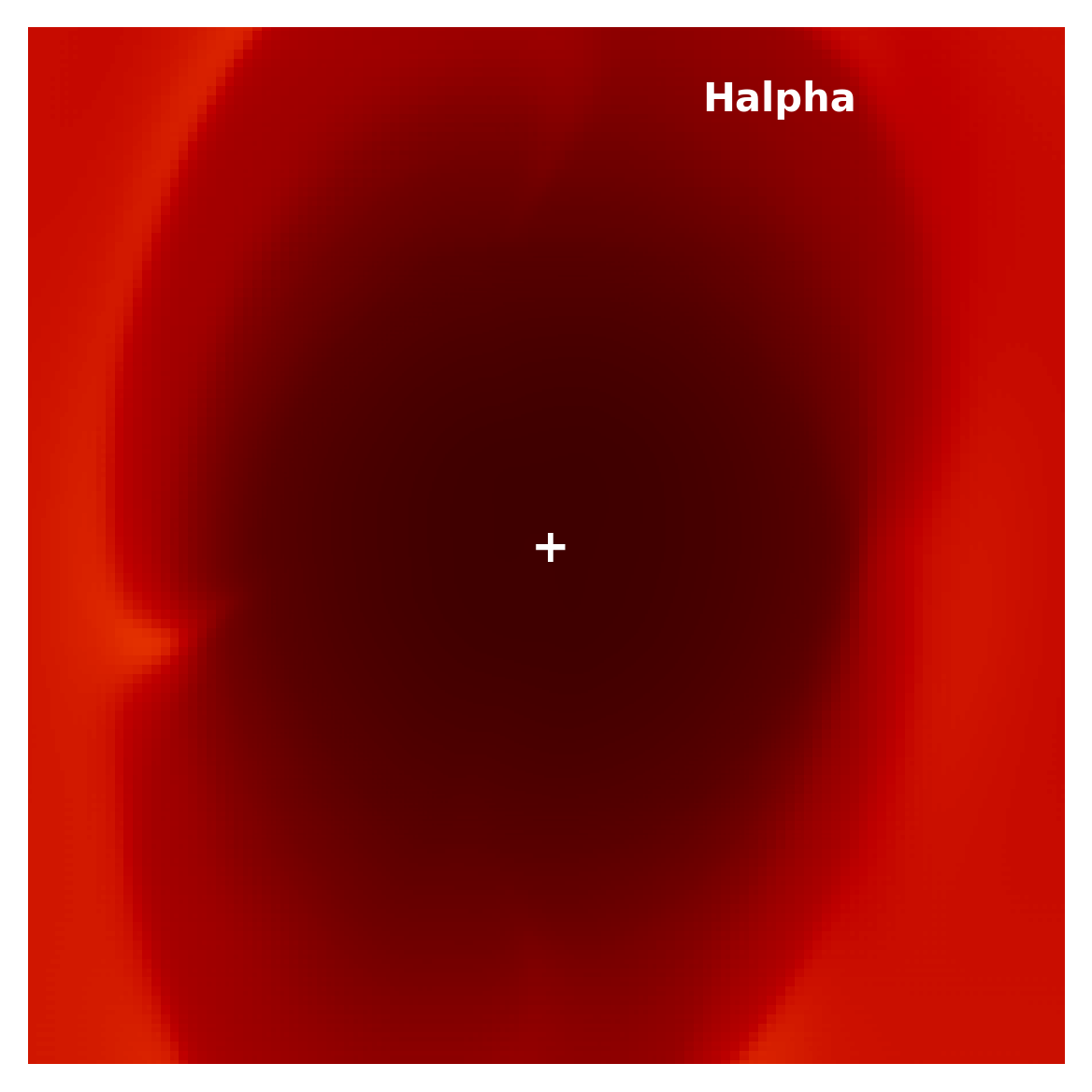}
	\includegraphics[width=.138\textwidth]{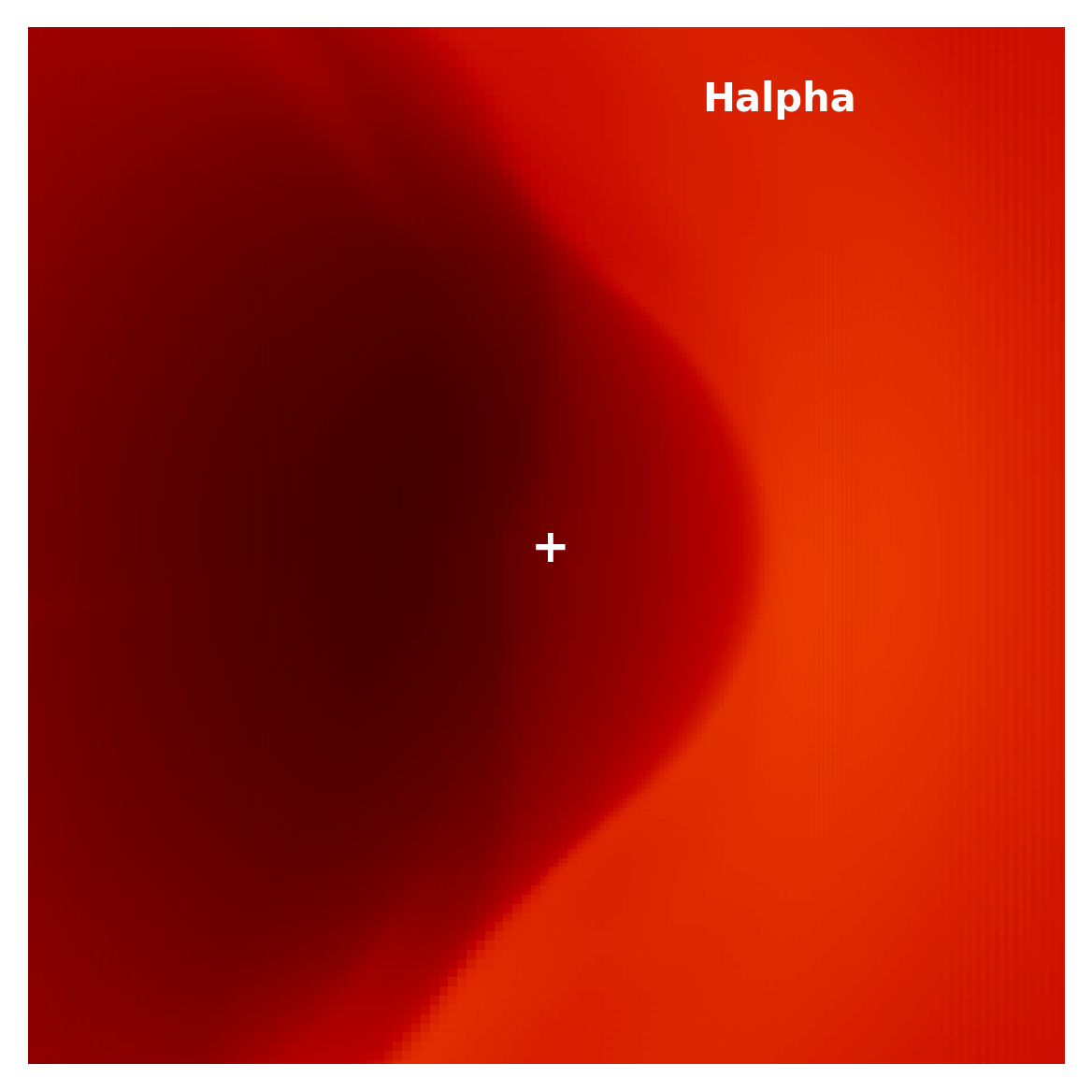}
	\includegraphics[width=.138\textwidth]{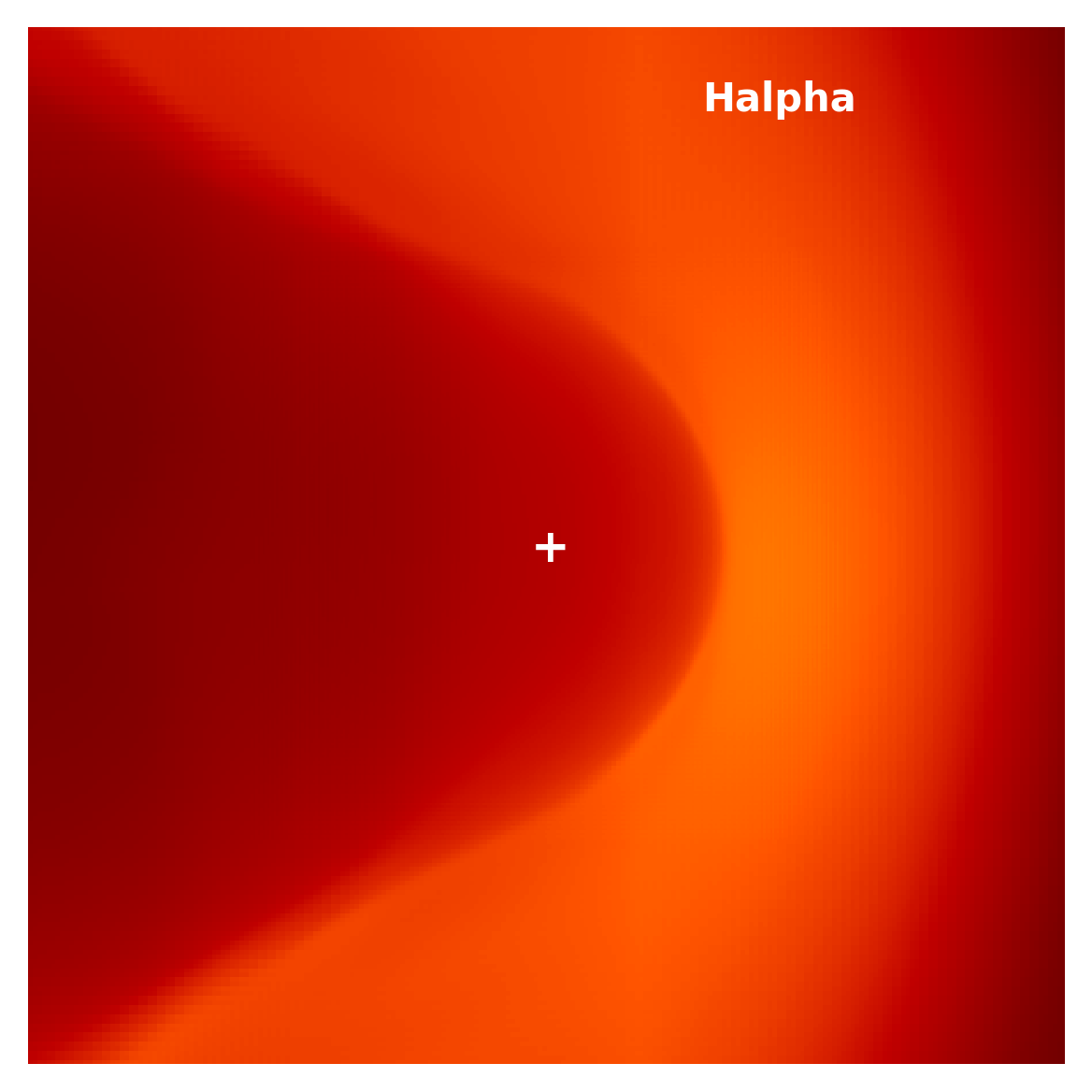}
	\includegraphics[width=.138\textwidth]{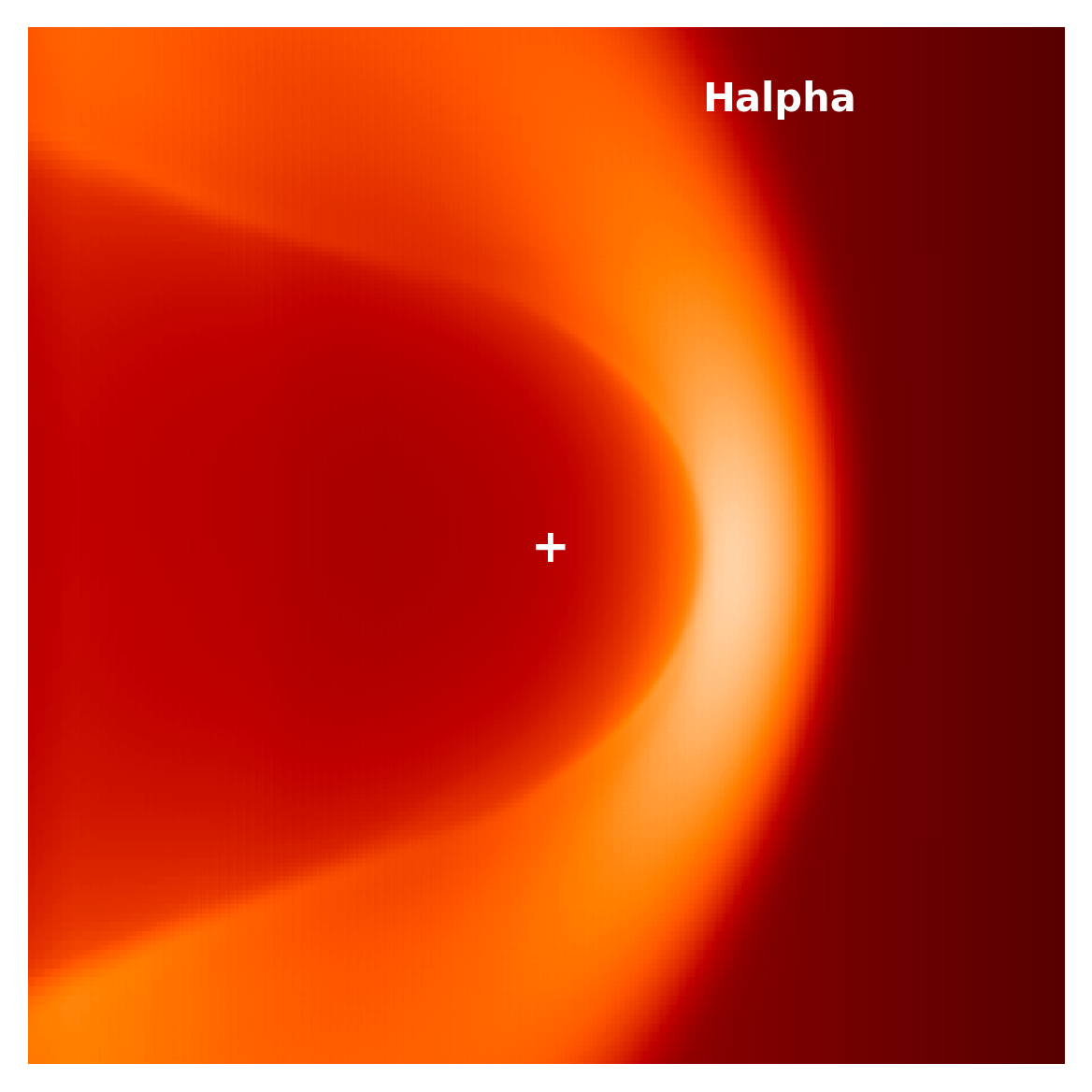}
	\includegraphics[width=.138\textwidth]{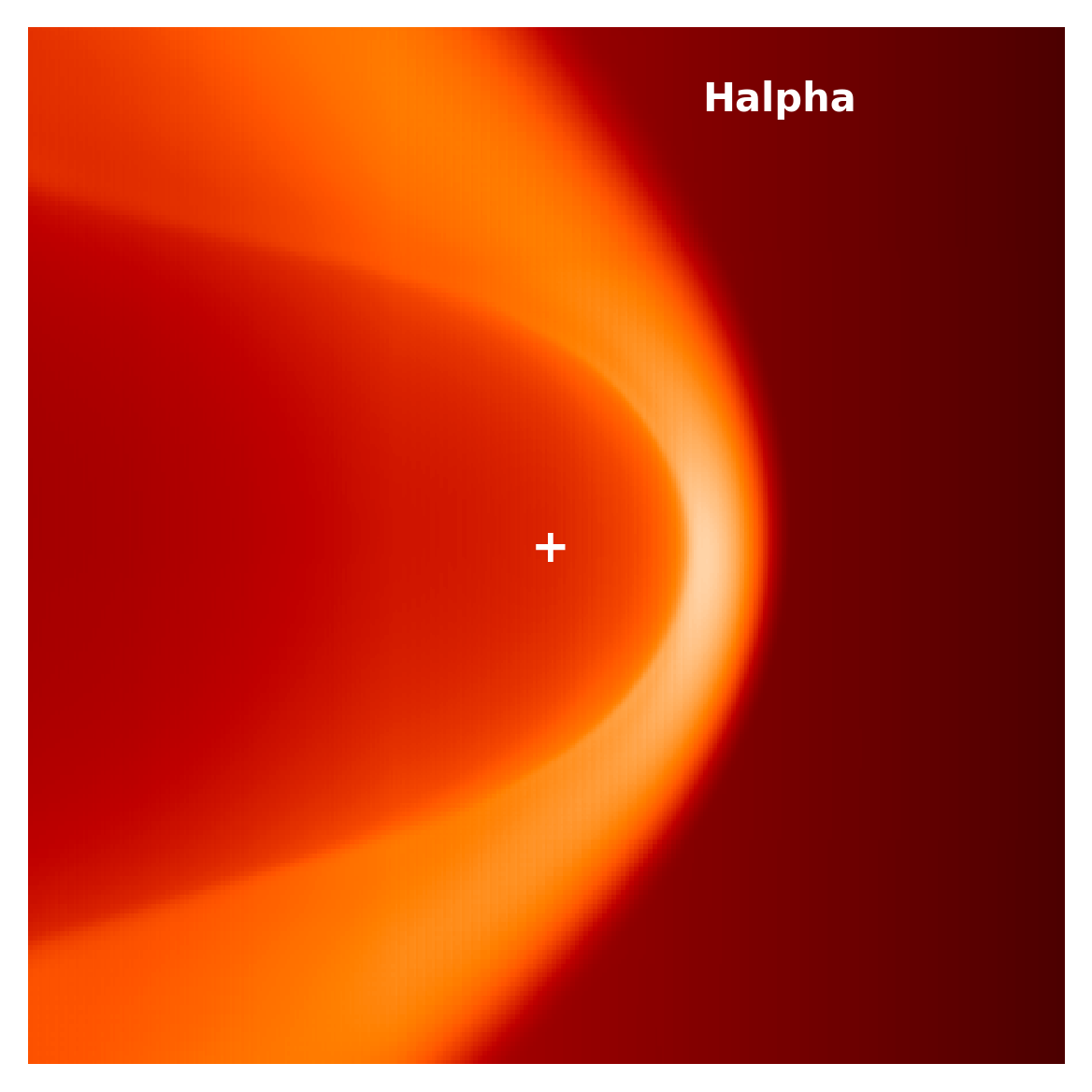}
	\includegraphics[width=.138\textwidth]{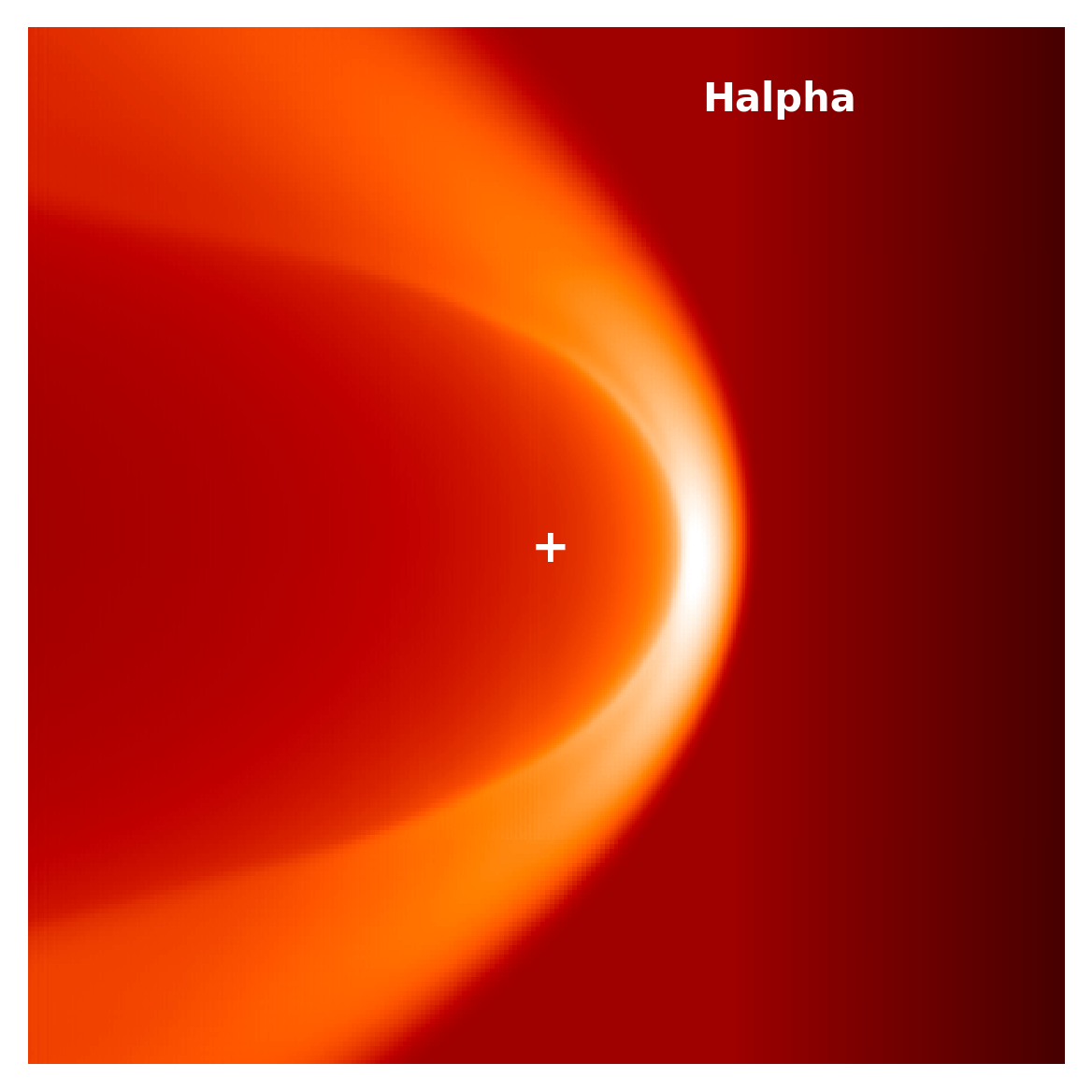}
	\includegraphics[width=.138\textwidth]{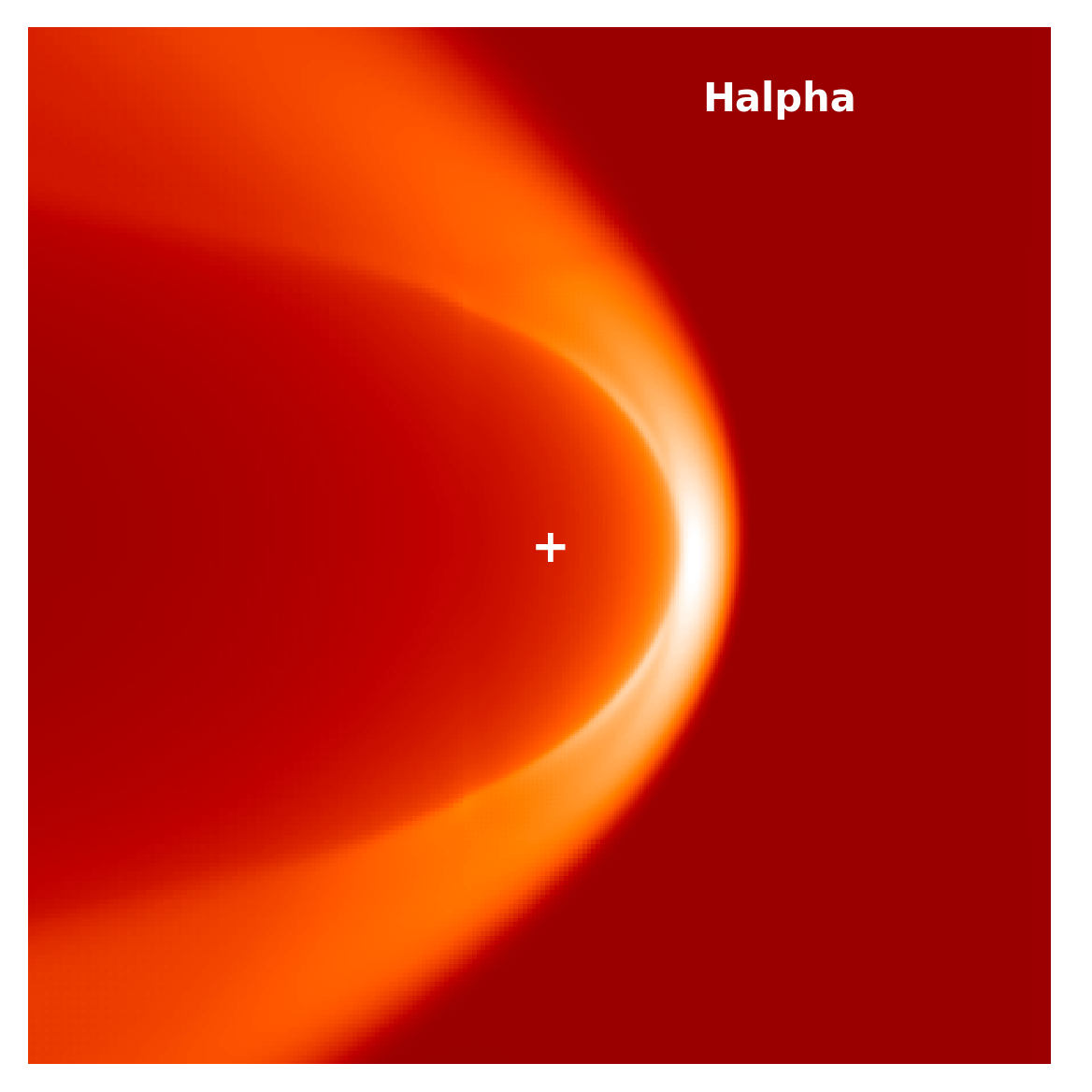} \\
	\includegraphics[width=.138\textwidth]{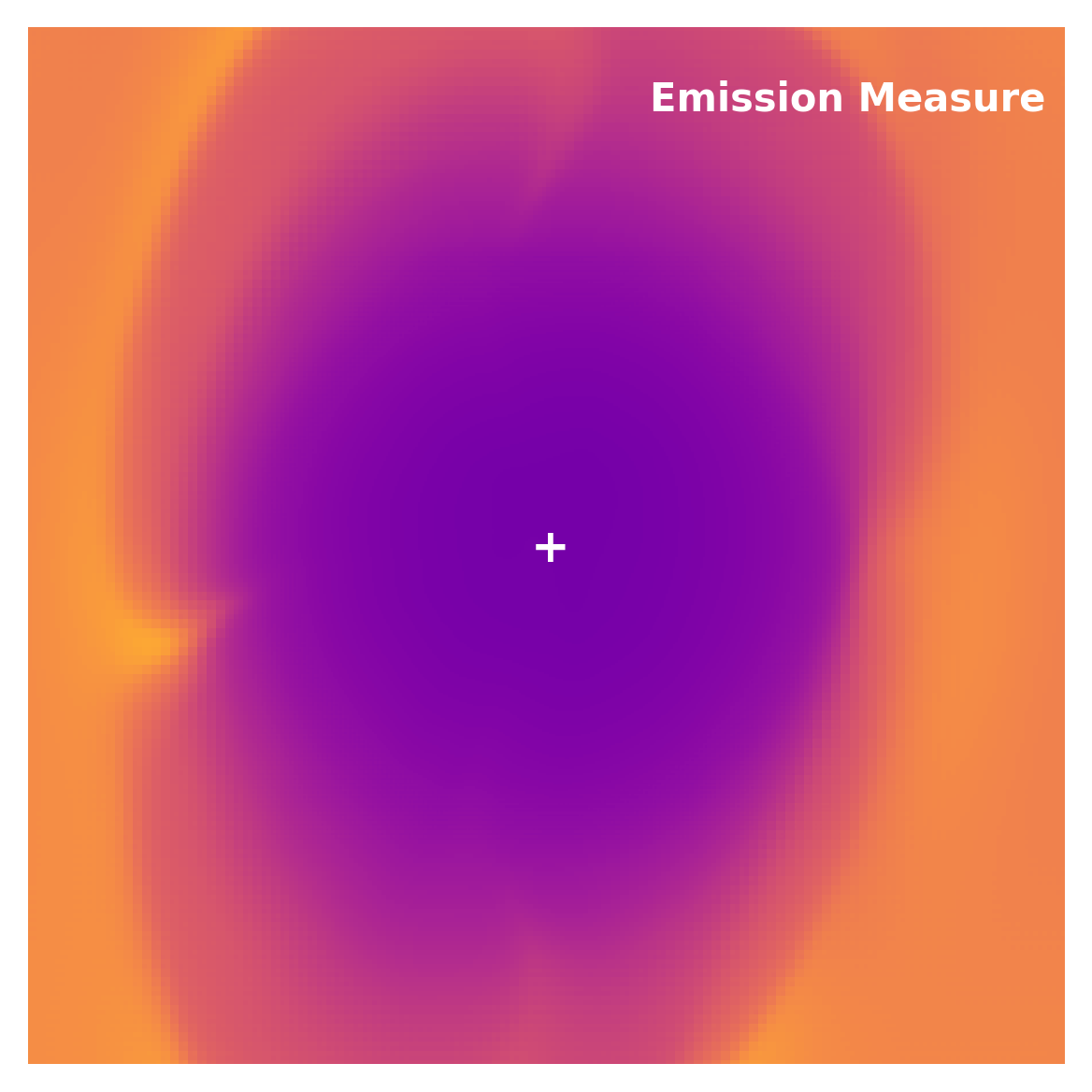}
	\includegraphics[width=.138\textwidth]{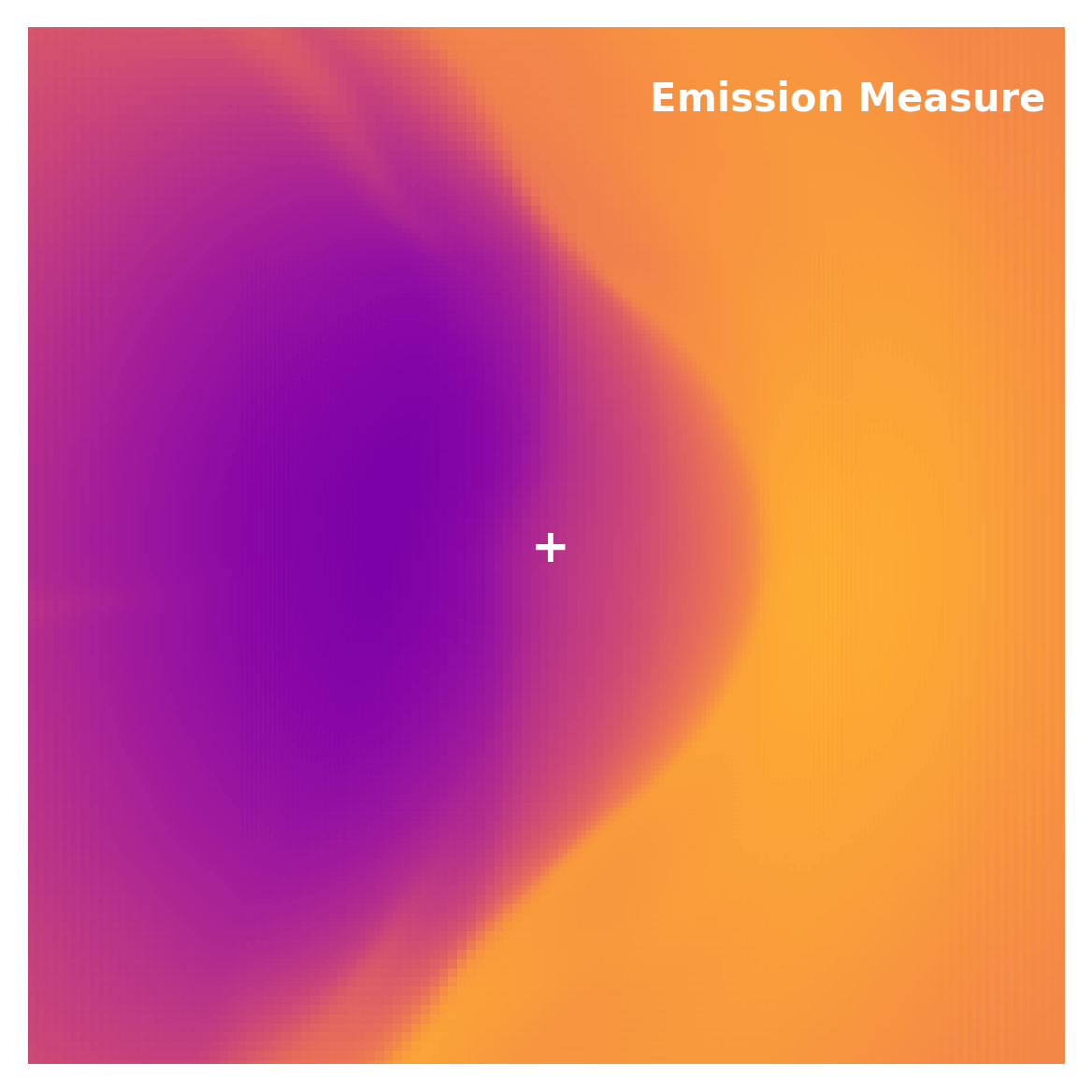}
	\includegraphics[width=.138\textwidth]{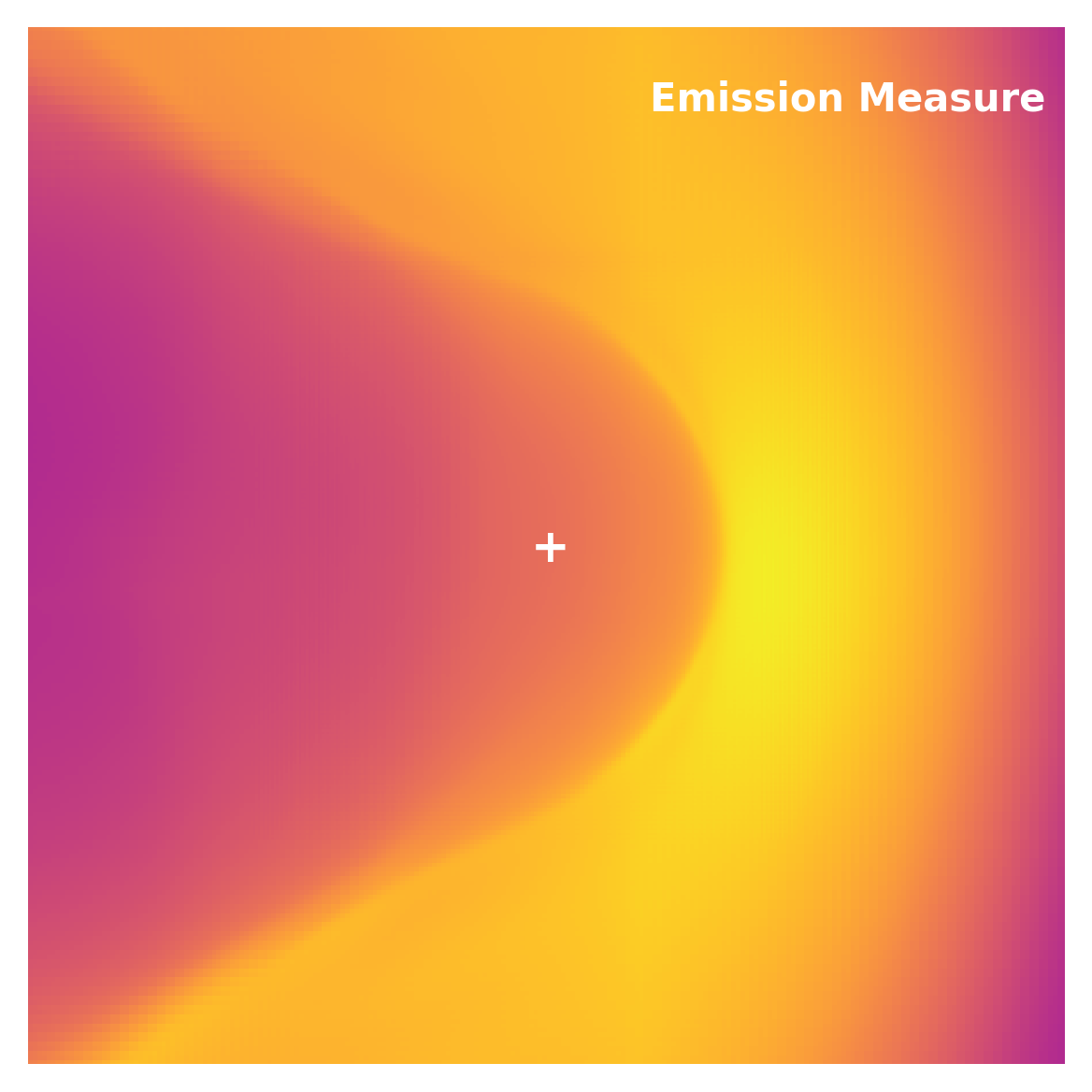}
	\includegraphics[width=.138\textwidth]{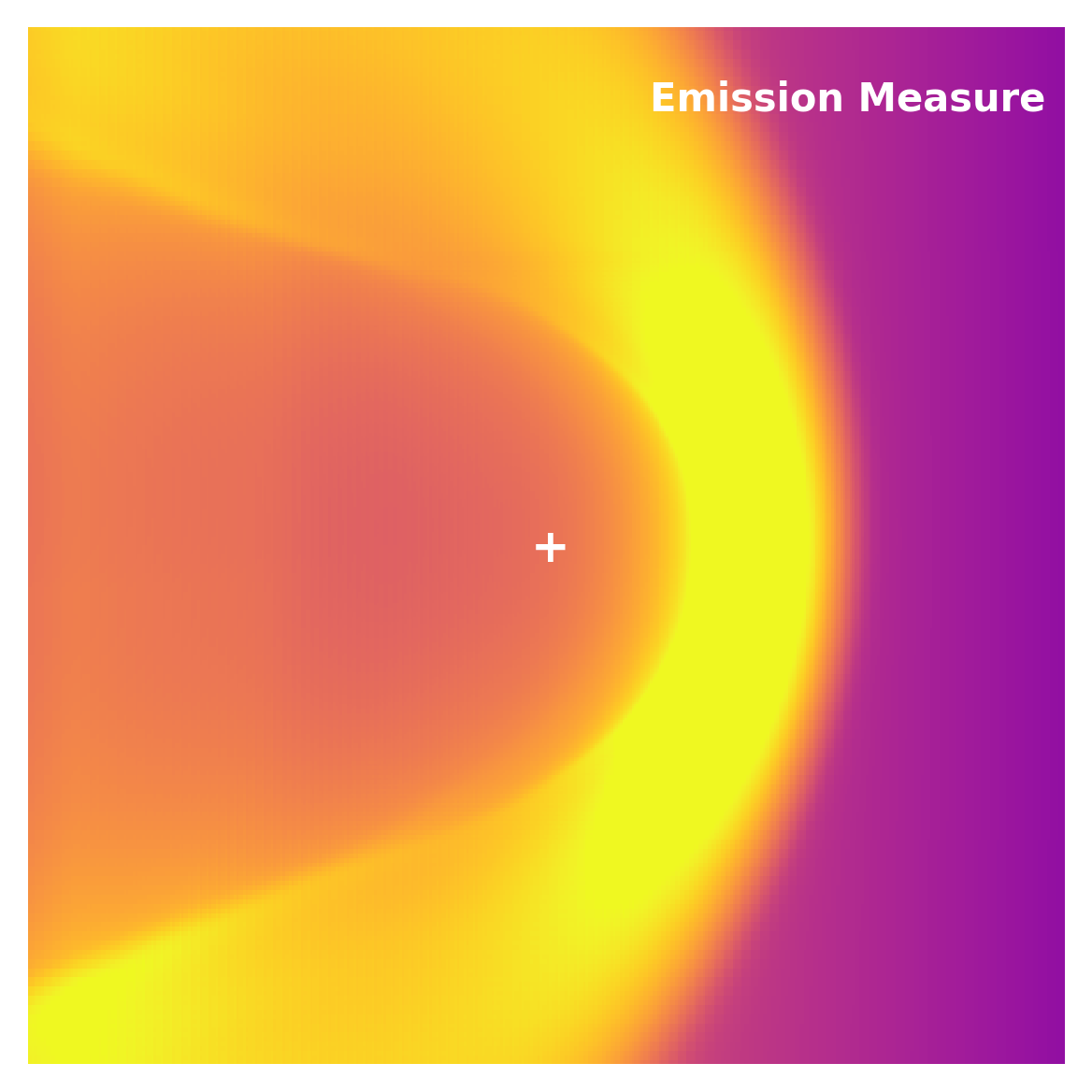}
	\includegraphics[width=.138\textwidth]{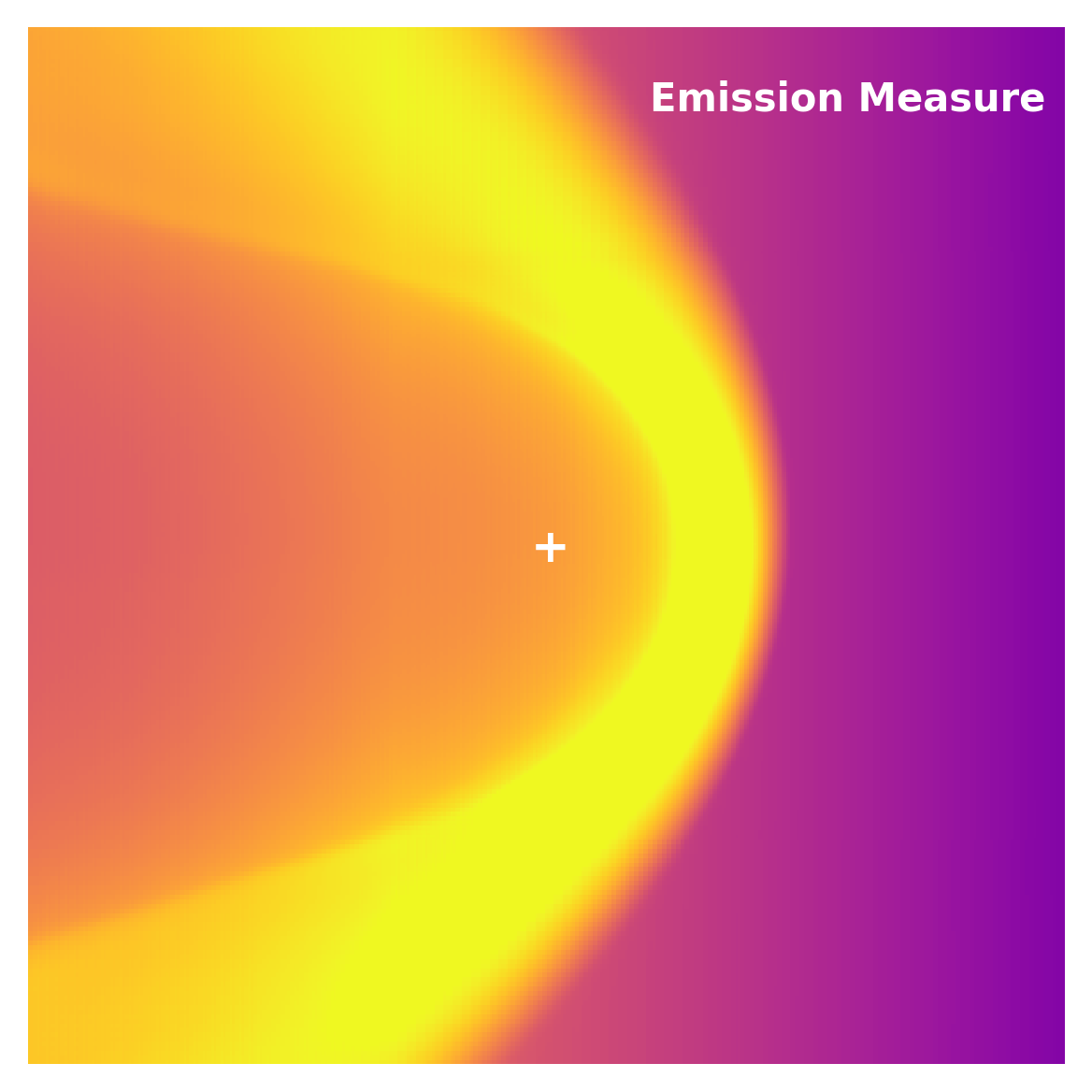}
	\includegraphics[width=.138\textwidth]{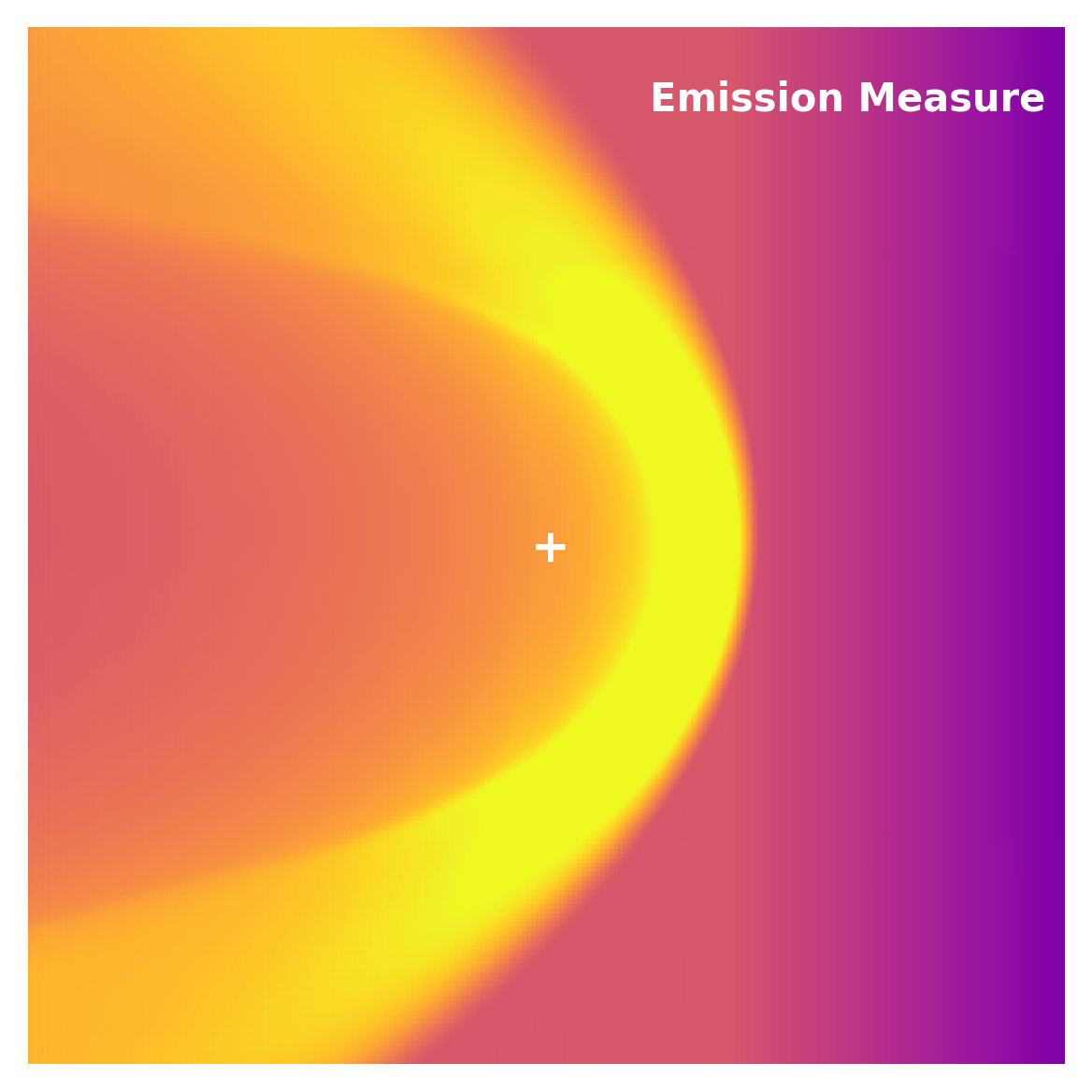}
	\includegraphics[width=.138\textwidth]{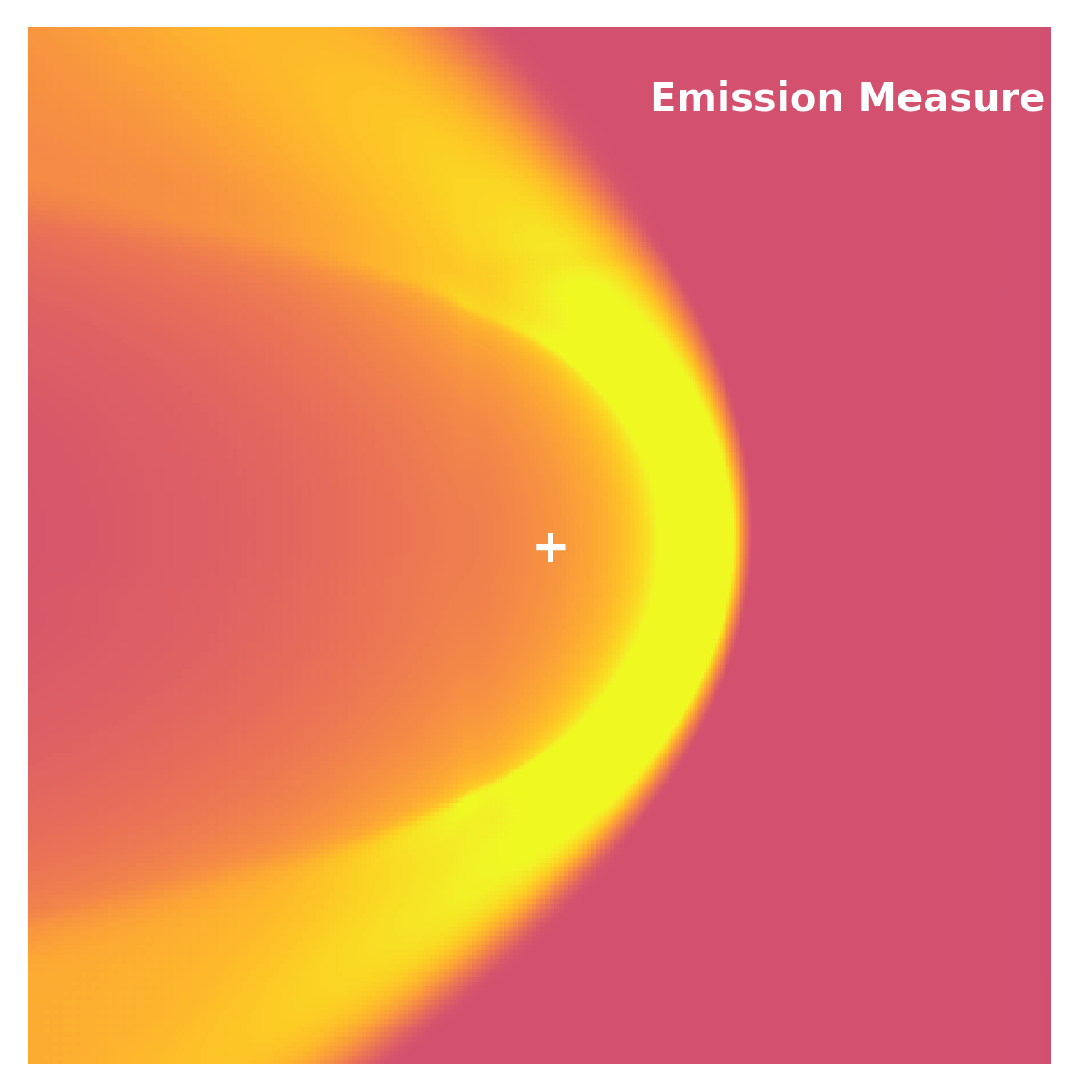} \\
	\includegraphics[width=.138\textwidth]{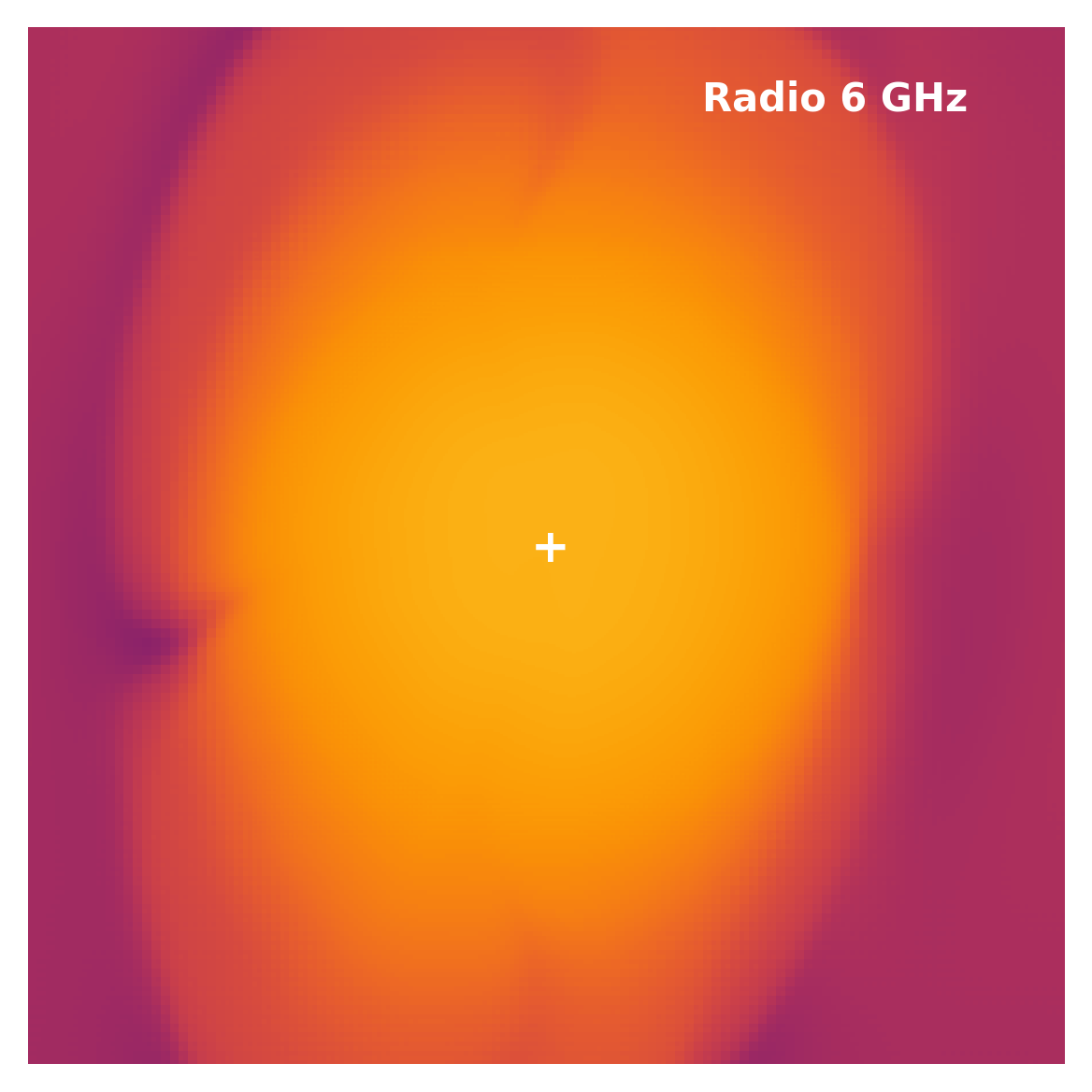}
	\includegraphics[width=.138\textwidth]{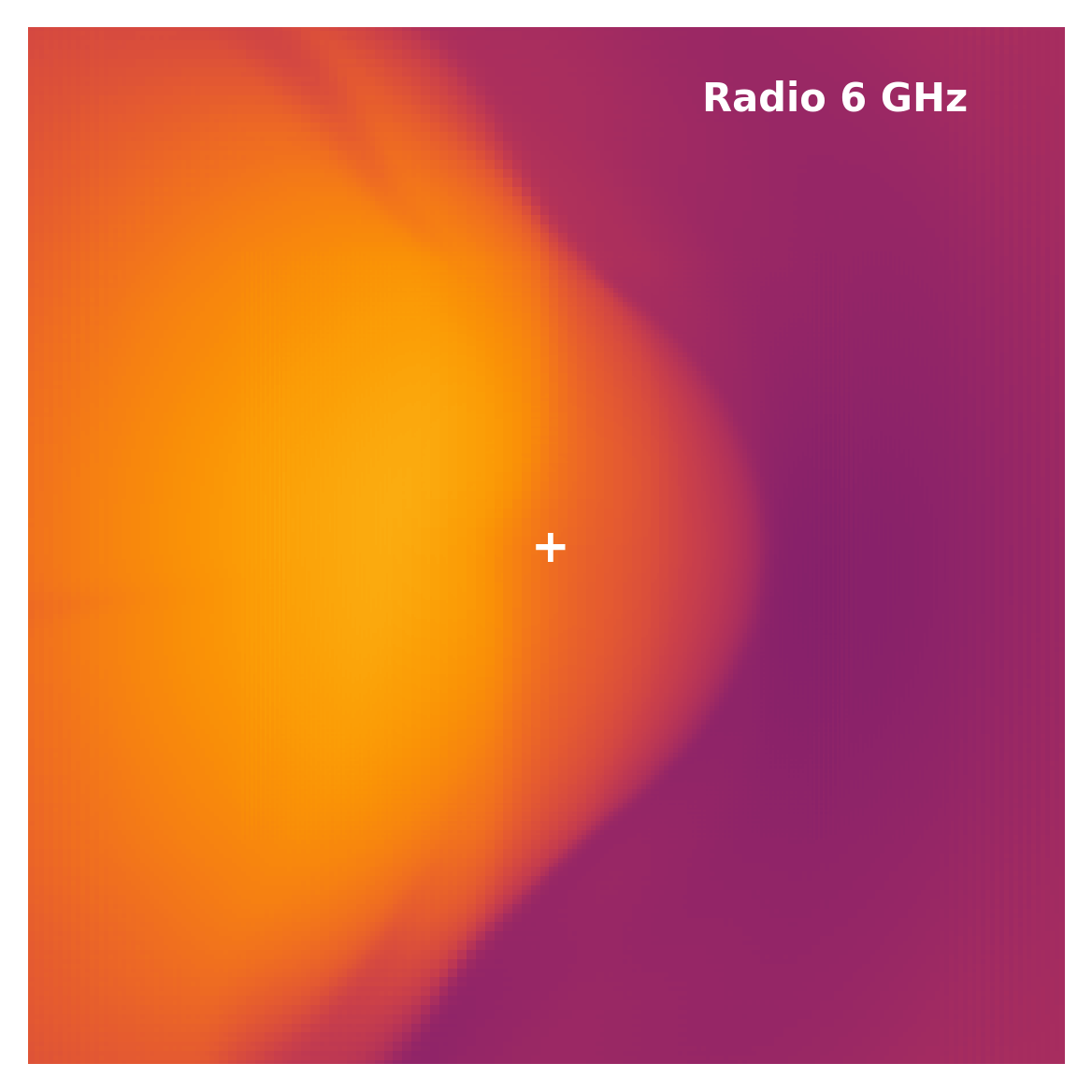}
	\includegraphics[width=.138\textwidth]{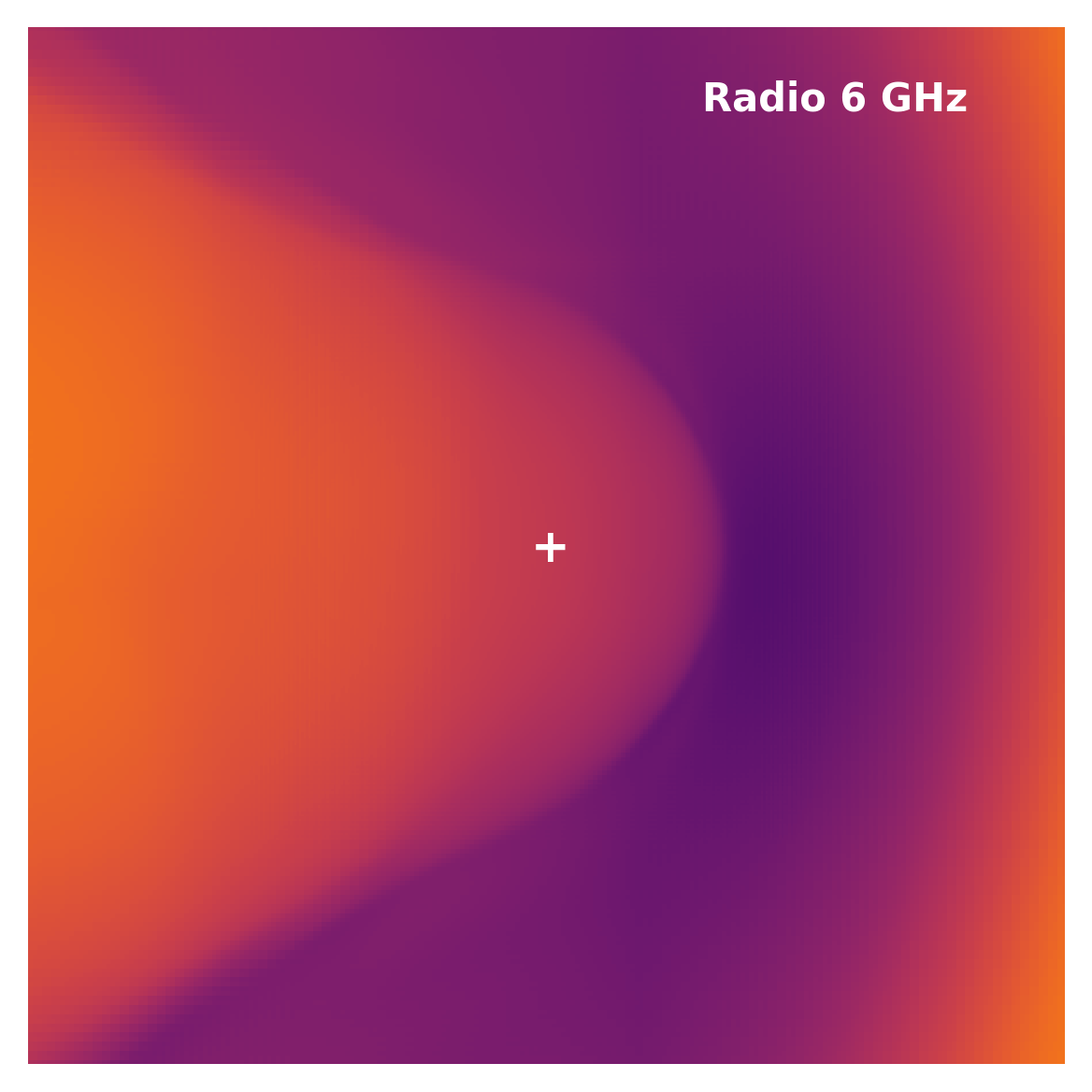}
	\includegraphics[width=.138\textwidth]{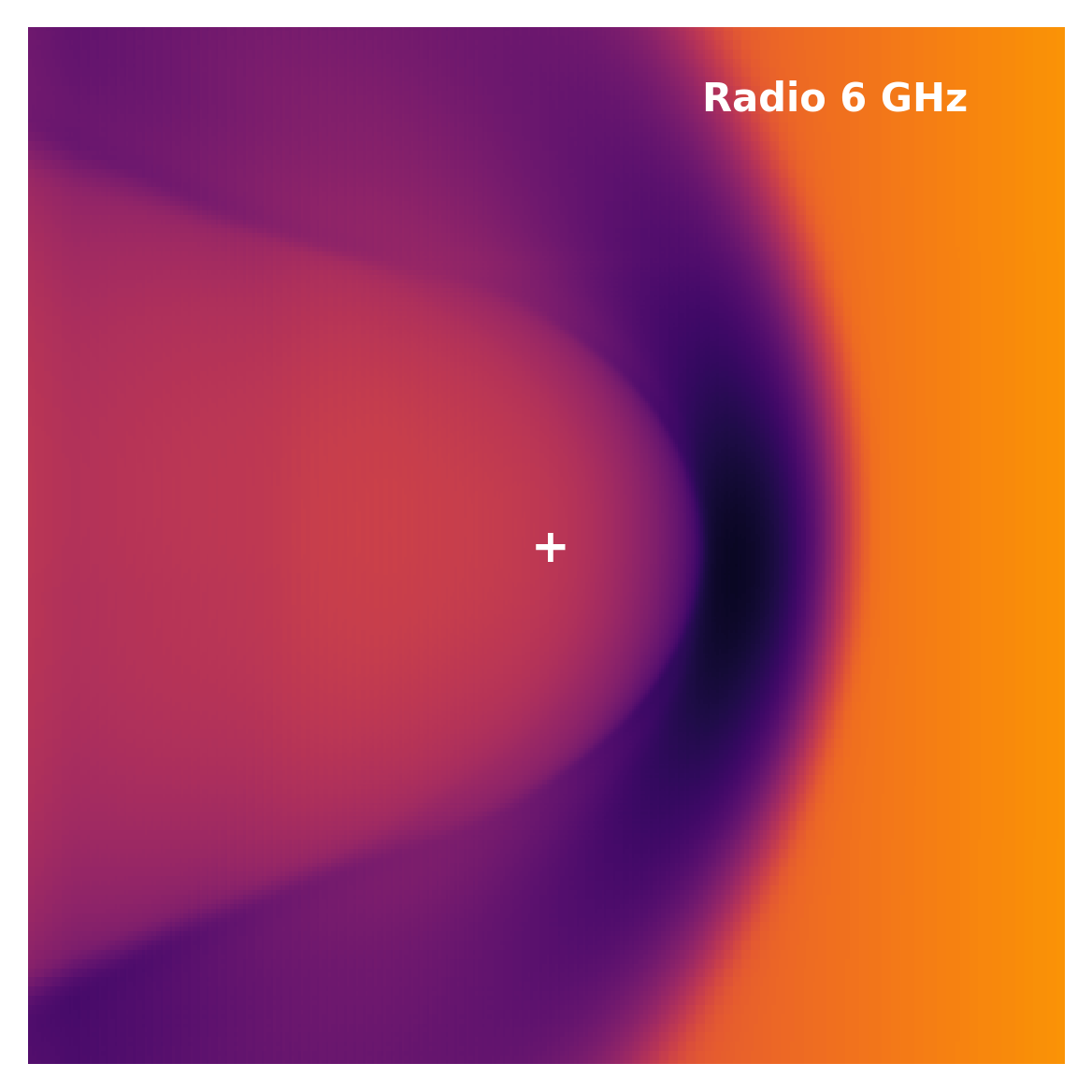}
	\includegraphics[width=.138\textwidth]{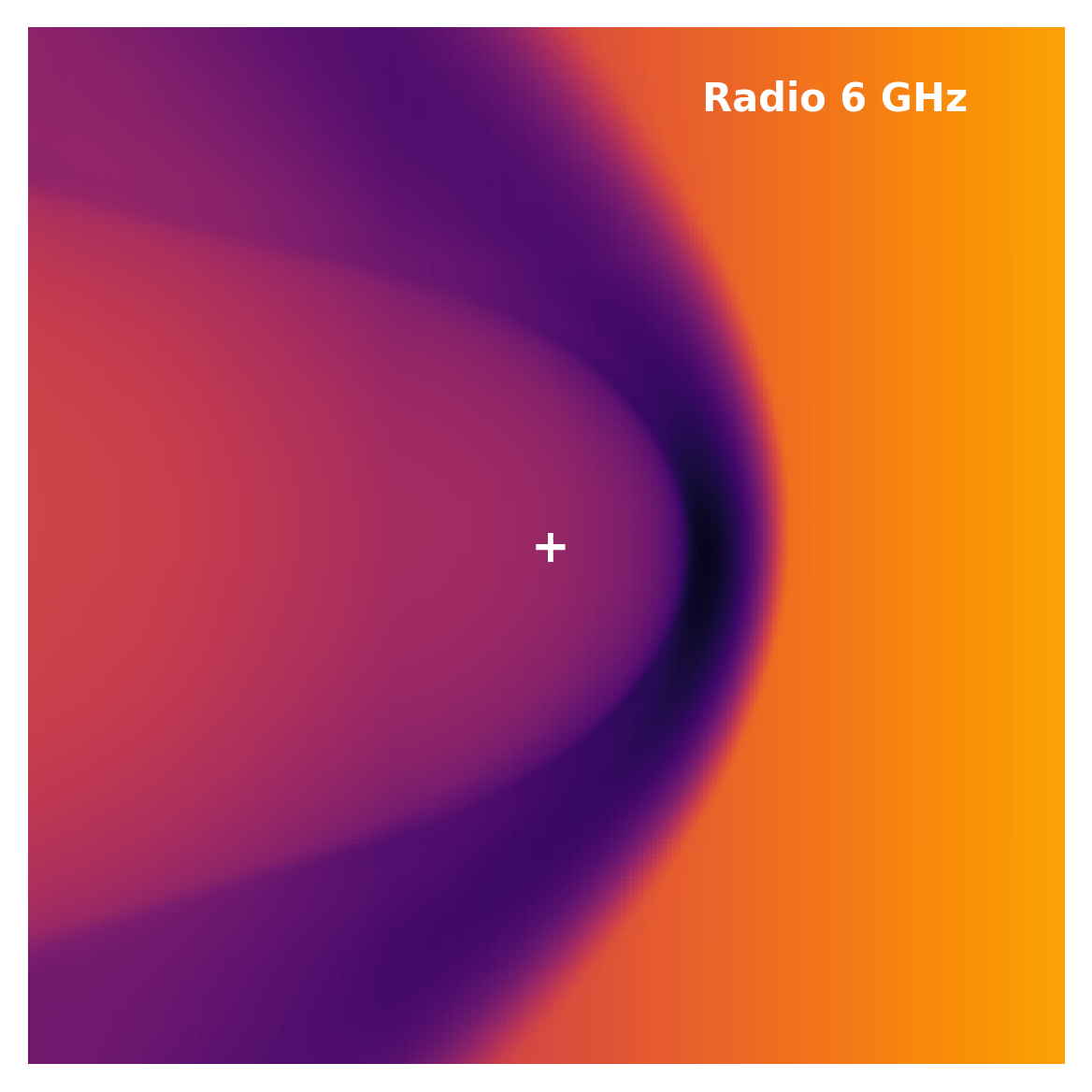}
	\includegraphics[width=.138\textwidth]{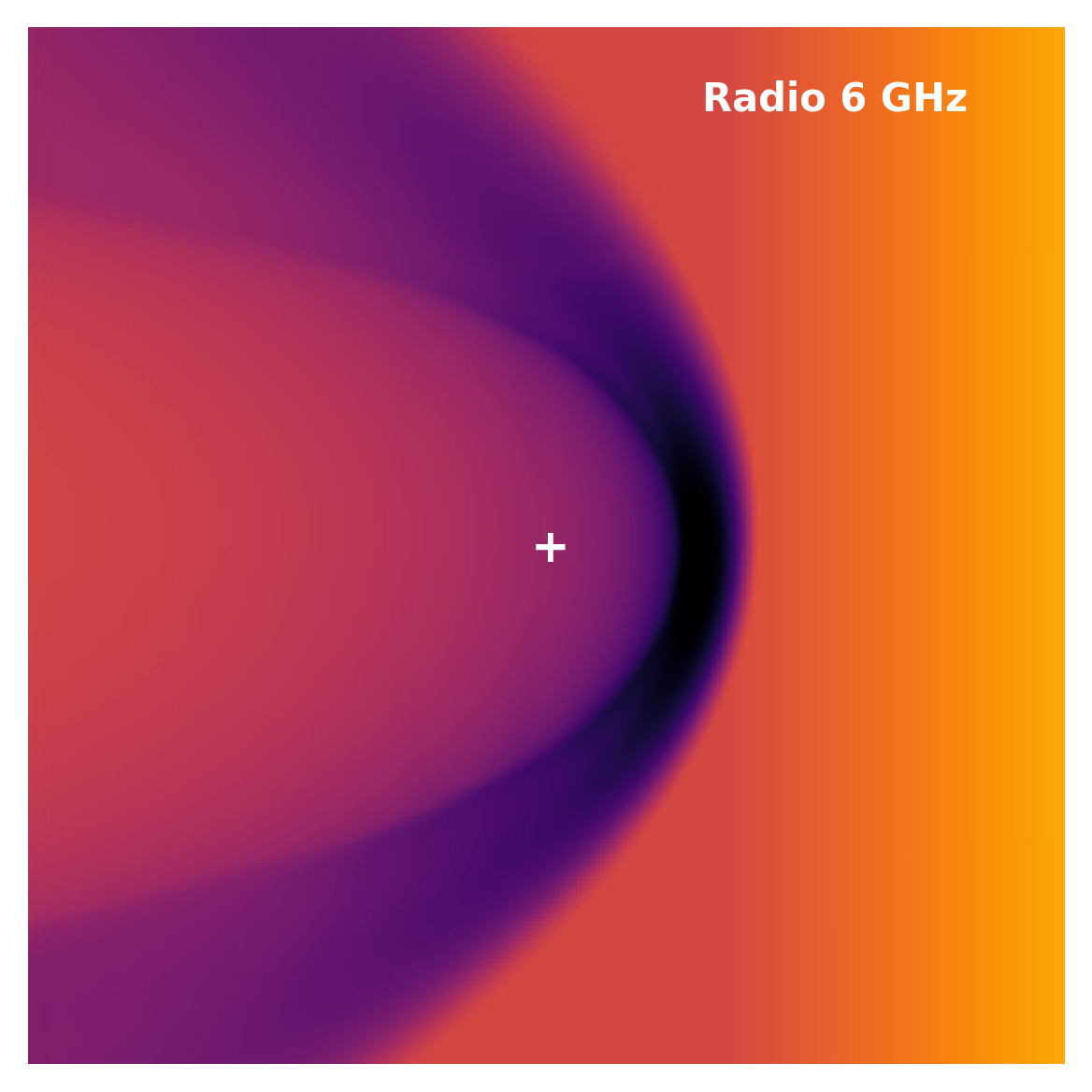}
	\includegraphics[width=.138\textwidth]{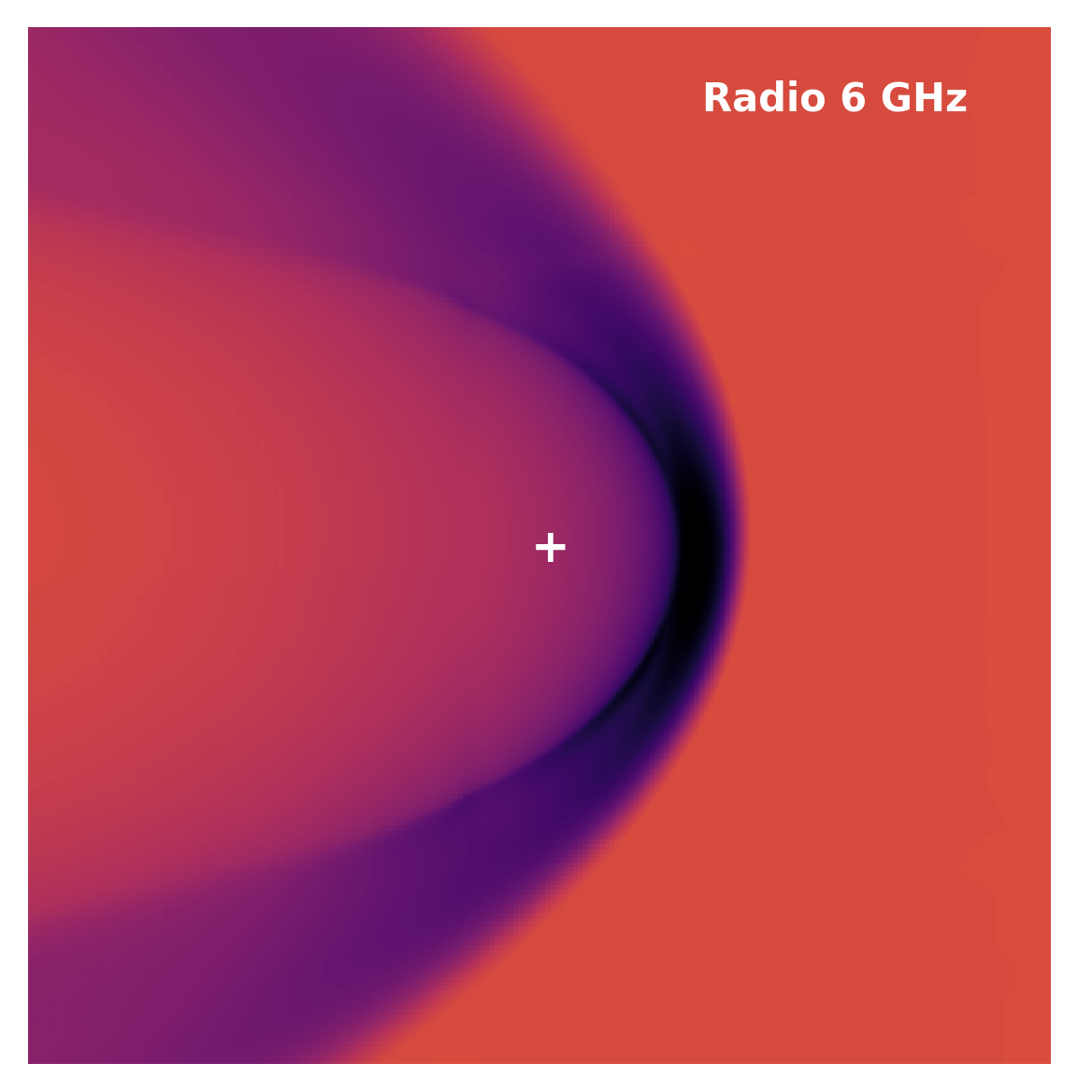} \\
	\includegraphics[width=.138\textwidth]{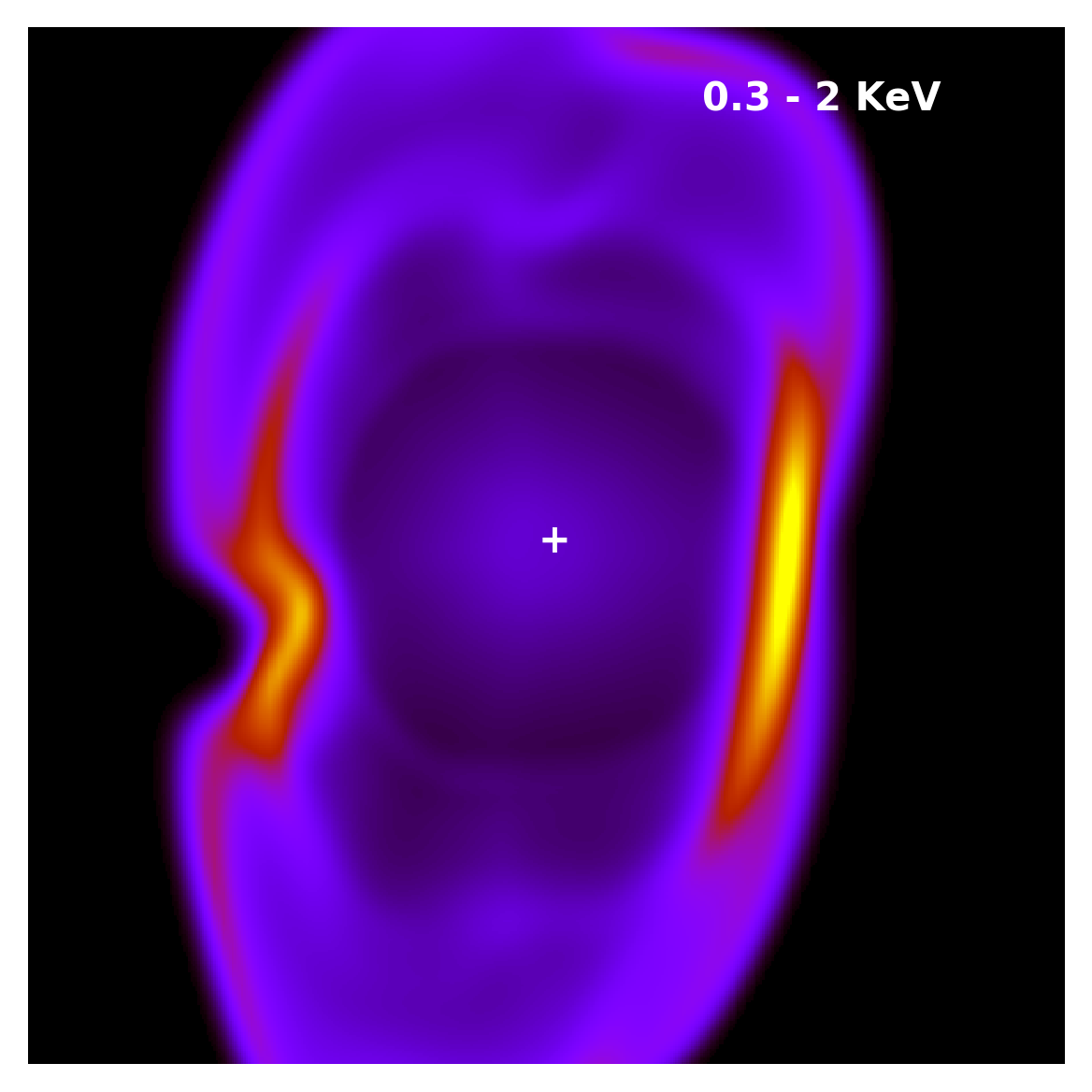}
	\includegraphics[width=.138\textwidth]{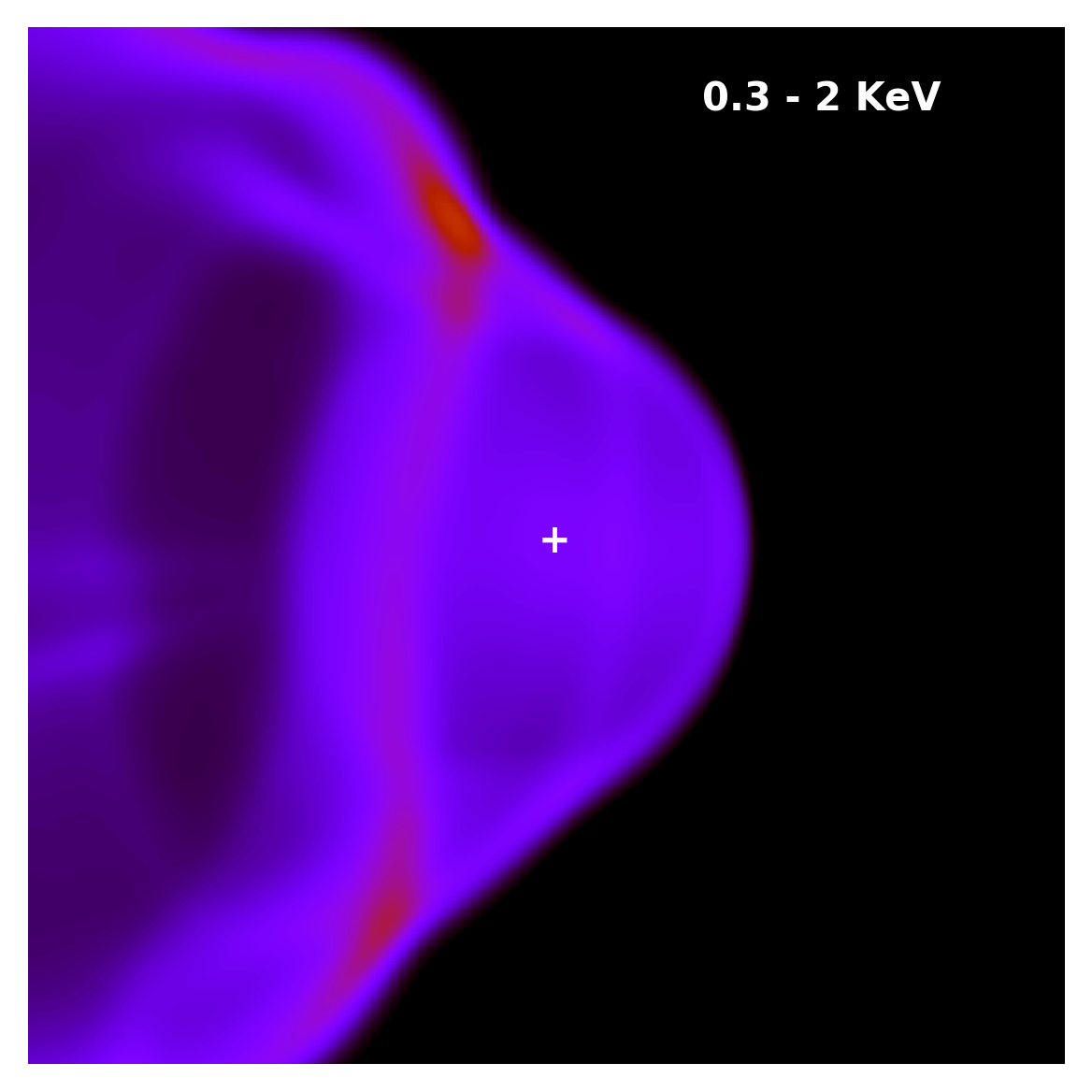}
	\includegraphics[width=.138\textwidth]{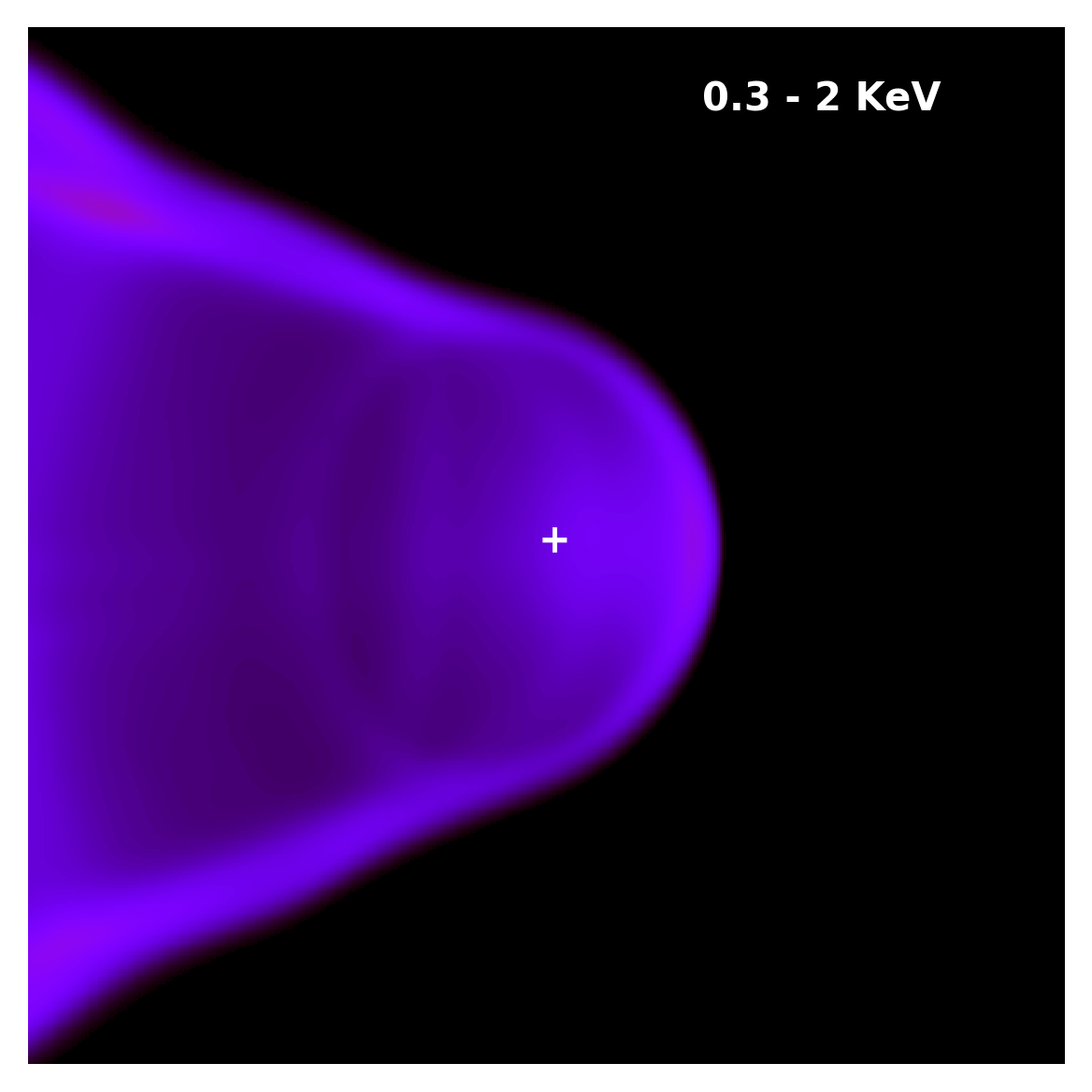}
	\includegraphics[width=.138\textwidth]{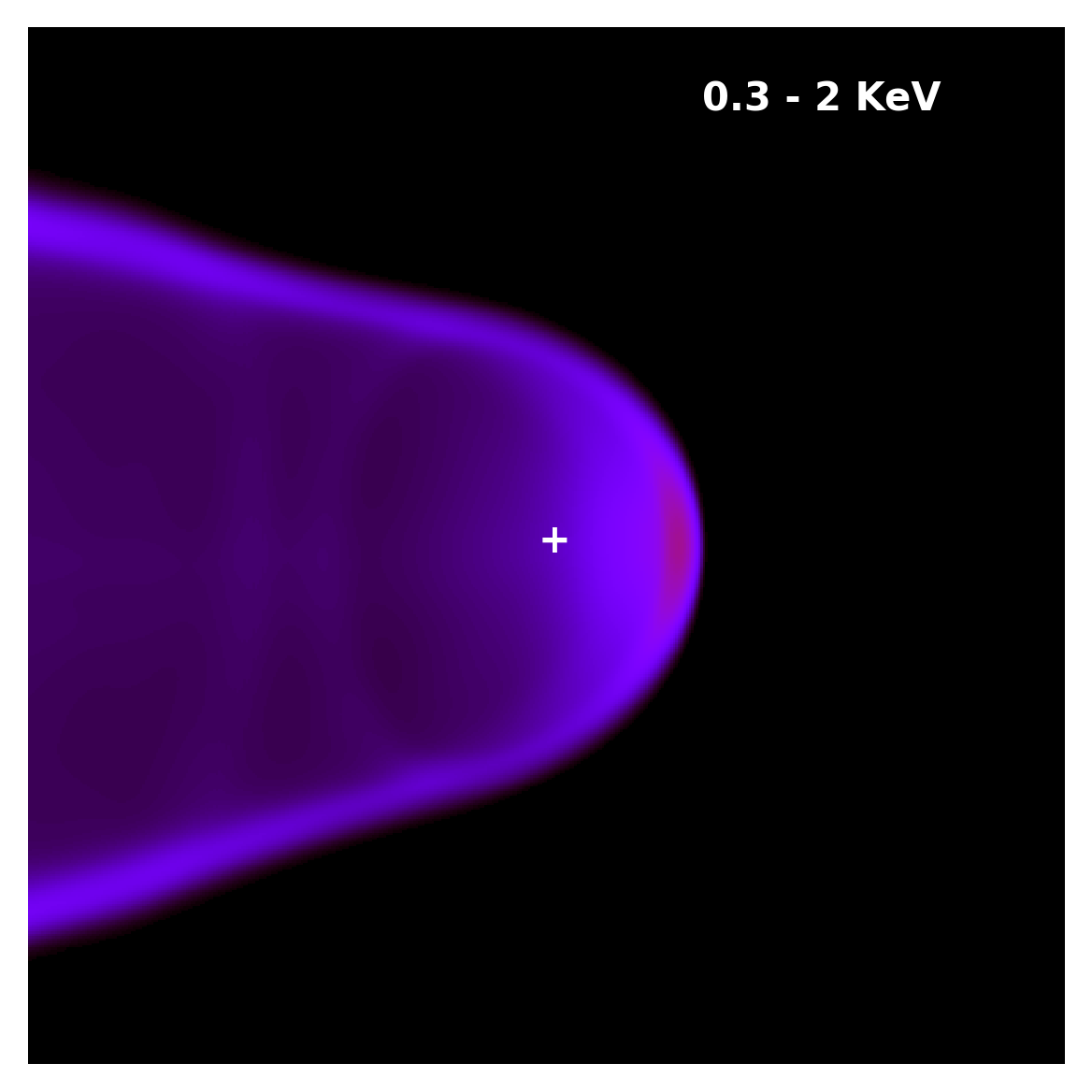}
	\includegraphics[width=.138\textwidth]{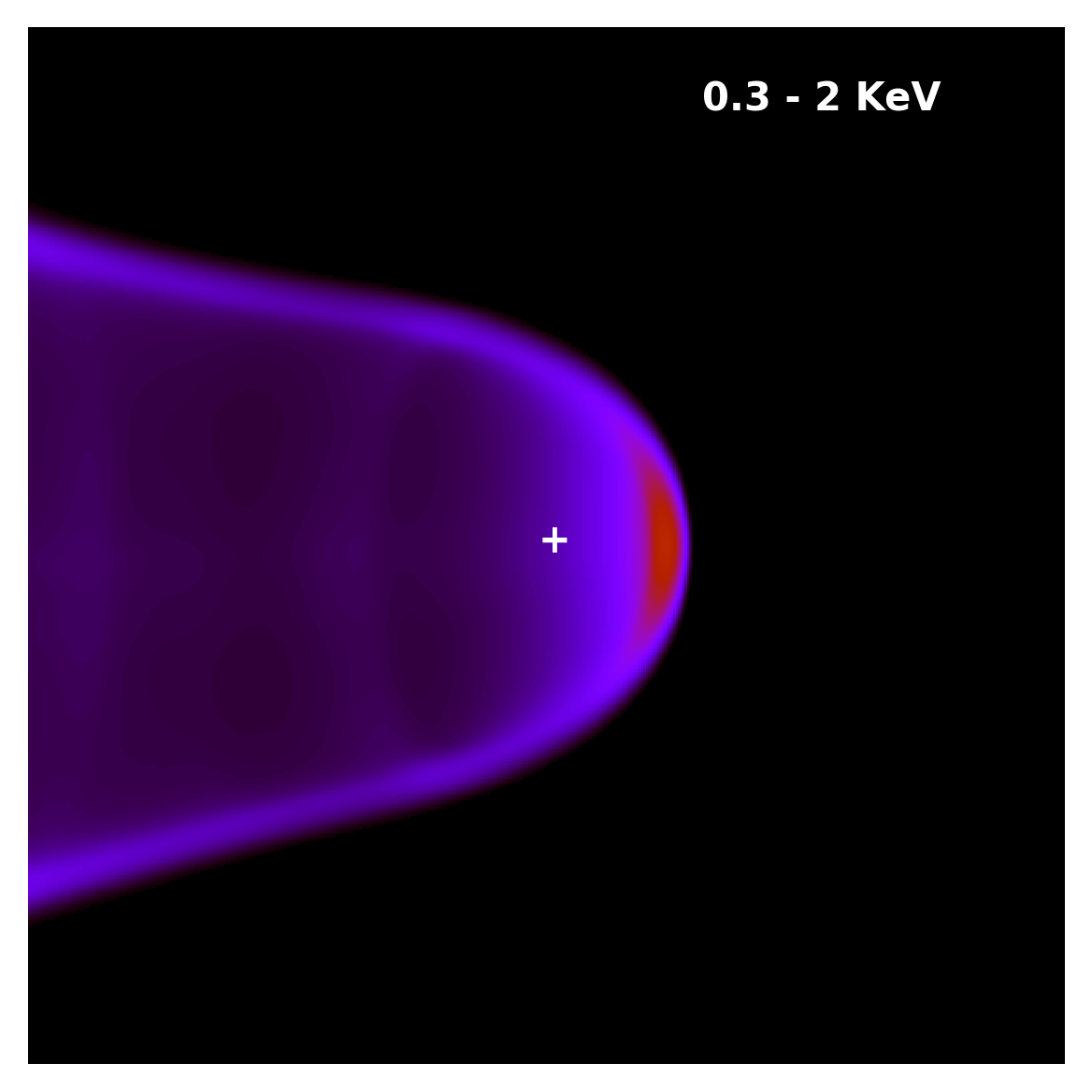}
	\includegraphics[width=.138\textwidth]{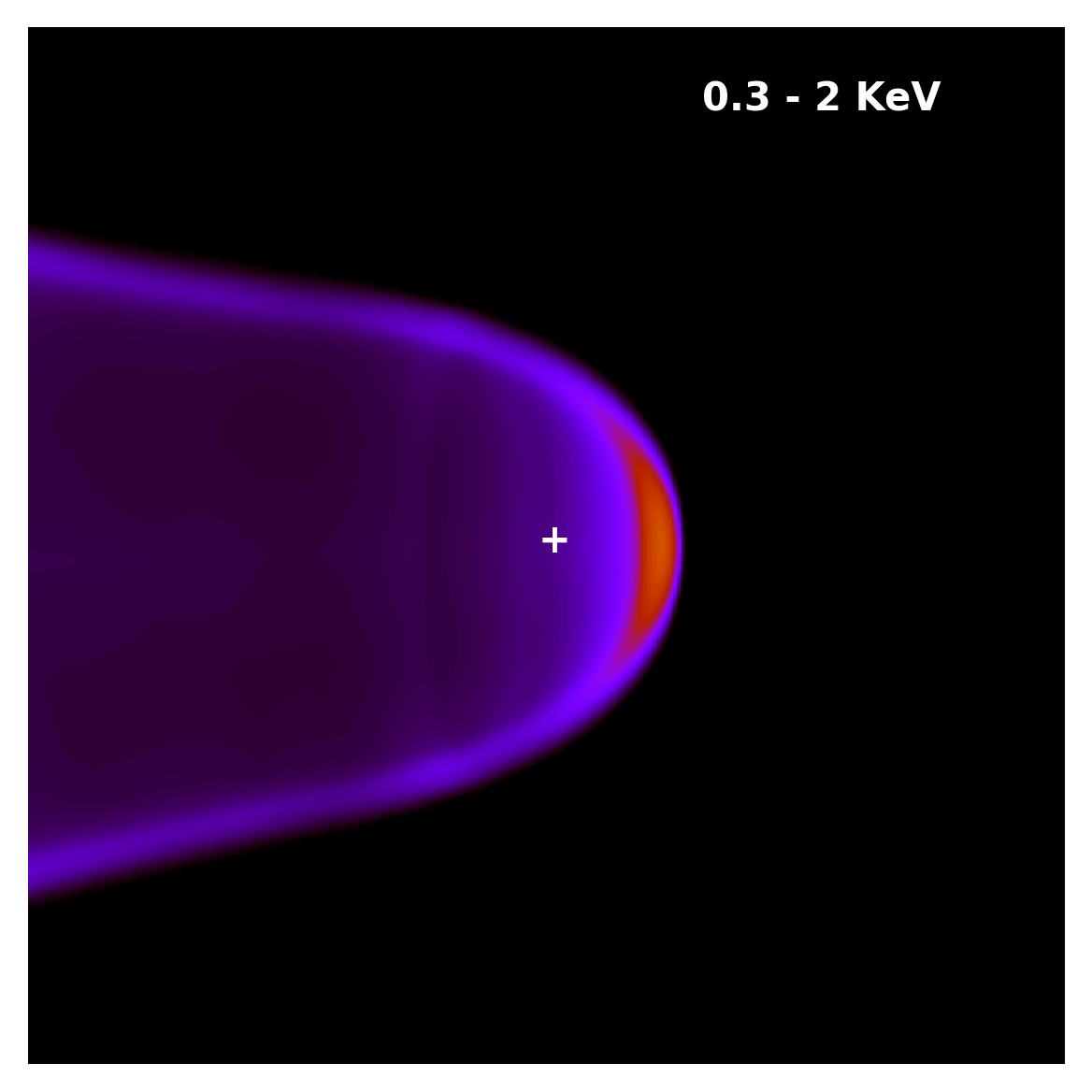}
	\includegraphics[width=.138\textwidth]{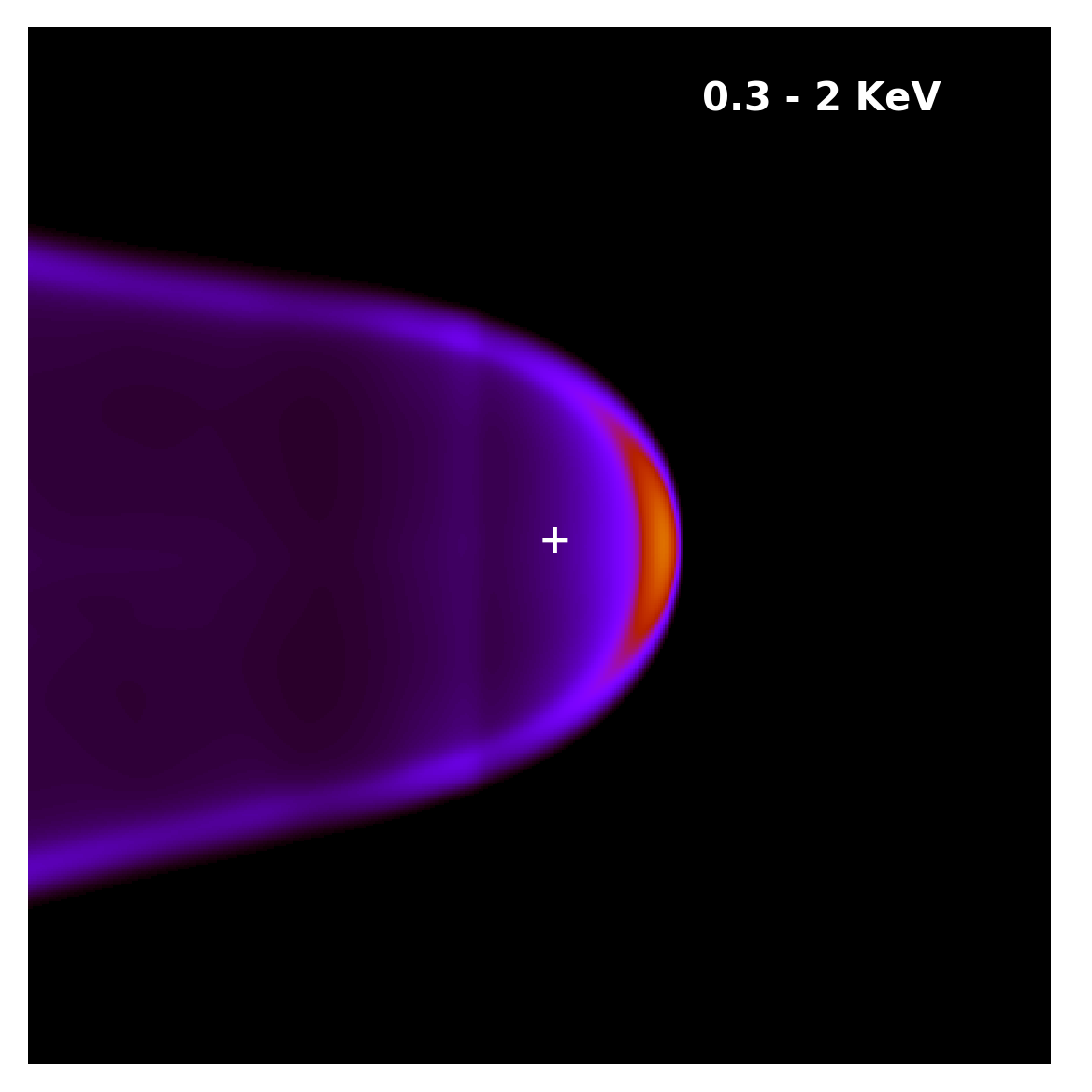} \\
	
	\caption{A snapshot from the Z01 simulation was used to generate synthetic images of $\zeta$~Ophiuchi at angles from $0^{\circ} - 90^{\circ}$ with respect to the direction of stellar motion. The first row shows the 24$\mu$m emission, second row shows the 70$\mu$m emission, third row shows the H$\alpha$ emission, fourth row shows the Emission Measure, the fifth row shows the 6Ghz Radio, and the sixth row shows the soft X-ray emission. From left to right, the projection angles are $0^{\circ}$, $15^{\circ}$, $30^{\circ}$, $45^{\circ}$, $60^{\circ}$, $75^{\circ}$, and $90^{\circ}$ with respect to the x-axis. Each image is centred on the star (black/white cross) and is  roughly 1.0$\times$1.0 parsec.}
	\label{angles_zeta}
\end{figure*}

Fig.~\ref{angles_zeta} shows synthetic emission maps from the Z01 simulation projected at angles 0$^\circ$-90$^\circ$ with respect to the velocity vector of the star (from left to right), and for the observational tracers (from top to bottom) 24\,$\mu$m thermal dust emission, 70\,$\mu$m thermal dust emission, H$\alpha$, Emission Measure, 6\,GHz Bremsstrahlung flux and soft X-rays (0.3-2\,keV).

\end{document}